\documentclass[
twocolumn,
amsmath,amssymb,
aps,
longbibliography
]{revtex4-2}
\usepackage[utf8]{inputenc}
\usepackage{amsmath}
\usepackage{amsfonts}
\usepackage{amssymb}
\usepackage{braket}
\usepackage{graphicx}
\usepackage{float}
\usepackage[percent]{overpic}
\usepackage{tikz}
\usepackage{bm}
\usetikzlibrary{arrows,decorations.pathmorphing,backgrounds,positioning,fit,petri,decorations.pathreplacing,decorations.markings}
\usepackage{xcolor}
\definecolor{mygreen1}{rgb}{0, 0.4, 0}
\usepackage[margin=0.7in]{geometry}

\usepackage{pifont}
\usepackage{hyperref}
\hypersetup{
	colorlinks=true,
	linkcolor=blue,
	filecolor=magenta,
	urlcolor=cyan,
	citecolor=blue
}

\tikzset{
	on each segment/.style={
		decorate,
		decoration={
			show path construction,
			moveto code={},
			lineto code={
				\path [#1]
				(\tikzinputsegmentfirst) -- (\tikzinputsegmentlast);
			},
			curveto code={
				\path [#1] (\tikzinputsegmentfirst)
				.. controls
				(\tikzinputsegmentsupporta) and (\tikzinputsegmentsupportb)
				..
				(\tikzinputsegmentlast);
			},
			closepath code={
				\path [#1]
				(\tikzinputsegmentfirst) -- (\tikzinputsegmentlast);
			},
		},
	},
	mid arrow/.style={postaction={decorate,decoration={
				markings,
				mark=at position .5 with {\arrow[#1]{stealth}}
	}}},
}

\definecolor{myblue1}{rgb}{0.9, 0.9, 1}

\begin{document}
	
	\title{Excitations in the higher lattice gauge theory model for topological phases I: Overview}
	\author{Joe Huxford and Steven H. Simon}
	\affiliation{Rudolf Peierls Centre for Theoretical Physics, Clarendon Laboratory, Oxford OX1 3PU, UK}
	
	\begin{abstract}		
		 In this series of papers, we study a Hamiltonian model for 3+1d topological phases introduced in [Bullivant et al., \textit{Phys. Rev. B}, 2017], based on a generalisation of lattice gauge theory known as ``higher lattice gauge theory". Higher lattice gauge theory has so called ``2-gauge fields" describing the parallel transport of lines, in addition to ordinary 1-gauge fields which describe the parallel transport of points. In this series we explicitly construct the creation operators for the point-like and loop-like excitations supported by the model. We use these creation operators to examine the properties of the excitations, including their braiding statistics. These creation operators also reveal that some of the excitations are confined, costing energy to separate that grows linearly with the length of the creation operator used. This is discussed in the context of condensation-confinement transitions between different cases of this model. We also discuss the topological charges of the model and use explicit measurement operators to re-derive a relationship between the number of charges measured by a 2-torus and the ground-state degeneracy of the model on the 3-torus. From these measurement operators, we can see that the ground state degeneracy on the 3-torus is related to the number of types of linked loop-like excitations. This first paper provides an accessible summary of our findings, with more detailed results and proofs to be presented in the other papers in the series. 
	\end{abstract}
	
		\maketitle
	
	\tableofcontents

		\section{Introduction}

		Outside of the phases of matter described by Landau symmetry breaking classification \cite{Landau}, there exist so-called topological phases of matter \cite{Wen1989, Wen1990a, Wen2013}. These topological phases, which include the celebrated fractional quantum Hall systems \cite{Tsui1982, Wen1990, Stern2008, Chakraborty1995, Sarma1997}, are characterised by long-range entanglement between their local degrees of freedom \cite{Wen2013, Chen2013, Chen2010}. While the fact that these phases cannot be described by symmetry breaking is itself interesting, topological phases can also possess rather unique properties as a result of this long-range entanglement. For example, these long-range entangled topological phases may have a ground state degeneracy even in the absence of additional symmetry \cite{Mesaros2013, Kitaev2003} (when also considering phases with enforced symmetry, the classification of topological phases becomes more rich and includes so-called symmetry protected and symmetry enriched topological phases \cite{Chen2013}). This ground-state degeneracy depends on the topology of the manifold on which the topological phase is placed (e.g., such a phase may have no ground state degeneracy on the sphere, but have a degeneracy on the torus), with this degeneracy being resistant to local perturbations \cite{Mesaros2013, Kitaev2003}. This feature may allow such topological phases to serve as quantum memories \cite{Kitaev2003, Dennis2002, Terhal2015, Brown2016}, because encoding information in the topologically protected degenerate subspace makes the information resistant to local noise \cite{Kitaev2003, Bridgeman2016}.

		Further intriguing properties of the long-ranged entangled topological phases are revealed when we consider excitations. In 2+1d, the entanglement structure allows these phases to support \textit{anyonic} excitations, which are generalizations of the more familiar bosons and fermions. Moving two anyons around each-other can induce non-trivial transformations, even at large distances \cite{Leinaas1977, Wilczek1982, Arovas1984} (for the interested reader, we note that there are many works giving pedagogical introductions to anyon physics, such as Ref. \cite{Pachos2012}, Ref. \cite{Rao1992} and Appendix E in Ref. \cite{Kitaev2006}). Because these transformations (called braiding relations) do not depend on local details, it is believed that these excitations can be used for fault-tolerant quantum computation \cite{Nayak2008, Lahtinen2017}, should sufficiently stable phases and excitations be constructed. In three spatial dimensions, while any point-like excitations must be fermionic or bosonic \cite{Doplicher1971, Doplicher1974, Nayak2008, Rao1992}, topological phases can admit loop braiding, such that point-like or loop-like excitations transform non-trivially when passed \textit{through} a loop-like excitation \cite{Wang2014, Alford1992}. This can be thought of as a generalization of the Aharanov-Bohm effect \cite{Ehrenberg1949, Aharanov1959}, where braiding a point-like electron around a loop- or string-like magnetic flux tube results in a phase depending on the magnetic field enclosed.

		In order to provide a setting where the unique properties of topological phases can be studied in detail, it is convenient to use exactly solvable toy models \cite{Kitaev2003, Lin2014, Lin2021, Levin2005}. While these models may not resemble those used to describe real materials \cite{Lin2021}, or only describe the renormalization group fixed point of their phase \cite{Cheng2017}, they provide representatives for a large class of phases of matter \cite{Lin2021}. This means that such constructions can be used to probe (and attempt to classify \cite{Levin2005}) which kinds of phases can exist. The toy models have Hamiltonians that are constructed out of commuting projector operators, which allows the quasiparticle excitations to be found exactly. Of these constructions for 2+1d topological phases, two of the most successful are the Levin-Wen string-net model \cite{Levin2005} and Kitaev's Quantum Double model \cite{Kitaev2003} (which is related to discrete gauge theory that had priorly been discussed in Refs. \cite{Bais1992} and \cite{WildPropitius1999}). The Kitaev Quantum Double class of models includes the toric code as its simplest case, which appears to have a practical application as a robust way to store qubits \cite{Kitaev2003}. Indeed one approach to building quantum computers uses so-called surface codes, which take inspiration from the toric code \cite{Fowler2012} and which have recently been experimentally realized on a small scale \cite{Andersen2020, Satzinger2021ReducedScience}. The string-net construction is more general than Kitaev's Quantum Double model, and has been conjectured to cover all phases that can be represented by commuting projector models in 2+1d in the absence of an additional symmetry \cite{Heinrich2016}, when generalized appropriately from the original construction in Ref. \cite{Levin2005} (see Refs. \cite{Lin2014, Lin2021, Hahn2020, Lake2016, Runkel2020} for such generalizations). In both of these classes of models, it is well understood how to find the ground-state degeneracy \cite{Kitaev2003, Y.Hu2012} and the properties of the excitations, such as braiding statistics \cite{Kitaev2003, Lin2021, Levin2005}.

		In the 2+1d commuting projector models, a useful way of obtaining information on the underlying topological theory is to find the operators, known as ribbon operators, that create and move the quasiparticle excitations \cite{Kitaev2003}. This approach was used in Ref. \cite{Kitaev2003} to study the excitations in Kitaev's Quantum Double model, and in Ref. \cite{Levin2005} for the string-net model. As well as classifying the quasiparticles, these ribbon operators can be used to find the braiding relations of the quasiparticles, by taking appropriate commutation relations of the ribbon operators. Furthermore, in Ref. \cite{Bombin2008} a method was developed for constructing operators to measure topological charge, which is a conserved charge that can exist without symmetry, by using closed ribbon operators. By applying this method to a modified version of the Quantum Double model which describes a condensation-confinement transition, the charges which condense and the charges that confined during the transition were identified \cite{Bombin2008}. It is clear then that these ribbon operators provide a wealth of information about the topological phase under study.

		The models that we have discussed so far describe topological phases in two spatial dimensions. However, there are also many toy models for topological phases in three spatial dimensions. Existing commuting projector Hamiltonian models include the twisted gauge theory model \cite{Wan2015, Wang2015, Bullivant2019}, which is a generalized version of the Quantum Double model in 3+1d and is based on the Dijkgraaf-Witten topological field theory \cite{Dijkgraaf1990}; a class of models developed from Unitary G-crossed Braided Fusion Categories (UGxBFCs) \cite{Williamson2017}; the Walker-Wang models \cite{Walker2012, Keyserlingk2013, Chen2015, Wang2017}, which are 3+1d generalizations of the Levin-Wen string-net models \cite{Walker2012}; and the higher lattice gauge theory models \cite{Bullivant2017,Delcamp2018, Bullivant2020,Bullivant2020b}, based on a generalization of lattice gauge theory (and related to the Yetter quantum field theory \cite{Yetter1993}). For the twisted gauge theory model in particular, there has been significant study of the properties of the ground state \cite{Wan2015} and the excitations, including their braiding properties \cite{Wang2015, Bullivant2019}. However, the general approach to studying these 3+1d models has been different from the approach used for models in two spatial dimensions. While in the 2+1d case, the use of ribbon operators to obtain the properties of the excitations is common, in the 3+1d case an explicit construction of the ribbon and membrane operators (the higher dimensional counterparts to ribbon operators, which produce loop-like excitations) can be difficult. There are some examples of such explicit constructions for the twisted gauge theory models in three spatial dimensions, such as in Refs. \cite{Else2017} and \cite{Jiang2014}, but less so for other models. Instead, indirect methods like dimensional reduction \cite{Wang2014, Wang2015} and tube algebras \cite{Bullivant2019} are often used. These methods are certainly useful, but seem to offer a less complete picture of the excitations than a direct construction. Given the success of ribbon operator approaches in 2+1d, and these examples of membrane operators in the 3+1d twisted gauge theory model, we would like to be able to apply similar approaches to other 3+1d models. In this work we will do precisely that, with one of the models discussed above.

		In this series of papers we study a model \cite{Bullivant2017} based on higher lattice gauge theory \cite{Pfeiffer2003, Baez2010}, which can be defined in arbitrary dimension but which we will study in two and three spatial dimensions. Higher lattice gauge theory is a generalization of lattice gauge theory, where there is a second gauge field which describes the parallel transport of the ordinary 1-gauge field across surfaces. This type of higher gauge theory (both on the continuum and in the lattice) has seen significant prior study in the context of topological phases. In Refs. \cite{Gukov2013, Kapustin2014}, related TQFT constructions were used to describe confinement in regular gauge theories, while in Ref. \cite{Kapustin2017} the corresponding TQFT was treated as a theory in its own right. In Ref. \cite{Bullivant2017}, a Hamiltonian model was constructed which realizes higher lattice gauge theory, in the same way that Kitaev's Quantum Double model \cite{Kitaev2003} realizes lattice gauge theory. Ref. \cite{Bullivant2017} also explored several of the properties of these Hamiltonian models. For example, the ground state degeneracy was given in terms of the partition function of a topological quantum field theory (TQFT), the Yetter TQFT, and also explicitly computed for some examples. Then in Ref. \cite{Bullivant2020}, the excitations were studied using a tube algebra approach, through which the loop-like excitations in the model were classified and the simple types were counted. Furthermore, it was shown in Ref. \cite{Bullivant2020} that there is a relationship between the number of types of elementary excitation and the ground-state degeneracy of the model on a 3-torus. In addition, it was shown in Ref. \cite{Bullivant2018} that higher gauge theory could lead to loop-like excitations with non-trivial loop braiding statistics, and the associated representations of the loop braid group were found (the loop braid group describes the motions of loops \cite{Baez2007a, Damiani2017}), although this was not done in the Hamiltonian model but instead from more geometric reasoning about the fluxes and gauge transforms involved.

		However, until now there was no explicit construction of these excitations in the Hamiltonian model using ribbon and membrane operators, and the braiding statistics of the excitations in the Hamiltonian model have not been found. We aim to address this, and describe some of the other features of the excitations, in this work. To do so, we will explicitly construct the membrane and ribbon operators for the Hamiltonian model \cite{Bullivant2017} and use them to find the other properties of the excitations. We note that these models are particularly interesting to study in this way because they share a similar structure to lattice gauge theory models, which helps with the difficult task of directly constructing ribbon and membrane operators, and yet still exhibit features not seen in ordinary (1-gauge) gauge theory models, as we elaborate on shortly.

		Our main results in the 3+1d case are as follows. We construct the membrane and ribbon operators which produce the excitations for this model, in a broad subset of the higher lattice gauge theory models. We find that the basic excitations are either loop-like or point-like and that some of the point-like excitations are confined, with an energy cost to separate a particle from its anti-particle that grows linearly with the length of ribbon used to do so. This is described in terms of a condensation-confinement transition between different higher lattice gauge theory models, during which some of the loop-like excitations condense out, becoming topologically trivial. Then, using our direct construction of the ribbon and membrane operators, we find the (loop)-braiding relations of our excitations in terms of simple group-theoretic quantities. We find that the braiding is generally non-Abelian, so that our relations involve more than a simple accumulation of phase. Instead the excitations generally have an internal space, which can transform non-trivially under braiding, in addition to a conserved topological charge, which is not changed by the braiding. This topological charge is of significant interest, and so we also consider the charges present in the higher lattice gauge theory model. Extending the methods of Ref. \cite{Bombin2008} to 3+1d, we construct operators that can measure the topological charge present in a region. These measurement operators are made from closed membrane and ribbon operators applied on the boundary of the region in question, and the topology of this boundary determines what types of charge we can resolve. For example, the charge associated to point-like objects is measured by putting a sphere around that charge, similar to Gauss's law for electric charge. On the other hand, loop-like excitations require a surface with handles in order to detect their loop-like character. This is similar in concept to the tube algebra methods used in Ref. \cite{Bullivant2020}, which classify the boundary conditions of unexcited regions of space. Indeed, just like Ref. \cite{Bullivant2020} we find that the number of different charges that can be measured by a torus is equal to the ground state degeneracy of the model when placed on a 3-torus.

		\subsection{Structure of this series}
		
		Due to the large amount of algebra needed to fully describe and prove our results, we have divided the discussion into three parts. This work is the first of the series, so we feel that it would be valuable to provide a brief guide to the set of articles. In this work, we will provide a more informal and descriptive overview of our main results for the 3+1d model. We suggest that a general reader consider this work, before looking through the other papers in the series if they are interested in more detail, or are specifically interested in the 2+1d model.

		In the second paper \cite{HuxfordPaper2}, we consider the 2+1d version of the higher lattice gauge theory model. Perhaps the most interesting feature of this model is that, despite being in 2+1d, this model still hosts loop-like excitations. In addition to studying the topological content of the model, we demonstrate that in certain cases the loop-like excitations can be described as domain walls between different symmetry sectors. This idea that the 2+1d models can describe symmetry enriched topological phases is further expanded on when we map a subset of these models to another construction for such phases, the symmetry enriched string-net model from Ref. \cite{Heinrich2016}.

		In the final paper \cite{HuxfordPaper3}, we return to the 3+1d model to provide more detailed results. This includes an explicit presentation of the commutation relations of the ribbon and membrane operators that give us the braiding relations. We also directly construct the measurement operators for topological charge within a torus and a sphere, and find the point-like charge of the simple excitations of the model (including the loop-like ones).

		\subsection{Structure of this paper}
		
		In the rest of the introduction, we describe the model introduced in Ref. \cite{Bullivant2017} and introduce other important concepts from existing work. To introduce the model, we first discuss lattice gauge theory in Section \ref{Section_Lattice_Gauge_Theory}, then higher lattice gauge theory in Section \ref{Section_HLGT}. In Section \ref{Section_Hamiltonian_Model} we use these ideas to motivate the Hamiltonian model from Ref. \cite{Bullivant2017}, and explicitly define the model. Throughout the paper, we will be placing different conditions on the model to examine special cases, which we describe in Section \ref{Section_Special_Cases}. In the last section of the introduction, Section \ref{Section_Braiding_3D}, we describe what we mean by braiding statistics in 3+1d.

		After discussing the background for our work, we then move on to a description of our results. We start in Section \ref{Section_Properties_From_Gauge_Theory} by using ideas from gauge theory to motivate the excitations and some of their properties. Then in the rest of the paper we will look at how these properties (and additional features) arise in the lattice model. In Section \ref{Section_Part_One_Excitations} we construct the operators that create and move the various excitations. Then in Section \ref{Section_Part_One_Condensation_Confinement} we describe how some of the excitations are confined, with a cost to separate two of these particles that grows linearly with separation. We explain how, at least in certain cases, this can arise from a ``condensation-confinement" transition between different higher lattice gauge theory models. After this, in Section \ref{Section_Part_One_Braiding} we present the braiding relations of the excitations, describing the result of exchanging our excitations in various ways. Finally in Section \ref{Section_Part_One_Topological_Sectors} we discuss topological charge, a type of conserved charge realised by topological phases, and point out a relation between the allowed values of this charge and the ground-state degeneracy of our model.

		\subsection{Lattice gauge theory}
		\label{Section_Lattice_Gauge_Theory}
		
		In this section, we review lattice gauge theory and Kitaev's Quantum Double model. This material may be familiar to some readers, who may still wish to read it to familiarise themselves with the notation we use throughout. To describe a continuum gauge theory, the key ingredients are matter fields, gauge fields (which describe parallel transport of the matter fields) and gauge symmetry. As an example, we can consider conventional electrodynamics. In this case the matter fields describe charges, such as electrons, which couple to the usual gauge field. This gauge field describes parallel transport of the charged matter via the Aharanov-Bohm effect \cite{Ehrenberg1949,Aharanov1959}. Finally there is a gauge symmetry, which gives us gauge transforms that appear to change the values of the gauge and matter fields. However, states that are related by gauge transforms represent the same physical state and are simply different descriptions of the same physical system.

\begin{figure}[h]
			\begin{center}
				\includegraphics{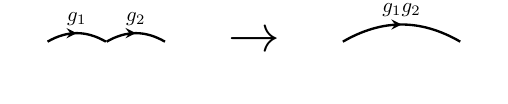}
				\caption{Composition of paths is described by group multiplication}
				\label{composition_paths}
			\end{center}
		\end{figure}

		While gauge theories are typically constructed in the continuum, the same ideas can be applied to a lattice gauge theory \cite{Montvay1994}. The first thing to consider is the physical space on which we consider the model, that is the lattice. Throughout this paper, we will use lattice in the more informal sense, referring to a collection of vertices and edges (and later plaquettes), without requiring a repeating structure (i.e., we consider a graph). Then we have to place the other ingredients of gauge theory into this discrete setting. The matter field is placed on the vertices of the lattice \cite{Montvay1994}, while the gauge field is placed on the (directed) edges of the lattice and determines the result of transport of matter along the edges \cite{Montvay1994}. However, for the purposes of this paper, we will not include matter as a dynamical field, and charges are instead represented by violations of the gauge symmetry. This leaves us only with the gauge field, which is valued in some discrete group $G$. The group structure means that two paths that lie end-to-end can be composed, with the field label of the resulting path given by group multiplication of the labels of the constituent paths \cite{Montvay1994}, as shown in Figure \ref{composition_paths}. If we wish to combine two paths that point in opposite directions, we must first reverse the orientation of one of them, so that they align. The group element associated to the reversed path is then the inverse of original group element. For example, if in Figure \ref{composition_paths} the path labelled $g_2$ pointed in the opposite direction, the combined path would instead have label $g_1g_2^{-1}$.
		
		\subsubsection{Gauge transforms}
		\label{LGT_Gauge_Transforms}
		
\begin{figure}[h]
			\begin{center}
				\includegraphics{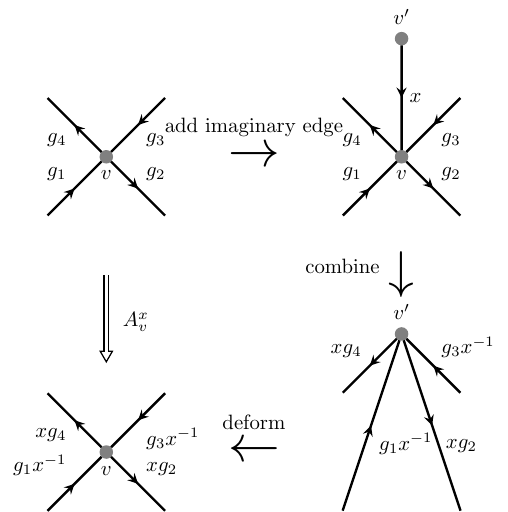}
			\end{center}
			\caption{The gauge transform on a vertex is equivalent to adding an imaginary edge at that vertex and then combining this edge into the diagram, or equivalently transporting the vertex along that edge.}
			\label{vertex_transform}
		\end{figure}

		Having considered the fields present in the model, we now look at the gauge symmetry. The gauge symmetry is included through a set of local operators that each act on the degrees of freedom near a vertex. Each operator is labelled by the vertex it acts on and an element of $G$, so that the gauge transform for a vertex $v$ and element $x \in G$ is denoted by $A_v^x$ \cite{Kitaev2003}. This transform affects the edges surrounding it by pre-multiplying the group element on each adjacent edge by $x$ if the edge is outgoing, and post-multiplying the element by $x^{-1}$ if the edge is incoming \cite{Kitaev2003, Montvay1994, Wilson1964, Durhuus1980}. We give an example of the action of the vertex transform in Figure \ref{vertex_transform}, from which we can see that this action is equivalent to adding an imaginary edge, labelled by $x$, to the vertex $v$ and parallel transporting the entire vertex along it. 
		
	This geometric picture reveals two important properties of the vertex transform. Firstly, the vertex transform only affects paths that start or end at that vertex, because a path passing through the vertex will travel both ways along the added edge. For example, in the top-left image in Figure \ref{vertex_transform} the path entering the vertex $v$ from the lower left and exiting $v$ through the lower right is labelled by the product $g_1 g_2$. In the bottom-left image of Figure \ref{vertex_transform}, which represents the state after the gauge transform, the same path is labelled by $g_1 x^{-1} x g_2=g_1g_2$. That is, the path label is unchanged by the gauge transform, because the path does not start or terminate at the vertex $v$. Secondly, note that applying two gauge transforms to the same vertex is the same as parallel transporting along two edges in sequence. This is equivalent to parallel transport of the vertex across a single path composed of the two edges, and so is the same as applying a single vertex transform with a label obtained by combining the labels of the two edges (and so combining the labels of the two original transforms). If we first apply a vertex transform $A_v^g$ and then another transform $A_v^h$, the label of the combined path introduced by the transforms is $hg$ (it is $hg$ rather than $gh$, due to the fact that the vertex is parallel transported against the direction of the edge, as seen in Figure \ref{vertex_transform}). Therefore we must have that $A_v^h A_v^g =A_v^{hg}$ \cite{Kitaev2003}.
		
		\subsubsection{Gauge-invariants}
		\label{Section_Gauge_Invariants_LGT}
\begin{figure}[h]
			\begin{center}
			\includegraphics{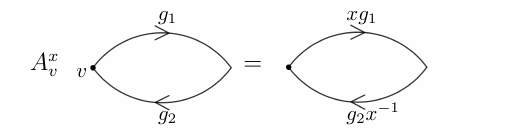}
				
				\caption{We consider the effect of a vertex transform $A_v^x$ on a closed loop starting at the vertex $v$. The path label in this case goes from $g_1g_2$ to $xg_1g_2x^{-1}$. That is, the path label of the closed loop is conjugated by $x$ under the action of the transform.}
				\label{vertex_transform_closed_loop}
				
			\end{center}
		\end{figure}

		Because states related by gauge transforms are equivalent, any physical quantity should be gauge-invariant. We can construct these gauge-invariant quantities from the closed loops of our lattice \cite{Wilson1964, Durhuus1980}. Under a gauge transform, the group element assigned to a closed loop is at most conjugated by the vertex transforms \cite{Durhuus1980}. Therefore the conjugacy class of that label is a gauge invariant quantity. For example, consider Figure \ref{vertex_transform_closed_loop}, which shows the action of a vertex transform $A_v^x$ on a closed loop starting at $v$. Initially the group element associated to the closed loop is $g_1g_2$. After applying the vertex transform it becomes $xg_1g_2x^{-1}$. This indicates that the group element is not generally a gauge-invariant quantity, but its conjugacy class is.
		
		As an example of the importance of such closed loops, we can consider the case of electromagnetism. Here we have a $U(1)$ gauge symmetry, so that the edges in our lattice would be labelled by phases. There is a physical process where we take a charge $q$ around a closed loop in the presence of a magnetic field described by a vector potential $\vec{A}$. In the continuum theory this leads to the Aharanov-Bohm effect \cite{Ehrenberg1949, Aharanov1959}, where the wavefunction accumulates a phase of $\theta = q \oint \vec{A} \cdot \vec{dl}$. This phase is the label we would give our closed loop in the lattice model. Using Stoke's theorem, the Aharanov-Bohm phase can be related to the magnetic flux through the surface enclosed by the loop. The phase is a gauge invariant quantity, as is required by the fact that this phase can be measured in interference experiments and thus is a physical quantity.

		These gauge-invariant quantities allow us to differentiate between physically distinct states. For instance, many gauge configurations can be reduced to the trivial configuration, where every edge is labelled by $1_G$ (the identity in the group $G$), by applying gauge transforms. The state where the edges are all labelled by the identity describes trivial parallel transport, and so the states related to this trivial state by gauge transforms must also be trivial. This indicates that in these equivalent states the apparently non-trivial edge labels only describe a change of basis, rather than a physical change under parallel transport. On the other hand, if a state has any closed loops with non-trivial path label, then (because the conjugacy classes of closed path labels are gauge invariant) the state cannot correspond to this trivial state. Therefore, in such a state the parallel transport across the edges must describe both a change of basis and some physical ``flux", analogous to the magnetic flux in electromagnetism, which differentiates it from the trivial case.

		\subsubsection{The quantum double model}
		\label{Section_Quantum_Double}

		Lattice gauge theory can be used to build a model for topological phases, known as Kitaev's Quantum Double model \cite{Kitaev2003}. The lattice represents the spatial dimensions of the models, while a Hamiltonian controls the time evolution. In order to construct the Hamiltonian, we first demote gauge invariance to an energetic constraint by adding an energy term to the Hamiltonian for each vertex that enforces the symmetry. We also add an energy term at each plaquette that penalizes plaquettes with non-trivial boundary paths. The Hamiltonian is \cite{Kitaev2003}
		$$H = - \sum_{\text{vertices, }v} A_v - \sum_{\text{plaquettes, }p} B_p.$$
		Here we have $$A_v = \frac{1}{|G|} \sum_{g \in G}A_v^g,$$ where the $A_v^g$ are the gauge transforms from earlier and $|G|$ is the number of elements in the discrete group $G$. $A_v$ is therefore an average over all gauge transforms at vertex $v$. The operator $A_v$ is a projector \cite{Kitaev2003}, because 
		\begin{align}
		A_v A_v &= \frac{1}{|G|^2} \sum_{g \in G} \sum_{h \in G} A_v^g A_v^h = \frac{1}{|G|^2} \sum_{g \in G} \sum_{h \in G} A_v^{gh} \notag \\
		& = \frac{1}{|G|^2}\sum_{g \in G} \sum_{gh \in G} A_v^{gh} = \frac{1}{|G|} \sum_{g \in G} A_v \notag \\
		& = A_v. \label{Equation_Quantum_Double_Vertex_Projector}
		\end{align}
		
		As a projector, $A_v$ has eigenvalues of zero and one, with the eigenvalue of one corresponding to states which are gauge symmetric at that vertex (because the gauge transforms leave such states unchanged). $A_v$ enters the Hamiltonian with a minus sign, so the gauge-invariant states are lower in energy.

\begin{figure}[h]
			\begin{center}
			\includegraphics{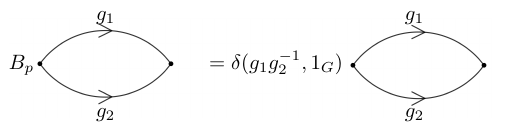}
				
				\caption{The plaquette term $B_p$ gives 1 if the closed path forming the boundary of the plaquette $p$ is $1_G$ (i.e., if it is flat) and 0 otherwise. For the example plaquette shown in this figure, where the edges are in states labelled by $g_1$ and $g_2$, that means that acting with the plaquette term gives a non-zero result only if $g_1g_2^{-1}=1_G$.}
				\label{Plaquette_term_bigon_1}
				
			\end{center}
		\end{figure}

		The other term in the Hamiltonian, $B_p$, acts on the edges around a plaquette $p$. It leaves states where the boundary of the plaquette is labelled by the identity unchanged and returns zero for other states. As an example, consider Figure \ref{Plaquette_term_bigon_1}, which illustrates the action of the plaquette term $B_p$ on a simple plaquette made from two edges (a bigon). In this case the boundary path label is given by $g_1g_2^{-1}$, and so the plaquette term returns the state if $g_1g_2^{-1}=1_G$. $B_p$ is clearly a projector just like the vertex term, with the eigenvalue of one corresponding to states with trivial flux around the plaquette (we say the plaquette satisfies flatness in these states). Again, $B_p$ enters the Hamiltonian with a minus sign, so that these trivial flux states are lower in energy. The trivial flux label $1_G$ is in a conjugacy class on its own, meaning that it is unchanged by gauge transforms. This means that the operator $B_p$ is built out of gauge-invariant quantities and therefore commutes with the gauge transforms. All of the terms in the Hamiltonian are projectors and they all commute, so this is an example of a commuting projector model. This structure to the Hamiltonian enables the model to be solved exactly. The excitations are charge-like (excitations of the vertex term), flux-like (primarily excitations of the plaquette term, though they may also excite a vertex term), or some combination of the two \cite{Kitaev2003}. These excitations are called electric if they are charge-like, magnetic if they are flux-like and dyonic if they are a combination. As we will see later, some of these properties will carry over to the higher lattice gauge theory Hamiltonian model.

		\subsection{Higher lattice gauge theory}
		\label{Section_HLGT}
		
\begin{figure}[h]
 	\begin{center}
 		\includegraphics{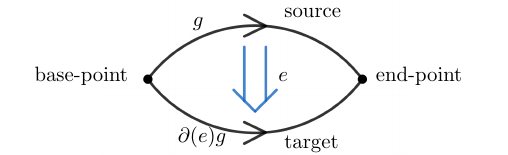}
 		\caption{Just as a path is associated to parallel transport of points, so is a surface associated to parallel transport of a path. The initial position of the path is called the source and the final position is called the target. The parallel transport of a path over a surface labelled by $e \in E$ results in the path element $g$ gaining a factor of $\partial(e)$, where $\partial$ is a group homomorphism from $E$ to $G$.}
 		\label{surface_labelled_1}
 		
 	\end{center}
 \end{figure}

 In lattice gauge theory we consider parallel transport along paths, and label paths by group elements to allow composition of paths. That is, we label geometric objects, the paths, with algebraic objects, the group elements. A natural generalization is to label more types of geometric objects. We still label the paths with elements of a group $G$ (this is the 1-gauge field, or the 1-holonomy of that path \cite{Bullivant2017}). However, we now also label the surfaces with elements in a second group, $E$. We refer to this field as the 2-gauge field. As we will see shortly, parallel transport will involve various mappings between the groups $E$ and $G$. If paths describe the parallel transport of points, then surfaces describe the parallel transport of paths, that is of the 1-gauge fields \cite{Baez2010}. We can view this pictorially as shown in Figure \ref{surface_labelled_1}. The blue double arrow on the surface enclosed by the paths represents the transport of one path (the source) into another (the target) \cite{Bullivant2017}. Both of these paths must be specified in order to give the surface a label, which is called the 2-holonomy \cite{Bullivant2017} for that surface. The two paths (source and target) both start at a common vertex, called the start-point of the surface, and end at a common vertex, called the end-point. As indicated in Figure \ref{surface_labelled_1}, the parallel transport over a surface labelled by $e$ causes the source to gain a factor of $\partial(e)$, where $\partial$ is a group homomorphism from $E$ to $G$ (i.e., a map that preserves the group multiplication), so that for $e \in E$, $\partial(e) \in G$ \cite{Baez2010}. Normally the labels of the two paths on either side of the surface are independent variables, but if the label of the source is related to that of the target by this parallel transport rule then the surface is called fake-flat. These fake-flat surfaces play an important role in the theory. Fake-flatness replaces the trivial flux condition for the Quantum Double model and will determine the low energy space in the topological model. In the rest of this section, we will therefore discuss such fake-flat surfaces unless otherwise mentioned.

\begin{figure}[h]
		 	\begin{center}
		 		\includegraphics{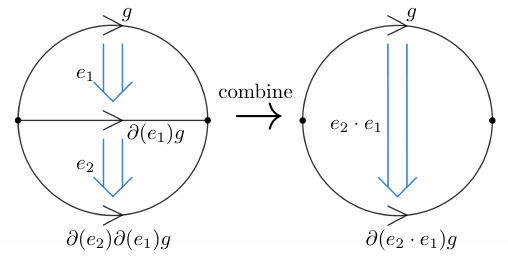}
		 		\caption{Consider two surfaces over which we can sequentially transport a path, such as the ones in the left side of the figure. In this case we first transport the top path over the upper surface (across the arrow) to the middle location (the straight path) and then over the lower surface to the bottom position. We can express the same process as transport over a single surface, made from a combination of the two individual surfaces, as shown in the right figure. This is called vertical composition of the surfaces.}
		 		\label{vertical_composition}
		 	\end{center}
		 \end{figure}

		 In the same way that we can compose paths that lie end-to-end, so may we combine adjacent surfaces. In fact, surfaces can be composed in two ways. Firstly, they may be combined vertically \cite{Pfeiffer2003, Baez2010}, as shown in Figure \ref{vertical_composition}. Vertical composition corresponds to the case where we perform two parallel transportations of a path (the top path in Figure \ref{vertical_composition}) in sequence (first moving it to the middle position in the figure and then to the bottom). We can combine these two steps to describe the two parallel transportations as parallel transport along a single, combined, surface. In order to compose the two surfaces in this way, the target of the first surface must match the source of the second one. After composition, the source of the combined surface is the source of the first surface and the target of the combined surface is the target of the second surface.
		 
		 We map vertical composition of two surfaces onto the group multiplication, with the first surface label on the right and the second on the left, following the convention in Ref. \cite{Bullivant2017}. As shown in Figure \ref{vertical_composition}, requiring the label of the bottom path to be the same on both sides of Figure \ref{vertical_composition} gives the consistency condition $\partial(e_2 \cdot e_1)= \partial(e_2)\partial(e_1)$, which is why $\partial$ must be a group homomorphism. This ensures that the effect of transporting the edge along one surface, labelled by $e_1$, and then another surface, labelled by $e_2$, is the same as transporting the edge along the combined surface (labelled by $e_2 e_1$). 
		 
\begin{figure}[h]
			\begin{center}
			\includegraphics{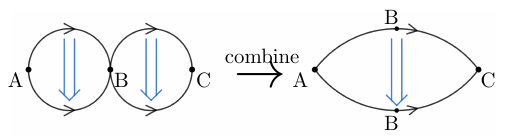}

			\end{center}
			\caption{In addition to vertical composition of surfaces, as shown in Figure \ref{vertical_composition}, we can consider horizontal composition of two surfaces that lie side by side. In this case, the individual surfaces describe parallel transport of two paths (A-B and B-C) which can be composed, while the combined surface describes the parallel transport of the paths after they have been composed (into A-C). While we show the resulting surface as a simple bigon (2-gon) for convenience, the composition does not change the shape of the constituent surfaces on the lattice, and so the point B remains on the boundary of the combined surface. The two points labelled B in the right-hand side are the same and so should be glued together.}
			\label{horizontal_composition_unlabelled}
		\end{figure}

		We may also combine the surfaces horizontally \cite{Pfeiffer2003, Baez2010}, as shown in Figure \ref{horizontal_composition_unlabelled}. This horizontal combination corresponds to the case where we have two paths lying end to end, which we can parallel transport separately. However, we can also combine the two paths into one, before transporting them across a single surface.

\begin{figure}[h]
			\begin{center}
				\includegraphics{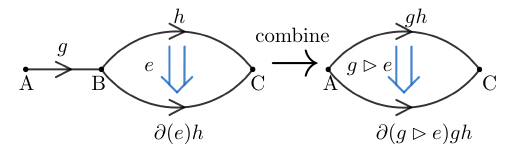}
				\caption{Here we consider ``whiskering" a surface (the bigon with base-point B and end-point C) along a path, which is described by the map $\rhd$. Just as for horizontal composition, the shape of the final surface should match the combined shape of the path and surface, though we have represented the final surface as a simple bigon. The point B remains on the boundary of the final surface, in both the source and target. Note that the base-point of the surface changes from B in the first image to A in the second. Requiring the lower path from $A$ to $ C $ to have the same label after combining the edge with the surface via whiskering gives the condition that $g\partial(e)h = \partial(g \rhd e)gh$, so that $\partial(g \rhd e)=g\partial(e)g^{-1}$. }	
				\label{whiskering_algebraic}	
			\end{center}
		\end{figure}

		As a special case of horizontal combination, we have the case where the first path is not parallel transported across any surface. This lets us combine a surface with a path. As an example, such a situation is shown in Figure \ref{whiskering_algebraic}. In Figure \ref{whiskering_algebraic}, we combine the edge that runs from A to B with a surface, by treating the edge and its inverse (the inverse is the same edge, but with reversed direction) as bounding an infinitesimally thin surface and then using horizontal composition. This process of combining a surface with a path is known as whiskering \cite{Baez2010}, and can also be thought of as moving the base-point of a surface (in the case shown in Figure \ref{whiskering_algebraic}, the base-point of the surface is initially at B, but is moved to A). Because parallel transport of objects along paths is described by the group $G$, the whiskering must be described by an action of $G$ on $E$. This action is given by a map $\rhd$ from $G$ to the endomorphisms on $E$ \cite{Bullivant2017, Baez2010} (endomorphisms are homomorphisms from a group to itself). That is, given an element $g$ of $G$, the object $g \: \rhd$ is then a map from $E$ to itself. We write $g \: \rhd$ acting on an element $e$ of $E$ as $g \rhd e$. In Figure \ref{whiskering_algebraic}, we see how this map $\rhd$ is involved in whiskering. If we move the base-point of a surface (initially B in the Figure) across an edge labelled by $g$, against the direction of the edge, then the surface label changes from $e$ to $g \rhd e$. On the other hand, moving the end-point (C in the figure), rather than the base-point, has no effect on the surface label.
		
\begin{figure}[H]
			\begin{center}
			\includegraphics{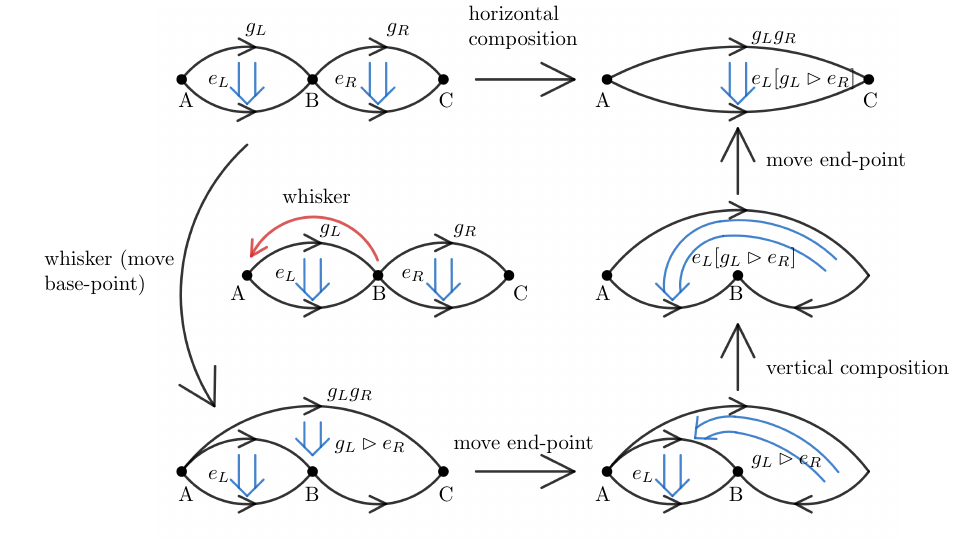}
				\caption{The requirement that the different ways of combining a diagram must be consistent means that we can express the horizontal composition of two surfaces in terms of whiskering and vertical composition. Consider the top-left image, consisting of a left and right surface with label $e_L$ and $e_R$ respectively. We wish to combine these two surfaces to obtain the surface in the top-right image. There are two ways around the diagram that lead from the top-left to the top-right image, and these are required to give the same result for the label of the final surface. The first way is horizontal composition of the two surfaces, which gives us an unknown label for the final surface that we wish to find. The second way around the diagram involves the other processes that we do have algebraic expressions for. We can therefore use this to find the label resulting from horizontal composition. The first step (represented by the downwards arrow) is to whisker the right-hand surface so that it has the same base-point (A) as the left surface. This gives the right surface a label $g_L \rhd e_R$, where $g_L$ is the label of the path from A to the original base-point B, as shown in the bottom-left image. The next step is to move the end-point of the right-surface from C to B, to match the end-point of the left surface, as shown in the bottom-right image. This has no effect on the label of the surface. Then the target of the right surface is the same as the source of the left surface (the path from A to B with label $g_L$) and so the two surfaces can be combined via vertical composition, giving a surface with label $e_L[g_L \rhd e_R]$ as shown in the centre-right image. Finally, moving the end-point from B to C (the original end-point of the right surface) gives us the same surface that we would have from horizontal composition, with label $e_L[g_L \rhd e_R]$. This is therefore the label resulting from horizontal composition.}
				
				\label{horizontal_composition_labelling}
			\end{center}
		\end{figure}

		For the diagrams that we have considered to give a consistent theory, the different ways of combining the elements of the diagram must be consistent. One consequence of this is that we can find the result of horizontal composition of two surfaces, by combining the rules for vertical composition and whiskering. Consider Figure \ref{horizontal_composition_labelling}, which shows a diagram involving the horizontal composition of two surfaces (on the top line), where the left and right surfaces are labelled by group elements $e_L$ and $e_R$ respectively. We can reproduce this horizontal composition with a series of other manipulations, which takes us the other way around the diagram. These other processes involve changing the base-point and end-point of surfaces, as well as vertical composition, all of which we already know how to perform. Applying these manipulations (as explained in Figure \ref{horizontal_composition_labelling}), we find that the label resulting from horizontal composition must be $e_L [g_L \rhd e_R]$.

\begin{figure}[H]
			\begin{center}
				\includegraphics{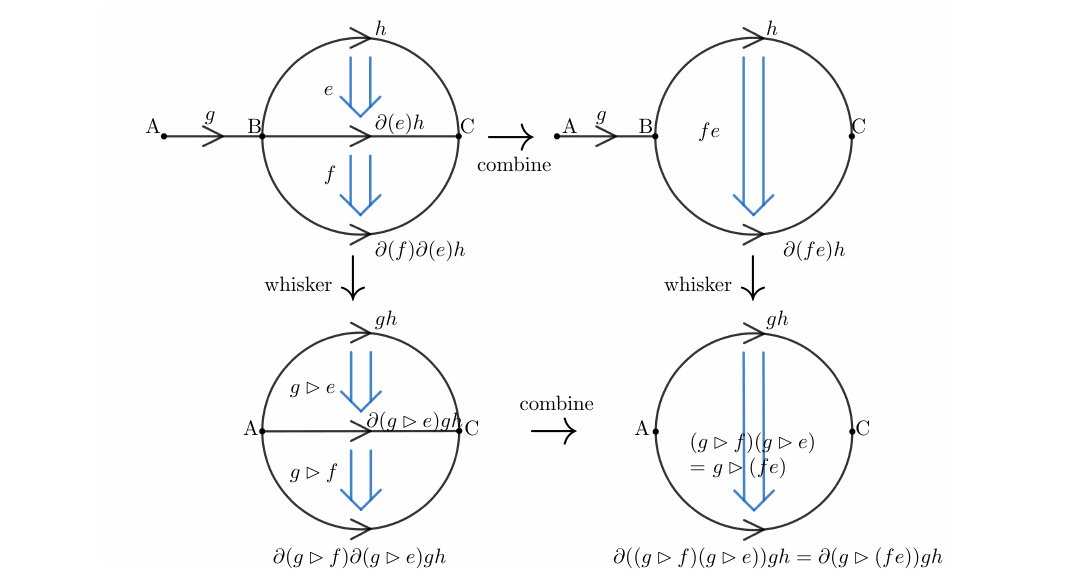}
				\caption{Requiring consistency of whiskering with the vertical composition of surfaces demands that the map $g \: \rhd$, for arbitrary $g \in G$, is a group homomorphism on $E$. That is $(g \rhd f) (g \rhd e)= g \rhd (fe)$ for $g \in G$ and $e,f \in E$. This can be seen from the figure, because consistency demands that the diagram commute, i.e., the two routes from the top-left image to the bottom-right image should give the same result.}
				\label{consistency_whiskering_vertical}
				
			\end{center}
		\end{figure}

\begin{figure}[H]
			\begin{center}
				\includegraphics{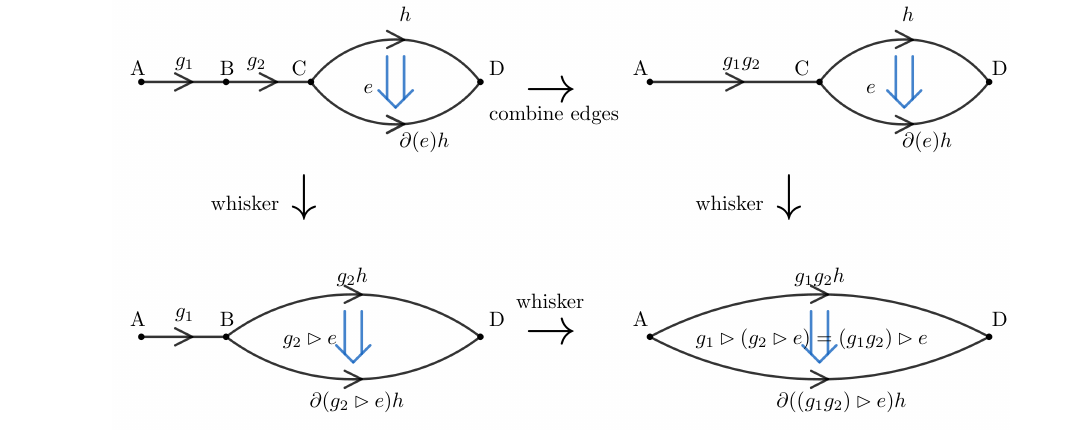}
				\caption{We require that whiskering is consistent with the composition of edges. That is, combining two edges together and then whiskering a surface along the combined edge should give the same result as whiskering that surface by one edge and then the other in sequence. This gives us the mathematical condition $g_1 \rhd (g_2 \rhd e)=(g_1g_2) \rhd e$, which is the condition that the map $\rhd :G \rightarrow \text{End} (E)$ be a group homomorphism on $G$.}	
				\label{consistency_whiskering_edge_composition}
			\end{center}
		\end{figure}

	Requiring the consistency of various diagrams also enforces certain restrictions on the algebraic objects we have already discussed. For example, if we have a diagram with three surfaces to combine, the order in which we combine the surfaces should not matter. This restricts our multiplication of surface labels to be associative. Because our vertical composition is described by group multiplication in the group $E$, this associativity is immediately guaranteed by the group properties without any additional conditions on the group. However, there are additional constraints that must be satisfied by the maps $\partial$ and $\rhd$. Requiring the consistency of whiskering with vertical composition of surfaces and composition of paths (see Figures \ref{consistency_whiskering_vertical} and \ref{consistency_whiskering_edge_composition} or Ref. \cite{Baez2010} for more detail) gives us the following conditions for all $g$, $h \in G$ and $e$, $f \in E$ \cite{Bullivant2017}:
		\begin{align}
		g \rhd (ef) &= (g \rhd e) \: (g \rhd f) \label{Equation_rhd_condition_1}\\
		g \rhd (h \rhd e) &= (gh) \rhd e \label{Equation_rhd_condition_2}. 
		\end{align}
		These are the conditions for a group action of $G$ on $E$. That is, these conditions mean that $\rhd$ is a homomorphism from $G$ to the endomorphisms on $E$, where endomorphisms are group homomorphisms from $E$ to itself. Furthermore, because these endomorphisms are invertible (from Equation \ref{Equation_rhd_condition_2}, $g^{-1} \: \rhd$ is the inverse of $g \: \rhd$), they are automorphisms.

\begin{figure}[H]
			\begin{center}
				\includegraphics{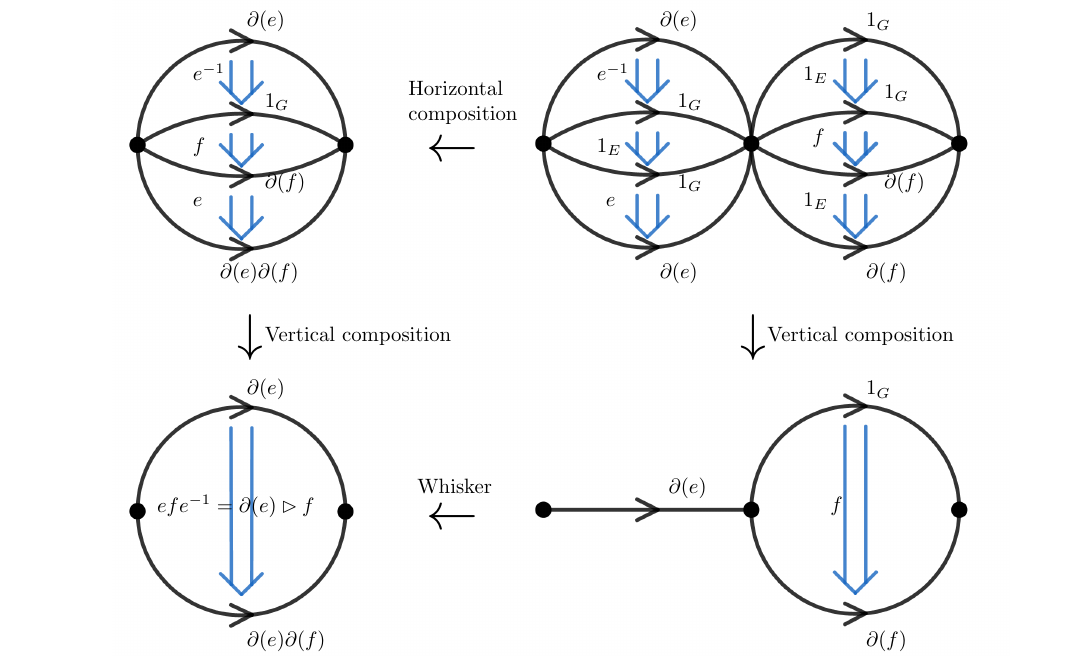}
				\caption{The second Peiffer condition can be derived from demanding that this figure be consistent. Starting at the top-right diagram, we may combine the left and right parts of the diagram to obtain the top-left diagram, using the rules for horizontal composition given in Figure \ref{horizontal_composition_labelling}. We can then use vertical composition to obtain the bottom-left diagram, which should have a label of $efe^{-1}$. However, we could also have performed vertical composition on the top-right diagram to obtain the bottom-right diagram, before whiskering to obtain the bottom-left diagram. In that case the surface label is $\partial(e)\rhd f$. Consistency therefore demands that $efe^{-1} = \partial(e) \rhd f$. }
				\label{consistency_interchange}
				
			\end{center}
		\end{figure}

		As illustrated in Ref. \cite{Baez2010} (though note that different conventions are used in this reference and in particular group multiplication describes horizontal composition of surfaces), consistency of whiskering with other diagrams (see Figures \ref{whiskering_algebraic} and \ref{consistency_interchange}) also demands that \cite{Pfeiffer2003, Baez2002}
		\begin{align}
		\partial( g \rhd e)&= g \partial(e) g^{-1} \label{P1}\\
		\partial(e) \rhd f &= efe^{-1}. \label{P2}
		\end{align}
		These two conditions are known as the Peiffer conditions \cite{Bullivant2017}. The algebraic structure $(G,E, \partial, \rhd)$ satisfying all of these conditions (Equations \ref{Equation_rhd_condition_1}, \ref{Equation_rhd_condition_2}, \ref{P1} and \ref{P2} in addition to $\partial$ being a group homomorphism) is known as a crossed module.
		
		\begin{center}
			\noindent\fcolorbox{black}{myblue1}{%
				\parbox{0.9\linewidth}{%
					\textbf{Definition 1:}
					A crossed module is a collection $(G, E, \partial, \rhd)$, where $G$ and $E$ are groups, and $\partial:E \rightarrow G$ and $\rhd:G \rightarrow \text{Aut}(E)$ are group homomorphisms satisfying the Peiffer conditions Equations \ref{P1} and \ref{P2}.}}
		\end{center}

		 In order to familiarize the reader with these crossed modules, we describe a handful of examples here.

		\textbf{Example 1:}

		One example of a crossed module is $(G,G, \text{id, ad})$, where $G$ is any finite group, id is the identity map and ad maps $g \in G$ to conjugation by $g$ \cite{Bullivant2017}. That is, we have $\partial(e)=e$ and $g \rhd e = geg^{-1}$. This clearly satisfies the first Peiffer condition because
		$$ \partial (g \rhd e) =g \rhd e = geg^{-1} = g \partial(e) g^{-1}.$$
		It also satisfies the second condition as $$ \partial (e) \rhd f =e \rhd f= efe^{-1}.$$ This crossed module describes a model where all of the excitations are either confined or carry trivial charge.

		\textbf{Example 2:}

		Another example is $(G,\set{1_E}, \partial \rightarrow 1_G, \rhd \rightarrow \text{id})$ \cite{Bullivant2017}. That is, we take the group $E$ to be trivial. Then $\partial$ maps the element of $E$ to the identity of $G$ and $g \: \rhd$ is the identity map on $E$ (clearly these are the only allowed $\partial$ and $\rhd$ when $E$ is $\set{1_E}$). We have that 
		$$\partial (g \rhd e) =1_G =g g^{-1} =g 1_G g^{-1}= g \partial(e) g^{-1}$$
		so the first Peiffer condition is satisfied. Furthermore
		$$\partial (e) \rhd f = 1_E= efe^{-1},$$
		because the only element of $E$ is the identity, so the second Peiffer condition is also satisfied. This special case recovers lattice gauge theory, because the surfaces all have trivial label and so we can just neglect to label them.

		\textbf{Example 3:}

		A third, more interesting, example is $(\mathbb{Z}_2, \mathbb{Z}_3, \partial \rightarrow 1_G, \rhd)$ \cite{Bullivant2017}. We take the elements of $G=\mathbb{Z}_2$ to be $1_G$ and $-1_G$ and the elements of $E=\mathbb{Z}_3$ to be $1_E$, $\omega_E$ and $\omega^2_E$. Then we define $\rhd$ by $1_G \rhd e =e$ and $-1_G \rhd e = e^{-1}$ (where $\omega_E^{-1}= \omega_E^2$). This satisfies the requirement of having a group structure on the elements of $G$ (as described in Equation \ref{Equation_rhd_condition_1}) because applying two $-1_G \: \rhd$ maps in sequence gives
		\begin{align*}
		-1_G \rhd (-1_G \rhd e)&=-1_G \rhd e^{-1} = e\\
		& = 1_G \rhd e\\
		& = (-1_G \cdot -1_G) \rhd e,
		\end{align*}
	while the other conditions for the group structure involve $1_G \: \rhd$ and are satisfied because $1_G \: \rhd$ is the identity map. The individual maps $g \: \rhd$ are also endomorphisms as required. For $-1_G \: \rhd$ we have
		\begin{align*}
		-1_G \rhd (e_1 e_2)&=e_2^{-1} e_1^{-1}=e_1^{-1} e_2^{-1}\\
		& = (-1_G \rhd e_1) (-1_G \rhd e_2),
		\end{align*} 
		where we used the fact that $E$ is Abelian to swap the order of multiplication in the second line. This indicates that $-1_G \rhd $ is a group homomorphism on $E$. $1_G \rhd$ is also a homomorphism because it is the identity map. Therefore $\rhd$ is indeed a group action of $G$ on $E$. Next, we will check that the Peiffer conditions are satisfied. We have $\partial(g \rhd e)=1_G= \partial(e) gg^{-1}= g \partial(e)g^{-1}$ (using $\partial(e)=1_G$ and the fact that the group $G$ is Abelian). Finally $\partial(e) \rhd f= 1_G \rhd f = f = fee^{-1} =efe^{-1}$, where we used that $E$ is Abelian. Because all of the consistency conditions are satisfied, this is indeed a valid crossed module. This $(\mathbb{Z}_2, \mathbb{Z}_3, \partial \rightarrow 1_G, \rhd)$ crossed module can be generalized slightly by replacing $\mathbb{Z}_3$ with $\mathbb{Z}_n$, where $n$ is an odd integer, with $-1_G \: \rhd$ still acting as inversion.
		
		\onecolumngrid

	\twocolumngrid
	
	\subsubsection{Composing general surfaces}

	\label{Section_composing_surfaces}
\begin{figure}[h]
		\begin{center}
			\includegraphics{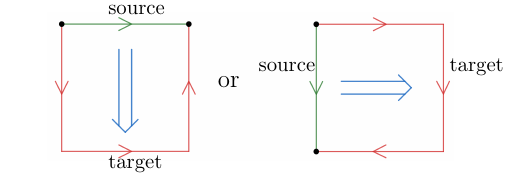}
		\end{center}
		\caption{The same surface can correspond to different 2-holonomies, depending on which parts of the boundary of that surface are designated as the source and target. For example, in the left image the source is the top edge and the target the bottom three edges, so the 2-holonomy corresponds to a process where we transport the upper edge into the bottom three. On the other hand, in the right-hand figure the 2-holonomy corresponds to the process where we transport the left edge into the other three. We expect the labels assigned to these processes to be different but related, as we describe shortly.}
		\label{2-holonomy_square_2}
	\end{figure}

\begin{figure}[h]
		\begin{center}
		\includegraphics{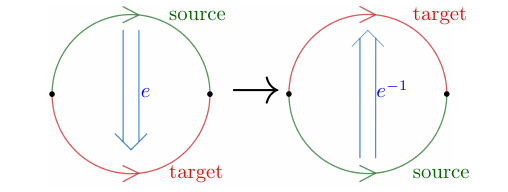}
			\caption{We can change the source and target of a surface by following a set of rules that tells us how the label of that surface should change. The first rule allows us to swap the source and target of a surface. If the plaquette has a 2-holonomy of $e$, then swapping the source and target changes the 2-holonomy to $e^{-1}$.}
			\label{2-holonomy_reverse_direction_2}
			
		\end{center}
	\end{figure}

	So far, we have considered how we may combine surfaces when their sources and targets are compatible. We can combine two surfaces using vertical composition when the target of one surface matches the source of the other. However we may also need to combine adjacent surfaces for which the sources and targets are not compatible. To understand this, we should first look in more detail at how we interpret the 2-holonomy in the case of a fixed lattice. The group element assigned to a surface corresponds to parallel transport of a particular path over that surface. However, we can also pull other paths over that same surface. For example, consider a square, with different paths denoted as the source or target, as shown in Figure \ref{2-holonomy_square_2}. In the left diagram, we consider the process where we transport the top edge (which is the source for the surface) into the bottom three (which form the target). However, as indicated in the right diagram, we could also transport the left edge into the right three over the same surface. Despite corresponding to the same square in space, the label in $E$ associated with these two parallel transports is different in general. We therefore need to know how the label changes when we change the transport process. The first thing we can do is to swap the source and target \cite{Bullivant2017}. The resulting plaquette label is just inverted \cite{Bullivant2017}, as shown in Figure \ref{2-holonomy_reverse_direction_2}.

\begin{figure}[h]
		\begin{center}
				\includegraphics{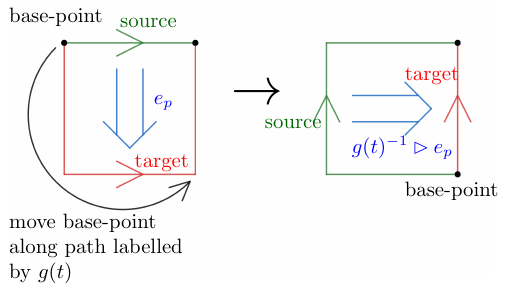}

		\end{center}
		\caption{We can also move the base-point of our surface along the boundary of that surface, which adds or removes edges from the start of the source (and removes or adds those edges to the target). This results in a $\rhd$ action on the surface label.}
		\label{2-holonomy_move_base-point_2}
	\end{figure}

\begin{figure}[h]
		\begin{center}
 		\includegraphics{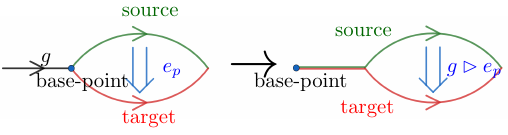}

		\end{center}
		\caption{We can whisker a surface by moving its base-point away from the original boundary of that surface. In the right image, the red and green section (which is the black path from the left image) is part of both the source and target.}
		\label{whiskering_3_2}
	\end{figure}

	Next, we can move the base-point around. We can either move it along the plaquette (as shown in Figure \ref{2-holonomy_move_base-point_2}), or away from the plaquette \cite{Bullivant2017} (as shown in Figure \ref{whiskering_3_2}). In either case, the surface label changes from its original label $e_p$ to $g(t)^{-1} \rhd e_p$, where $t$ is the path along which we move the base-point and $g(t)$ is the group element assigned to that path \cite{Bullivant2017}.

\begin{figure}[h]
		\begin{center}
		\includegraphics[width= \linewidth]{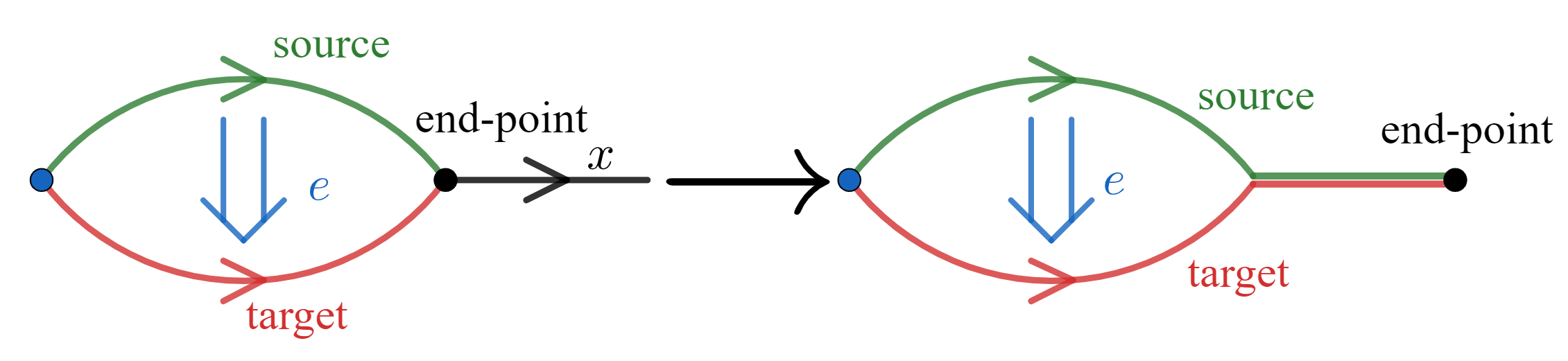}
			
			\caption{We can move the end-point (the black dot) away from the original boundary of the surface, thereby adding edges (in this case the black path from the left image) to the boundary. These edges do not enclose any area and appear once in the source and once in the target (in the right image, the rightmost edge is part of both the source and target). Moving the end-point in this way does not change the label of the surface, unlike moving the base-point.}
			\label{move_end_point_2}
		\end{center}
	\end{figure}

	We can also move the end-point, either along the plaquette, or away from it \cite{Bullivant2017}. Either way, the surface label is unchanged \cite{Bullivant2017}. This latter move (as shown in Figure \ref{move_end_point_2}) allows us to add additional edges to the boundary of the surface, though these additional edges enclose no area. Though in Figure \ref{move_end_point_2} the edges are added near the end-point, we can add these additional edges anywhere on the surface's boundary. If these edges are not added at the end-point, the added edges appear twice consecutively in the source or target and are travelled in opposite directions for their two appearances, meaning that they do not contribute to the path element of the source or target (because adding a path $t$ to the surface in this way contributes $g(t)g(t)^{-1}=1_G$ to the source or target). If the edges are added at the end-point, they contribute equally to the end of the source and target. Either way, their total contribution to the group element associated to the surface boundary (the 1-gauge value assigned to the path around the surface) is trivial. For example, in Figure \ref{move_end_point_2} adding the edge of label $x$ to the end-point takes the path label of the boundary from
	$$g(\text{boundary})= g(\text{source})g(\text{target})^{-1}$$
	to
	\begin{align*}
		g(\text{source}) x (g(\text{target})x)^{-1}&= g(\text{source})xx^{-1}g(\text{target})^{-1}\\ 
		&=g(\text{source})g(\text{target})^{-1}\\
		&=g(\text{boundary}),
	\end{align*}
	whereas adding such an edge in the middle of the source or target would lead to similar cancellation within $g(\text{source})$ or $g(\text{target})$.

\begin{figure}[h]
		\begin{center}
			\includegraphics[width=\linewidth]{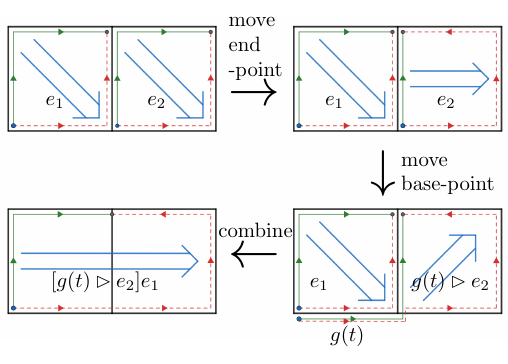}

			\caption{In order to combine two adjacent surfaces whose sources (shown in green) and targets (shown as red dashed lines) are not compatible, we need to manipulate the end-point and base-points of the surfaces first. In the first step, we move the end-point of the second surface (with label $e_2$), then in the second step we move the base-point of that surface. After doing this, the target of the first surface matches the source of the second, so we can combine them.}
			\label{surface_combination_example}
		\end{center}
	\end{figure}

	Now we consider an example of how we can use the rules we have discussed so far to combine two surfaces when their sources and targets are not immediately compatible. In Figure \ref{surface_combination_example} we show two such adjacent surfaces. For each surface, the source is represented by the solid green line and the target by the dashed red one, and we have displaced the source and target slightly away from the edges of the graph (shown in black) for clarity. In order to match the target of the first surface (with surface label $e_1$) to the source of the second (labelled by $e_2$), we first move the end-point of the second surface, as shown in the top-right of Figure \ref{surface_combination_example}. Because moving the end-point of the surface does not affect its label, the second surface still carries a label of $e_2$. Next we move the base-point of the second surface to match that of the first, as shown in the bottom-right image. When we do this, we must whisker the second surface, so that the path $t$ (the edge at the bottom of the first surface) appears in both the source and target of the second surface (represented by the parallel red and green arrows below that edge). Upon doing so, the label of the second edge is changed to $g(t) \rhd e_2$, because $t$ is the path from the new base-point of the surface to the old one. By moving the base-point and end-points in this specific way, we ensure that the target of the first surface matches the source of the second (consisting of the bottom edge of the first surface and the edge separating the two surfaces), so we can compose the surfaces. This gives us a combined surface with label $[g(t) \rhd e_2]e_1$. In general, there may be many ways to combine a given set of surfaces into the same final surface (i.e., a final surface with the same source and target). These are guaranteed to be consistent only when the surfaces that we are combining satisfy an the fake-flatness condition, meaning that each surface obeys the parallel transport rules given in Figure \ref{surface_labelled_1}.
	
\subsubsection{A note about notation}

\begin{figure}[h]
	\begin{center}
		\includegraphics[width=\linewidth]{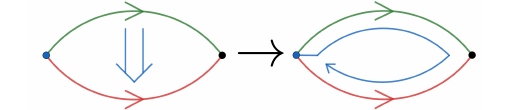}

		\caption{Instead of illustrating the orientation of the surface as an arrow between the source and target, we can draw it as an arrow that passes clockwise or anticlockwise around the surface, starting and ending at the base-point of the surface. The direction of this arrow matches the direction of the source.}
		\label{surface_switch_notation}
	\end{center}
\end{figure}

So far, when describing surfaces we have specified both the source and target of the surface. However, the fact that the label of a surface is unchanged when we move the end-point of the source and target means that we do not need to keep track of all of the information specifying a surface in order to be able to assign that surface a group label. This motivates us to consider a change of notation. Rather than specify the source and the target as two paths, with an arrow between them to highlight the parallel transport, we simply combine the source with the target by moving the end-point all the way along the target (so that the new source is now the original source composed with the inverse of the original target, and the new target is an empty path). This means that we now just have one path all the way around the surface. To specify this, we only need the start of that path (the base-point) and its orientation. Rather than draw an arrow, we indicate this as a circulation, as shown in Figure \ref{surface_switch_notation}. Due to the convenience of this notation, we will generally use it when we do not need to indicate the source and target of a surface explicitly.

		\subsubsection{Gauge transforms}
		\label{HLGT_Gauge_Transforms}
\begin{figure}[h]
			
			\begin{center}
				
				\includegraphics[width=0.9\linewidth]{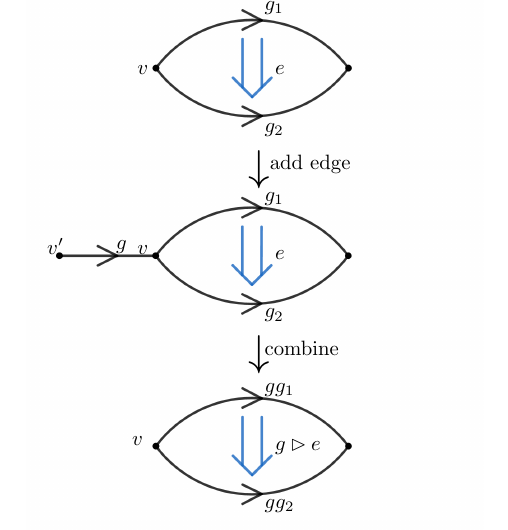}
				
				\caption{If the base-point of a plaquette is at $v$, then it is affected by the action of a gauge transform $A_v^g$ on that plaquette. The vertex transform is equivalent to parallel transport of the base-point (i.e., whiskering) and induces a $g \rhd$ action.}
				\label{vertex_transform_bigon}
			\end{center}
			
		\end{figure}

		Now that we have considered the fields and the parallel transport rules, we can describe the gauge transforms. There are two types of gauge transforms: those associated to the more familiar 1-gauge field and those associated to the 2-gauge field \cite{Bullivant2017}. We label the 1-gauge transforms associated to a vertex $v$ by $A_v^g$, where there is one such transform for each element $g$ of $G$. Just like for lattice gauge theory, $A_v^g$ acts on the degrees of freedom near the vertex $v$ in a way equivalent to parallel transport of the vertex $v$ along an edge of label $g^{-1}$ (or $g$ if we transport the vertex against the direction of the edge, as in Figure \ref{vertex_transform_bigon}). The effect on the edges around the vertex is therefore the same as in the lattice gauge theory case and so only paths that start or terminate on the vertex are affected by the gauge transform (see Section \ref{LGT_Gauge_Transforms} and Figure \ref{vertex_transform} in particular). The only difference is that now we must also consider parallel transport of surfaces along the edge, so that the vertex transform also affects the surface labels. This parallel transport can be performed by adding a new edge and vertex, which we proceed to combine with the rest of the lattice, as illustrated in Figure \ref{vertex_transform_bigon}. In the last step we relabel the vertex $v'$ to $v$ in order to match the original vertex, so that the lattice is the same at the end as it was before the transform, apart from changes to the group labels. We can recognise the middle diagram in Figure \ref{vertex_transform_bigon} as the whiskering diagram (see Figure \ref{whiskering_algebraic}), so combining the edge with the plaquette gives us a $g \: \rhd$ action on the plaquette label. This tells us that any surface with base-point at the vertex on which we apply the transform must be acted on by $g \: \rhd$. On the other hand, surfaces not based at that vertex are left unaffected \cite{Bullivant2017}. In summary, the 1-gauge transform acts on an edge $i$ or plaquette $p$ according to \cite{Bullivant2017}
		\begin{align}
		A_v^g: g_{i} &\rightarrow \begin{cases} gg_{i} &\text{ if $v$ is the start of $i$}\\
		g_ig^{-1} &\text{ if $v$ is the end of $i$}\\
		g_{i} &\text{ otherwise} \end{cases} \notag\\
		A_v^g : e_p &\rightarrow \begin{cases} g \rhd e_p &\text{if $v$ is the base-point of $p$}\\
		e_p &\text{otherwise.} \label{Equation_vertex_transform_1} \end{cases}
		\end{align}

\begin{figure}[h]
			\begin{center}
				\includegraphics{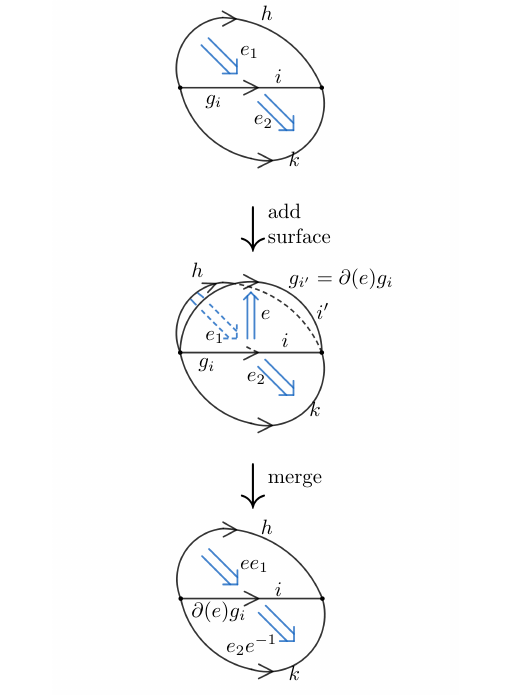}
				
				\caption{A 2-gauge transform $\mathcal{A}_i^e$ on an edge $i$, with initial label $g_i$, acts like parallel transport of that edge across an additional surface of label $e$. In the middle picture, the new surface points upwards, out of the plane of the other two surfaces. In the third picture, we combine this new surface with the others and then relabel the edge $i'$ (the target of the additional surface) to $i$.}
				\label{edge_gauge_transform}
			\end{center}
		\end{figure}

		In addition to these 1-gauge transforms, we also have 2-gauge transforms, which act on an edge and the surfaces that adjoin it \cite{Bullivant2017}. The 2-gauge transform on an edge $i$ and labelled by an element $e \in E$ (denoted by $\mathcal{A}_i^e$) acts like parallel transport of the edge along a surface labelled by $e$. That is, to find the action of a 2-gauge transform on a diagram we add a surface and combine the surface with the rest of the diagram, as shown in Figure \ref{edge_gauge_transform}. This fluctuates the plaquette labels surrounding an edge, as well as changing the edge label itself. This is similar to how the 1-gauge transform at a vertex fluctuates the edges around the vertex (along with any plaquettes based at that vertex).

	In Figure	\ref{edge_gauge_transform}, the base-point of each surface is also the start of edge $i$, which results in the simple expression for the edge transform given in that figure. To treat a more general case, we can use the rules for changing the base-point of a surfaces to move them to the start of edge $i$. Then we can perform the gauge transform $\mathcal{A}_i^e$ on this simple case before moving the base-points back to their original positions. Because moving the base-point has an $\rhd$ action on the plaquette label, this results in the plaquette label $e_p$ becoming $e_p (g \rhd e^{-1})$ or $(g \rhd e) e_p$ \cite{Bullivant2017} rather than just $e_p e^{-1}$ or $e e_p$, where $g$ is the label of the path on which we had to move the base-point. In order to define the edge transform, we therefore need a prescription for choosing this path.
	
\begin{figure}[h]
		\begin{center}
			\includegraphics{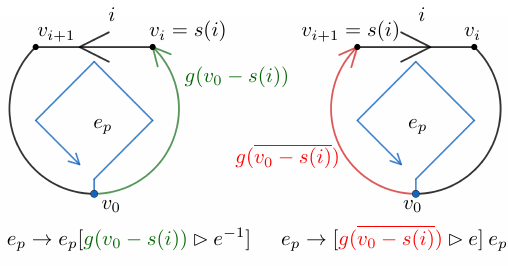}
			
			\caption{The path involved in the effect of the 2-gauge transform $\mathcal{A}_i^e$ on a plaquette $p$ depends on whether the edge $i$ is aligned with the $p$ (as in the left case) or anti-aligned (as in the right case). If the edge is aligned with the plaquette, then the path $(v_0-s(i))$ in the transformation of the plaquette label is aligned with $p$, whereas if $i$ is anti-aligned with $p$ then the path $(\overline{v_0-s(i)})$ appearing in the transformation is anti-aligned with $p$. Either way, the path is aligned with the edge $i$.}
			\label{pathsonplaquette2}
		\end{center}
	\end{figure}

\begin{figure}[H]
		\begin{center}
		\includegraphics[width=0.9\linewidth]{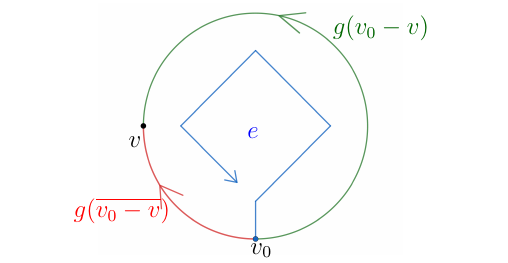}
			\caption{The two different paths from the base-point $v_0$ to the same vertex $v$ on a plaquette are shown in green and red. Paths that anti-align with the surface circulation (represented by the blue arrow in the centre) are indicated using overline notation.}
			\label{waysonplaquette2}
		\end{center}
	\end{figure}

	 Consider the path around one of the plaquettes affected by the transform, starting at the base-point $v_0$ of the plaquette and travelling along its boundary, aligned with its orientation. This path reaches the edge at a vertex that we call $v_i$, as shown in the left picture of Figure \ref{pathsonplaquette2}. The path up to this point is denoted by $g(v_0-v_i)$ \cite{Bullivant2017}. Now consider a path starting at the base-point of the plaquette, but travelling against the circulation of the plaquette. At some point this path will reach the other vertex on the edge, which we call $v_{i+1}$. This path is denoted by $g(\overline{v_0-v_{i+1}})$, where the overline is used to indicate that this path travels against the circulation of the plaquette \cite{Bullivant2017}. This overline notation is illustrated in Figure \ref{waysonplaquette2}, where we look at different paths around the plaquette to the same vertex. Then the action of the edge transform on each edge $i'$ and plaquette $p$ is \cite{Bullivant2017}:
	\begin{align}
		\mathcal{A}_i^e: g_{i'} &\rightarrow \begin{cases} \partial(e) g_{i'} &\text{ if $i=i'$}\\
			g_{i'} &\text{ otherwise} \end{cases} \notag \\
		\mathcal{A}_i^e : e_p &\rightarrow \begin{cases} e_p [g(v_0 - v_i) \rhd e^{-1}] &\text{if $i$ is on $p$ and } \\ &\text{ aligned with $p$}\\
			[g(\overline{v_0 - v_{i+1}}) \rhd e] e_p &\text{if $i$ is on $p$ and} \\ & \text{ aligned against $p$}\\
			e_p &\text{otherwise.} \end{cases} \label{Equation_edge_transform_definition_2}
	\end{align}
	The paths involved in the cases where the edge is aligned or anti-aligned with the plaquette are indicated in Figure \ref{pathsonplaquette2}. In either case the path terminates at the source of edge $i$, which is the adjacent vertex which the edge points away from (while the target is the vertex it points towards). This means that we can replace $v_i$ (in the aligned case) or $v_{i+1}$ (in the anti-aligned case) in the expression for the paths in Equation \ref{Equation_edge_transform_definition_2} with this source, $s(i)$.

		\subsubsection{Gauge-invariants}
		\label{Section_Gauge_Invariants_HLGT}
\begin{figure}[h]
			\begin{center}
				\includegraphics{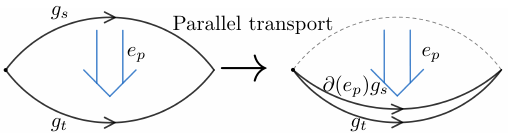}

			\end{center}
			\caption{The gauge invariant associated to plaquettes is modified to account for parallel transport of the paths over the plaquette.}
			\label{plaquette_invariant}
		\end{figure}

		In ordinary lattice gauge theory we could build gauge-invariant quantities out of closed loops. What are the appropriate quantities for higher lattice gauge theory? We can build gauge-invariants from the closed loops as before, but also from closed surfaces. For the closed loops, we need to modify the group element that labels them to account for parallel transport of paths over surfaces. Given a closed loop made of two paths, as shown in Figure \ref{plaquette_invariant}, to work out the group element for the loop, we need to transport the paths so that they are in the same location. This is necessary because the two paths may be defined with different gauge choices, with the conversion between the gauge choices performed by parallel transport. To obtain a gauge invariant, we will need to ensure that the two paths are described in the same gauge. The relevant transport is shown in Figure \ref{plaquette_invariant}.
		
		The parallel transport modifies the group element associated to the closed loop in Figure \ref{plaquette_invariant}, from $g_s g_t^{-1}$ to $\partial(e_p)g_s g_t^{-1}$. This quantity is the 1-flux or 1-holonomy for the closed loop. For a general surface, we replace $g_s g_t^{-1}$ with the label of the boundary of the surface. For a plaquette $p$, with boundary label $g_p$, the 1-flux is given by $\partial(e_p)g_p$ and we refer to this quantity as $H_1(p)$ \cite{Bullivant2017}. This label can be changed only within a conjugacy class by either the vertex transforms (as in Figure \ref{vertex_transform_bigon}) or the edge transforms (as in Figure \ref{edge_gauge_transform}), so those conjugacy classes are gauge invariant quantities \cite{Bullivant2017}.

		As an example, we can consider acting on the diagram in Figure \ref{plaquette_invariant} with a vertex transform. This gives us the situation shown in Figure \ref{vertex_transform_bigon}. From that figure, we see that the plaquette holonomy, which is initially given by $\partial(e)g_1g_2^{-1}$, transforms as
		\begin{align*}
		\partial(e) g_1 g_2^{-1} &\rightarrow \partial(g \rhd e) gg_1 g_2^{-1} g^{-1}\\
		&= g \partial(e) g^{-1} g g_1 g_2^{-1} g^{-1}\\ &\text{ (using the Peiffer conditions)}\\
		&= g \partial(e) g_1 g_2^{-1} g^{-1},
		\end{align*}
		which is only conjugation.

		In addition to closed paths, closed surfaces have their own gauge-invariants. The gauge invariant assigned to a closed surface can be found from the group label (2-gauge label) assigned to that closed surface, which may be obtained by using the rules for composing surfaces if that closed surface is comprised of multiple plaquettes. This label, which we call the 2-flux of that surface, is only changed within certain equivalence classes by the gauge transforms \cite{Bullivant2017} (as long as the constituent plaquettes satisfy fake-flatness). Again, the identity element is in a class on its own, so that trivial 2-flux is preserved by the transforms \cite{Bullivant2017}.

\begin{figure}[h]
			\begin{center}
				\includegraphics[width=0.6\linewidth]{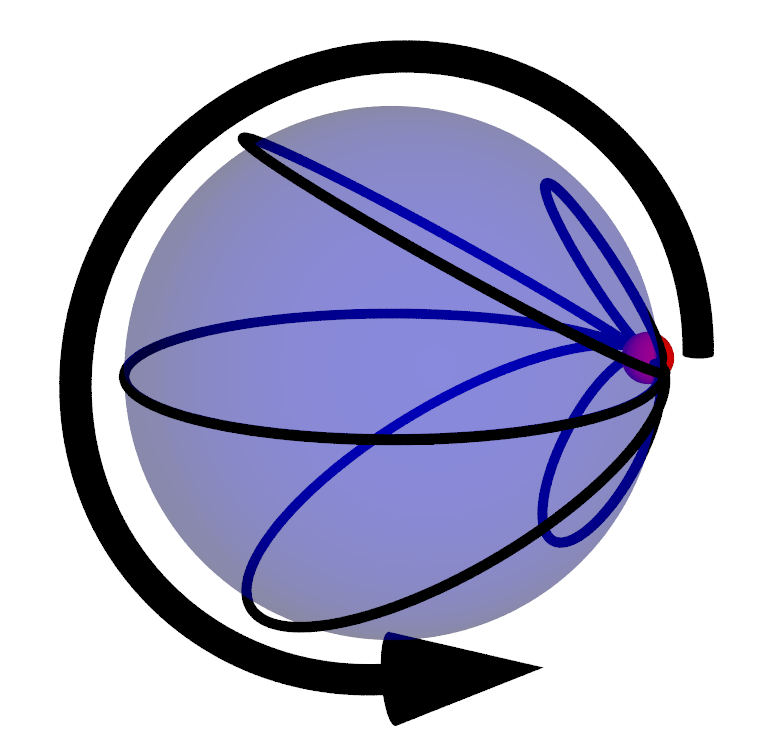}

				\caption{The 2-holonomy of a surface (in this case a sphere) can be measured by a transport process. A small loop is created at the base-point (the small red sphere), then dragged over the surface (the larger blue sphere), as indicated by the arrow.}
				\label{surfaceholonomy1}
			\end{center}
		\end{figure}

		In the same way that the 1-flux on a closed loop determines the result of a process where we move a charge around the loop, the 2-flux of a closed surface corresponds to a transport process. For a sphere at least, we can measure this 2-flux by nucleating a small loop at the base-point of that surface, before passing it over the surface and then contracting it again, as indicated in Figure \ref{surfaceholonomy1}. This reflects the fact that a spherical closed surface (which can be built from a series of open surfaces) can have empty source and target, and so can represent a transport process where we nucleate the loop at the start and collapse it at the end. For a surface such as a torus, with non-contractible cycles, the corresponding transport process may not involve nucleation and collapse.
		
		\subsection{Hamiltonian model}
		\label{Section_Hamiltonian_Model}

		Having considered higher lattice gauge theory, we can now define the Hamiltonian model based on it (as introduced in Ref. \cite{Bullivant2017}). The three spatial dimensions of the model are represented by a lattice, while the temporal dimension is continuous and time evolution is controlled by the Hamiltonian. As already alluded to, we label each edge of the lattice with an element of group $G$ and each plaquette with an element of group $E$ \cite{Bullivant2017}. Labelling every edge and plaquette gives a configuration (or colouration). These configurations then form a basis for the Hilbert space, so that a general state is a linear combination of the different labellings of the lattice. However, we have seen that a given plaquette can correspond to different transport processes depending on the source and target, so we need a way of specifying which transport process the assigned label corresponds to. As described in Section \ref{Section_composing_surfaces}, we can then use a set of rules to manipulate the source and target of a plaquette to find the label that would be associated to a different process. In order to have this unambiguous reference process, we define a ``canonical" position for the source and target paths of every plaquette when we set up the lattice. Because the label of the plaquette is invariant under changes to the end-point, it is sufficient to choose a base-point and orientation for each plaquette. We also need to choose an orientation for each edge. This can be done formally via a branching structure, which assigns every vertex in the lattice a unique integer vertex. The edges and plaquettes then inherit their data from the vertices involved \cite{Bullivant2017}. The details of this are not important for our discussion, so we will directly choose the canonical data (orientation and base-point) for each edge and plaquette. We will sometimes refer to this choice as the branching structure, or the decoration of the lattice. In Appendix \ref{Section_rebranching}, we demonstrate how the energy terms change under changes to this branching structure.

		To motivate the Hamiltonian considered by Bullivant et al. \cite{Bullivant2017}, we can take the same approach used for Kitaev's Quantum Double model. We first demote the gauge symmetries to energetic constraints, by including them as terms in the Hamiltonian:
		$$H = -\sum_{\text{vertices, }v} A_v - \sum_{\text{edges, } i}\mathcal{A}_i + ...$$
		Here $A_v$ is the average over gauge transforms at the vertex $v$: 
		\begin{equation}
		A_v = \frac{1}{|G|}\sum_{g \in G}A_v^g,
		\end{equation}
	where the action of the gauge transform $A_v^g$ is defined in Equation \ref{Equation_vertex_transform_1}. 
		
		As with Kitaev's Quantum Double model, the vertex transforms satisfy $A_v^g A_v^h = A_v^{gh}$ for any $g$, $h \in G$, which follows from their interpretation in terms of parallel transport (see Figure \ref{vertex_transform_bigon}). The relation $A_v^g A_v^h = A_v^{gh}$ results in $A_v$ being a projector \cite{Bullivant2017}, just as for the equivalent term in Kitaev's Quantum Double model (see Equation \ref{Equation_Quantum_Double_Vertex_Projector}). We can absorb vertex transforms into the corresponding vertex energy term, by which we mean that $A_v^x A_v =A_v$ for any $x \in G$, as we can demonstrate by expanding the vertex term and then using the algebra of the vertex transforms: 
		\begin{align*}
		A_v^xA_v &= A_v^x \frac{1}{|G|} \sum_{g \in G} A_v^g\\
		&= \frac{1}{|G|} \sum_{g \in G} A_v^{xg}\\
		&= \frac{1}{|G|} \sum_{x'=xg \in G} A_v^{x'}\\
		&=A_v.
		\end{align*}
		
		This means that the states (such as the ground states) which satisfy $A_v \ket{\psi}= \ket{\psi}$ are invariant under the individual vertex transforms, rather than just the energy term:
		\begin{align*}
		A_v^g \ket{\psi} &= A_v^g A_v \ket{\psi} \\
		&= A_v\ket{\psi} \\
		&= \ket{\psi}.
		\end{align*}
		Therefore the eigenvalue of one for the energy term corresponds to states which are gauge-invariant at that vertex.

		In a similar way to the vertex terms, the edge term $\mathcal{A}_i$ is the average over 2-gauge transforms at the edge $i$ \cite{Bullivant2017}: 
		\begin{equation}
		\mathcal{A}_i = \frac{1}{|E|} \sum_{e \in E} \mathcal{A}_i^e. \label{Equation_edge_term}
		\end{equation}
	  The edge terms can be combined in the same way as the vertex terms \cite{Bullivant2017}: 
		\begin{equation}
		\mathcal{A}_i^e \mathcal{A}_i^f =\mathcal{A}_i^{ef}.
		\end{equation}
		As with the vertex terms, this means that 
		\begin{equation}
		\mathcal{A}_i^e \mathcal{A}_i =\mathcal{A}_i. \label{Equation_edge_transform_absorb}
		\end{equation}
		This leads to the energy term $\mathcal{A}_i$ being a projector \cite{Bullivant2017}, with the eigenvalue of one corresponding to states that are 2-gauge-symmetric at that edge. The minus sign with which this term enters the Hamiltonian ensures that the energy term favours these gauge-symmetric states. 
		
\begin{figure}[h]
			\begin{center}
				\includegraphics{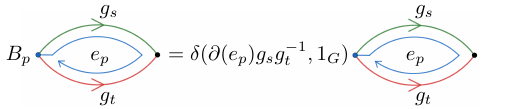}
				\caption{The plaquette term checks that the flux through a plaquette is trivial. In this case, the boundary of the plaquette is made of two edges, labelled by $g_s$ and $g_t$, and the boundary label is $g_sg_t^{-1}$. The flux through the plaquette is therefore $\partial(e_p)g_sg_t^{-1}$, where $e_p$ is the plaquette label, and the plaquette term checks whether this expression for the flux is trivial.}	
				\label{plaquette_term_bigon_1}
				
			\end{center}
		\end{figure}

		So far we have considered energy terms that enforce the 1-gauge symmetry and 2-gauge symmetry. Now we add terms that depend on quantities that are invariant under the two types of gauge transform. By building these terms from gauge-invariant quantities, we guarantee that the new terms commute with the gauge transforms. Recall from Section \ref{Section_Gauge_Invariants_HLGT} that there are gauge-invariant quantities associated to the closed cycles of the lattice. In particular, whether a cycle has a trivial group element or not is invariant under gauge transforms. We can therefore energetically penalize cycles that have non-trivial 1-flux. We do this with an energy term at each plaquette, which gives one if the plaquette has trivial flux and zero if the flux is non-trivial. As explained in Section \ref{Section_Gauge_Invariants_HLGT}, the 1-flux for a plaquette with label $e_p$ and path label $g_p$ for its boundary is given by $\partial(e_p)g_p$. The plaquette term therefore acts as $\delta(\partial(e_p)g_p, 1_G)$. An example of the plaquette energy term is shown in Figure \ref{plaquette_term_bigon_1}. The plaquette terms enter the Hamiltonian with a minus sign, which ensures that the lowest energy states have trivial flux on the plaquettes. We call plaquettes that satisfy this condition fake-flat \cite{Bullivant2017}.

\begin{figure}[h]
			\begin{center}
			\includegraphics{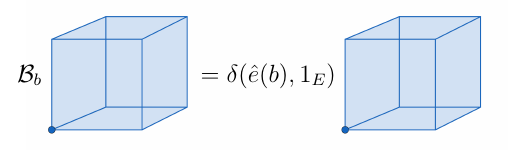}
				
				\caption{The blob energy term $\mathcal{B}_b$ checks whether the total surface label of the blob $b$, $\hat{e}(b)$, is the identity element or not. This surface label must be determined by using the rules for combining surface elements from Sections \ref{Section_HLGT} and \ref{Section_Hamiltonian_Model} to combine the plaquettes on the boundary of the blob. For example, when $\rhd$ is trivial the surface label is a product of the plaquette labels (with inverses if the orientation of the plaquette needs to be reversed to match the overall surface).}
				\label{blob_energy_term}
			\end{center}
		\end{figure}

		Finally, we consider the gauge-invariant quantity associated to the closed surfaces. In the same way as for closed cycles, we penalize closed surfaces with non-trivial 2-flux (2-holonomy). This is done with an energy term at each ``blob" (3-cell) \cite{Bullivant2017}. The blobs are the smallest three-dimensional volumes, such as the smallest cubes in a cubic lattice. For each blob, we have an energy term that checks the value of the surface of that blob, leaving it unchanged if that value is $1_E$ and giving zero otherwise \cite{Bullivant2017}, as shown in Figure \ref{blob_energy_term}. We denote the blob term associated to a blob $b$ by $\mathcal{B}_b$. The blob term also enters the Hamiltonian with a minus sign, so that the full Hamiltonian is given by \cite{Bullivant2017}
		
		\begin{equation}
			H = - \hspace{-0.1cm} \sum_{\text{vertices, }v} \hspace{-0.4cm} A_v \: - \sum_{\text{edges, } i} \hspace{-0.2cm} \mathcal{A}_i \: -\hspace{-0.3cm}\sum_{\text{plaquettes, }p} \hspace{-0.5cm} B_p \: - \sum_{\text{blobs, }b} \hspace{-0.2cm} \mathcal{B}_b. \label{Equation_Hamiltonian}
		\end{equation}
	Note that the model can also be defined in 2+1d, in which case there are no blob energy terms.

		Building the Hamiltonian out of gauge transforms and gauge invariant quantities should mean that the different energy terms commute. However, when fake-flatness is not satisfied, the blob terms may not actually be gauge invariant \cite{Bullivant2017}. In fact, the rules for combining surfaces become inconsistent and so the blob terms are ill-defined without some convention for how combination should be done. This means that the energy terms do not commute on the full Hilbert space and the model is not a commuting projector model (and so is not necessarily solvable). This problem will occur in models where $\rhd$ is non-trivial, i.e., models for which $g \rhd e \neq e$ in general. When $\rhd$ is trivial this complication does not occur and we have a commuting projector Hamiltonian regardless \cite{Bullivant2017}. One solution to this problem for non-trivial $\rhd$ is to define the blob terms to be zero when any of the nearby plaquette terms are not satisfied \cite{Bullivant2017}, similar to the approach taken for plaquette terms in the string-net model when the neighbouring vertices are not satisfied. However, there are some further complications when $\rhd$ is not trivial in the general case, and so we make some restrictions to the model in order to make it more manageable, as we discuss in Section \ref{Section_Special_Cases}.
		
		\subsection{Some special cases and consistency}
		
		\label{Section_Special_Cases}
		
\begin{figure}[h]
			
		\includegraphics{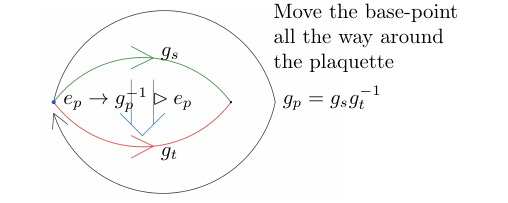}
			
			\caption{Moving the base-point of a plaquette all the way around the plaquette transforms the surface label from $e_p$ to $g_p^{-1} \rhd e_p$, where $g_p$ is the path label of the plaquette's boundary}
			\label{consistency_issue}
			
		\end{figure}

		As we mentioned in the previous section, for the most general crossed modules the Hamiltonian model has certain inconsistencies. As an example of this, consider the surface holonomy of a plaquette, $e_p$. We can move the base-point of the plaquette all the way around the plaquette and back to its initial position. This induces a change to the surface label of the plaquette, given by $e_p \rightarrow g_p^{-1} \rhd e_p$ \cite{Bullivant2017}, where $g_p$ is the path label of the boundary of the plaquette, as shown in Figure \ref{consistency_issue}. The base-point is back to the same position, and the surface appears to be the same, yet the label may have changed. The label does stay constant if the plaquette is fake-flat. In that case, the boundary label satisfies $g_p^{-1}=\partial(e_p)$ and so $g_p^{-1} \rhd e_p = \partial(e_p) \rhd e_p = e_p e_p e_p^{-1}=e_p$ \cite{Bullivant2017}, where we used the Peiffer condition Equation \ref{P2} in the second step. However, if fake-flatness is not satisfied then we cannot guarantee that the plaquette label is unchanged.

		 This is not the only issue arising from violating fake-flatness: as we describe in Section \ref{Section_rebranching_flip_edge} in the Appendix, the edge energy term also appears to become inconsistent with changes to the branching structure of the lattice. One approach for dealing with this problem is to enforce fake-flatness on the level of the Hilbert space, as a hard constraint rather than an energy term. This is the case most closely considered in the paper introducing this model \cite{Bullivant2017}. However, another possibility is to take $\rhd$ trivial, so that the base-point of the plaquette loses any meaning, but allow fake-flatness violations. If we use this condition, then all of the energy terms commute naturally, with no need to restrict the Hilbert space. In this case the model loses some of its complexity, due to the 1-gauge field having no way to act on the 2-gauge field. Some additional consequences of taking $\rhd$ trivial are that $E$ must be Abelian and that $\partial$ maps to the centre of $G$. The first condition, that $E$ is Abelian, comes from the second Peiffer condition (Equation \ref{P2} in Section \ref{Section_HLGT}), because $\partial(e) \rhd f = efe^{-1} \implies f=efe^{-1}$ so that any elements of $E$ commute with each-other. The second condition, that $\partial$ maps to the centre of $G$, comes from the first Peiffer condition (Equation \ref{P1} in Section \ref{Section_HLGT}), as $\partial(g \rhd e)=g\partial(e)g^{-1}$ becomes $\partial(e)=g \partial(e)g^{-1}$, so that $\partial(e)$ commutes with all elements of $G$. In this paper, both of these special cases ($\rhd$ trivial and restricting to fake-flat configurations) are considered. We also consider a third case where $E$ is Abelian and $\partial$ maps to the centre of $G$, but we do not enforce fake-flatness on the level of the Hilbert space or require $\rhd$ to be trivial. In this case, the inconsistencies we mentioned previously are still present, but are not as generic. For example, if a plaquette $p$ with label $e_p$ violates fake-flatness because the boundary label $g_p$ of the plaquette differs from $\partial(e_p)^{-1}$ only by an element $\partial(e) \in \partial(E)$, then moving the base-point of the plaquette around $p$ results in the plaquette label $e_p$ transforming to
		$$g_p^{-1} \rhd e_p = (\partial(e)^{-1}\partial(e_p)) \rhd e_p = (e^{-1}e_p) e_p (e^{-1}e_p)^{-1},$$
		which is just $e_p$ because $E$ is Abelian. This case, where the boundary label differs from $\partial(e_p^{-1})$ by an element in $\partial(E)$, is significant because it occurs when the fake-flatness violation is caused by a change to the plaquette label $e_p$, rather than changes to the edge labels. Such flatness-violating changes to the plaquette labels occur for a whole class of ribbon operators (the confined blob ribbon operators we describe in Section \ref{Section_Part_One_Condensation_Confinement}). We will see other simplifications that occur due to $E$ being Abelian and $\partial$ mapping to the centre of $G$ throughout the paper. Because we will refer to these restrictions, along with the other cases that we have considered in this section, many times in the following text, we summarize all of them in Table \ref{Table_Cases_intro}.

		\begin{table}[h]
			\begin{center}
				\begin{tabular}{ |c|c|c|c|c| } 
					\hline
					& & & &Full\\
					Case & $E$ & $\rhd$ & $\partial(E)$ &Hilbert \\ 
					& & & &Space\\
					\hline
					1 & Abelian & Trivial & $\subset$ centre($G$) & Yes\\ 
					2 & Abelian & General & $\subset$ centre($G$) & Yes\\ 
					3 & General & General & General & No \\
					\hline
				\end{tabular}
				
				\caption{A summary of the special cases of the model}
				\label{Table_Cases_intro}
			\end{center}
			
		\end{table}
		
		\subsection{Braiding relations in 3+1d}
		\label{Section_Braiding_3D}

		One of the important features that we are interested in is the braiding of the various excitations that we find in the higher lattice gauge theory model. While we anticipate that most readers will have at least some familiarity with the concept of braiding in 2+1d, it may be useful to give an overview of braiding in 3+1d, particularly where loop-like excitations are involved. In 2+1d, topological phases support excitations with exchange statistics that generalize the familiar Fermi and Bose statistics \cite{Leinaas1977, Wilczek1982, Arovas1984}, and which are described by the (coloured) braid group. On the other hand, in 3+1d the point-like particles can only have Fermi or Bose statistics \cite{Rao1992, Doplicher1971, Doplicher1974}. However, the presence of loop-like excitations means that we can still have interesting braiding statistics. The motion of loops can be described by the (coloured) loop braid group \cite{Bullivant2018, Baez2007a, Damiani2017} (considered under different names and contexts in papers such as Refs. \cite{McCool1986, Savushkina1996, Fenn1997}). The braid and loop braid groups are both examples of motion groups \cite{Dahm1962, Goldsmith1981}, which describe the motions of arbitrary objects, up to homotopy \cite{Baez2007a} (formally the objects should return to the same positions, perhaps swapping positions). 
		
\begin{figure}[h]
			\begin{center}
				\includegraphics{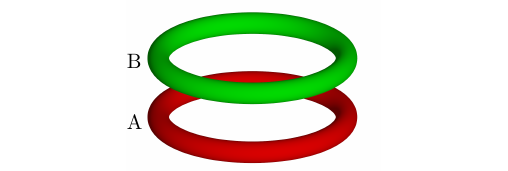}
				\caption{We first consider the motion of a pair of loops A and B, in the frame where the upper one (B) is held fixed. The motion of a set of loops can be described by such pairwise motions, together with flips of the two loops.} 
				\label{LoopStack}
				
			\end{center}
		\end{figure}

\begin{figure}[h]
			\begin{center}
			\includegraphics{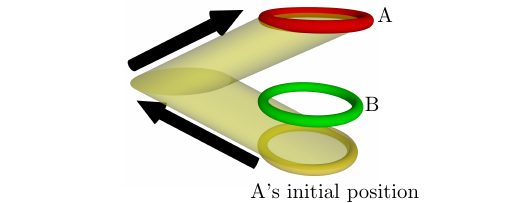}
				\caption{In the permutation move, we move one loop around the other so that neither loop passes through the other. The motion of the (red) loop A is represented by the (yellow) sheet, with the initial position of the moving loop represented by the lowest (yellow) torus.} 
				\label{ExchangeMoveIntro}
				
			\end{center}
		\end{figure}

\begin{figure}[h]
			\begin{center}
			\includegraphics{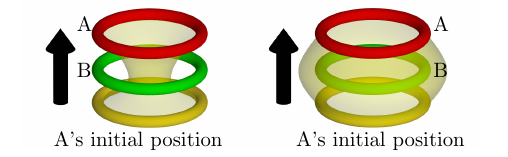}
				\caption{Schematic of the two braid moves. The sheet swept by the motion of the red loop A is shown in yellow, with the initial position of the moving loop shown as a yellow torus. In the left image, the loop A is passed up \textit{through} loop B. In the right image, the loop A passes over loop B, meaning that loop B passes down through loop A from its perspective. The right image can therefore be thought of as the inverse of the left one (analogous to the two types of crossing in the normal braid group) }
				\label{LoopMovesIntro}
				
			\end{center}
		\end{figure}

\begin{figure}[h]
			\begin{center}
			\includegraphics[width=0.7\linewidth]{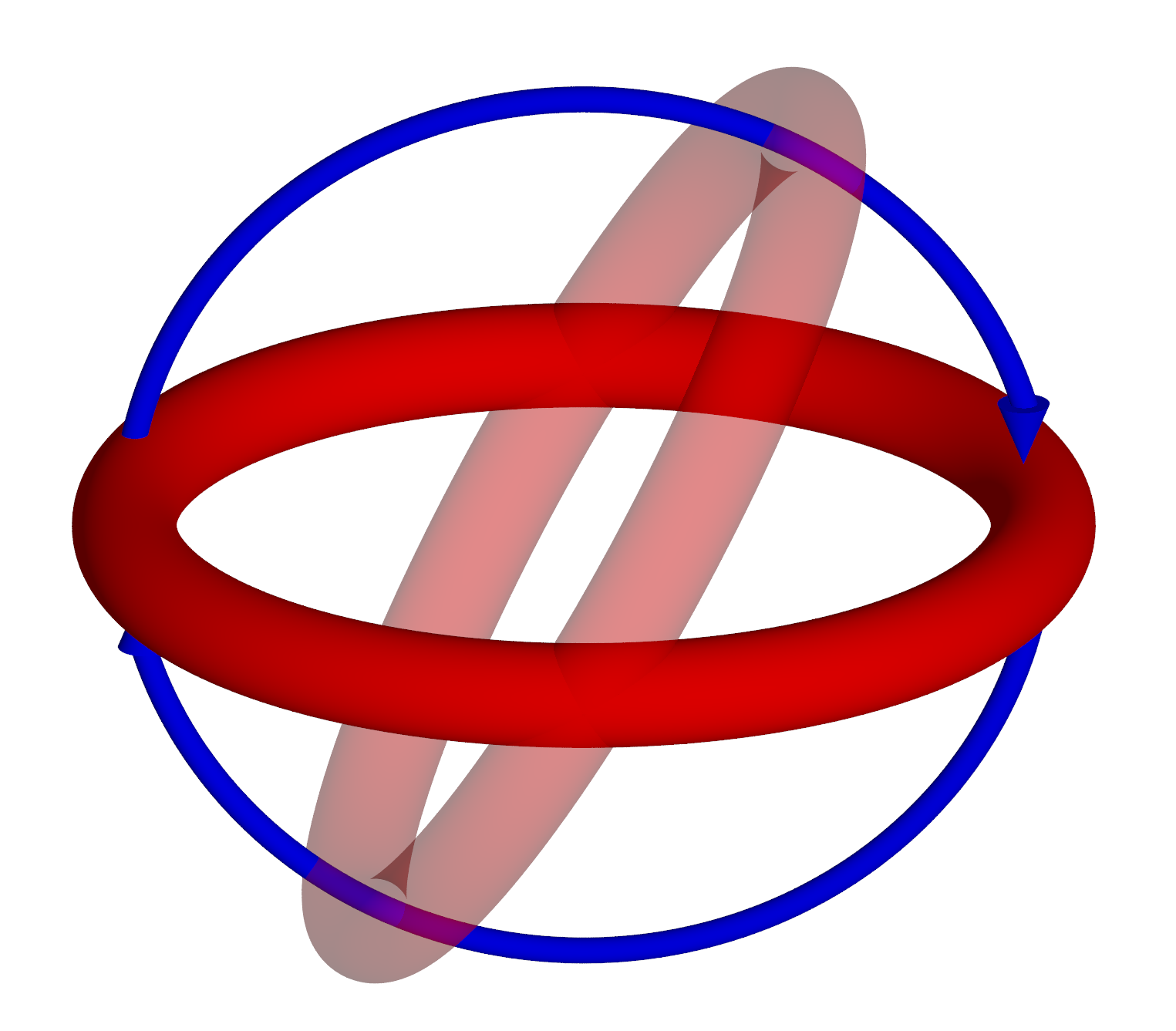}

				\caption{Unlike point particles, loop-like excitations have an orientation. We can reverse the orientation of a loop by rotating it by $\pi$ radians out of its plane, which leaves the loop in the same position but flipped over \cite{Damiani2017}.} 
				\label{LoopFlip}
				
			\end{center}
		\end{figure}

		The loop braid group is generated by a few simple motions. Firstly consider braiding that involves only two loop-like excitations, and for simplicity suppose that those two excitations are stacked vertically, as shown in Figure \ref{LoopStack}. Then if we want to move the lower loop up past the upper loop, there are several ways to do this. The first way, shown in Figure \ref{ExchangeMoveIntro}, is to simply move the lower loop around and past the upper one, so that neither loop passes through the other \cite{Bullivant2018, Damiani2017}. We refer to this motion as a \textit{permutation} move, because such moves generate the permutation group (symmetric group) \cite{Baez2007a} that would describe the motions of point particles in 3+1d. The second way, shown in the left side of Figure \ref{LoopMovesIntro}, is to move the lower loop \textit{through} the upper loop \cite{Bullivant2018, Damiani2017}. We call this a \textit{braiding} move and say that the lower loop has braided through the upper loop. The third way, shown in the right side of Figure \ref{LoopMovesIntro}, is to pull the lower loop \textit{over} the upper loop, which we can also think of as the upper loop passing through the lower loop (and so is the inverse of the previous motion). Another difference from point particles is that loops have an orientation, and so we must also allow a move that flips this orientation, as shown in Figure \ref{LoopFlip}. Then any motion of the two loops can be performed by a series of such moves (we say that these generate the loop braid group for the two excitations) \cite{Bullivant2018, Damiani2017}. More generally, we can have any number of loops, and the different motions of this set of loops can be performed using these pairwise moves.

		Generally we are interested in comparing two motions that result in the same final position of all of the excitations. For example, we could compare the result of the permutation move in Figure \ref{ExchangeMoveIntro} to the braiding move in Figure \ref{LoopMovesIntro}, or we could compare a motion that returns all particles to their initial positions to the trivial motion. Making a comparison between states where the excitations have the same final position is useful because it separates the topological content of the model from the geometric details. As we explain in Section \ref{Section_Part_One_Braiding}, we will be considering processes that involve the production of the loop-like excitations from the ground state, rather than just the movement of existing excitations, but the same principles discussed in this section hold regardless.

		In addition to loop-like particles, the lattice model supports point-like excitations. While braiding between two point-like particles in 3+1d is bosonic or fermionic \cite{Doplicher1971, Doplicher1974} as described earlier (and is exclusively bosonic in this model), the point-like excitations may braid non-trivially with the loop-like excitations of the model. In order to move a point-like particle past a loop excitation, we can either move the point-like excitation \textit{around} the loop-like excitation (analogous to Figure \ref{ExchangeMoveIntro} for two loops, if we replace the initially lower loop A with a point-like excitation), or we can move the point-like excitation through the loop-like one (analogous to the left-side of Figure \ref{LoopMovesIntro}, if we replace the lower loop A with a point-like excitation).

\section{Properties from gauge theory picture}
\label{Section_Properties_From_Gauge_Theory}

\subsection{Gauge theory}

Before we discuss the excitations that we find in the model in great mathematical detail, it will be instructive to give a more qualitative description of the excitations that we expect, using ideas from higher gauge theory. As a starting point, we shall briefly review the excitations in ordinary gauge theory, which are electric charges and magnetic fluxes. A clear exposition on these objects and their properties, in the 2+1d case, is given by Preskill's lecture notes on topological quantum computation \cite[Chapter~9]{Preskill2004} and an early description of non-Abelian magnetic fluxes is given in Ref. \cite{Bais1980} (see also Refs. \cite{Bais1992} and \cite{WildPropitius1999}). Here we will instead examine the 3+1d case, as described in (for example) Ref. \cite{Bucher1992}.

Electric charges are point particles labelled by irreducible representations of the group $G$ and excite the vertex gauge terms. The gauge transforms $A_v^g$ at a particular vertex form a group, with the product $A_v^g A_v^h =A_v^{gh}$, which is isomorphic to $G$. The Hilbert space therefore splits into subspaces that transform as irreducible representations (irreps) of $G$ under the action of the gauge transforms at every vertex. The trivial irrep corresponds to states that are gauge invariant at that vertex, which can be thought of as the absence of an electric charge. On the other hand, if a state transforms as some other, non-trivial, irrep at a particular vertex, then that vertex will be excited. This means that we expect to find excitations that carry some non-trivial irrep of $G$, with this irrep describing how the excitations transform under the gauge transforms. These excitations should be produced in pairs, with a particle and anti-particle. We denote such a pair, associated to irrep $R$, by $(R,a,b)$, where $a,b$ are the matrix indices of the representation and describe an internal space for the pair. The irrep determines the action of the vertex operator applied on one of the particles: 
\begin{equation}
A_v^g \cdot (R, a,b) = \sum_c [D^R(g^{-1})]_{ac} (R,c,b),
\label{Excitation_Representation_1}
\end{equation}
where $D^R(g)$ is the matrix representation of element $g$. The label $g^{-1}$, rather than $g$, is used to ensure that the action of the $A_v^g$ satisfies the composition rule $A_v^g A_v^h =A_v^{gh}$. We could equally have defined the action of $A_v^g$ to be right multiplication by $D^R(g)$ instead, which would also satisfy the composition law. In the current prescription, this right-multiplication instead describes the transformation of the anti-particle under a vertex transformation at its position. This transformation under the vertex transforms can also be used to tell us something about the transport properties of the excitation. In Section \ref{HLGT_Gauge_Transforms} we explained that the vertex transforms are equivalent to parallel transport. Therefore Equation \ref{Excitation_Representation_1} tells us how we expect these excitations to behave under parallel transport over an edge labelled by $g^{-1}$. Looking at Equation \ref{Excitation_Representation_1}, we see that there is mixing between states defined by different matrix indices, while the irrep is unchanged. This suggests that the electric charges carry some conserved charge labelled by the representation, while the matrix indices describe some non-conserved details.

\begin{figure}[h]
	\begin{center}
		\includegraphics[width=0.6\linewidth]{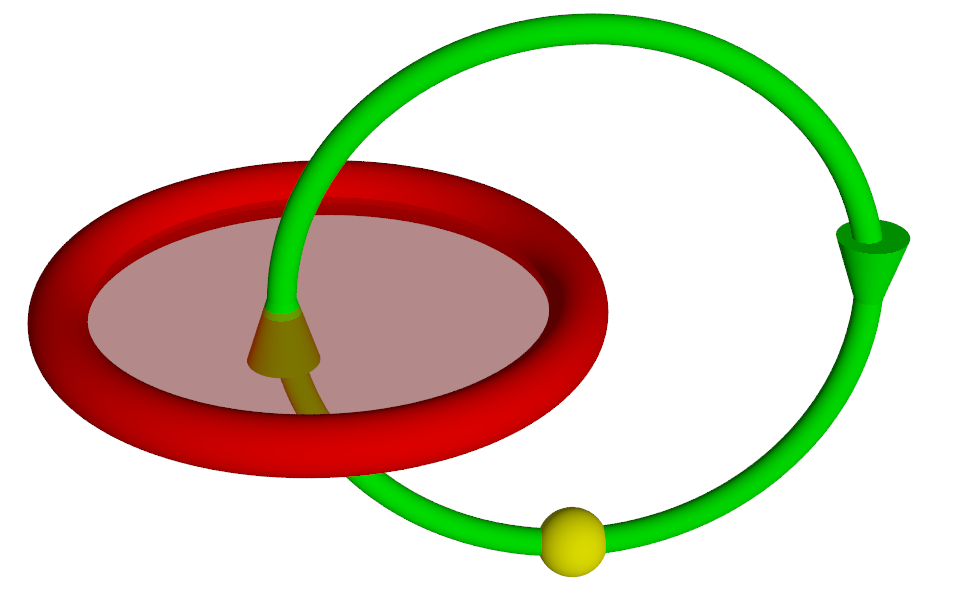}

		\caption{The magnetic excitation (thick red torus) is measured by a non-trivial closed loop linking with it, such as the thin green line. This closed loop begins at some start-point (shown as a yellow sphere) and the value of the flux that we measure depends on which start-point we choose, but the conjugacy class of the flux does not.}
		\label{Flux_tube_measurement}
	\end{center}
\end{figure}

In addition to the electric charges, lattice gauge theory hosts magnetic flux tubes. Recall from Section \ref{Section_Gauge_Invariants_LGT} that fluxes are associated with closed paths that have non-trivial labels. In that section, we drew an analogy to how a magnetic field leads to a non-trivial Aharanov-Bohm effect for taking a charge around a closed loop. In a 2+1d model, flux can be created by a point particle, which we can think of as being similar to a magnetic field penetrating our surface at a point. If this point particle generates a flux, then this flux should be measured by a closed loop that encloses the particle. Therefore we can describe this particle with the closed loop that measures the flux. However, a point particle cannot be sensibly described by a closed loop in a 3+1d topological theory. This is because any closed loop around a point particle can be smoothly deformed away from that particle and contracted to nothing, i.e., to a path with a trivial label. Therefore the label of the path cannot be a topological quantity if the excitation generating the flux is a point particle. Instead the magnetic flux particles should be closed loops (flux tubes). The flux generated by such a tube can be measured by a closed path that links with the flux tube, as shown in Figure \ref{Flux_tube_measurement}. Then there is no way to smoothly contract the measurement path without it intersecting with the flux tube, and so the label of the path can be a topological quantity.

\begin{figure}[h]
	\begin{center}
	\includegraphics{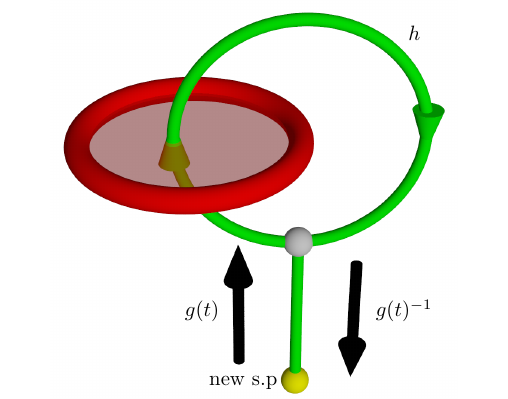}
		\caption{With the start-point ($s.p$) at the higher (grey) sphere, the flux label is $h$, however with the $s.p$ at the lower (yellow) sphere the flux is $g(t)hg(t)^{-1}$.}
		\label{flux_change_sp}
	\end{center}
\end{figure}

We can label a flux tube by the group element of the closed path that measures the flux. This label describes how a charge would evolve as it travels along that closed path. However, the value we assign to a path depends on the start-point of that path. To see this, we consider taking a particular closed path that links with that excitation and then changing its start-point, as shown in Figure \ref{flux_change_sp}. In order to traverse the new path, we must first travel the path joining the start-points, then the original closed path, and then back along the path joining the start-points. If the original path has label $h$, then the new one will have label $g(t)hg(t)^{-1}$, where $t$ is the path between the start-points. We therefore see that this new path has a label that is different from the label of the original path, but which lies in the same conjugacy class. This means that the conjugacy class describes the excitation in a robust way, but to obtain a full description of the flux tube we must also specify its element within that conjugacy class and the start-point from which we measure that value.

\begin{figure}[h]
	\begin{center}
	\includegraphics{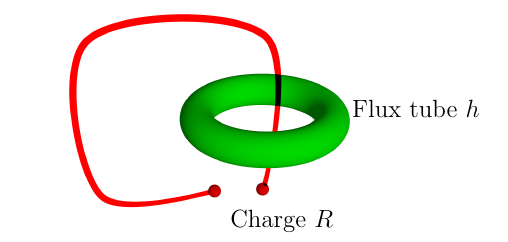}
		\caption{Schematic view of braiding a charge through a loop. The red line tracks the motion of the charge.}
		\label{chargethroughloop2}
	\end{center}
\end{figure}

This geometric picture for the two types of excitations also gives us their braiding relations. We already established how an electric excitation should transform as it moves through space. We can now creating a pair of electric excitations, labelled by an irrep $R$, and moving one on a closed path that links with a flux tube labelled by $h$, as shown in Figure \ref{chargethroughloop2}. If this closed path is the one used to define the flux, then the path label is given by the flux label of the flux tube, $h$. Therefore the object $(R,a,b)$ should become $\sum_c [D^R(h)]_{ac} (R,c,b)$ after the motion. On the other hand, if the path has a different start-point to the defining path of the flux, we should replace $h$ with some other element $g(t)hg(t)^{-1}$ to describe the transport of the charge to and from the start-point (along path $t$) as well as around the defining loop of the flux. This gives us the braiding relation
\begin{align*}
(R,a,b) &\rightarrow \sum_c \sum_d \sum_e [D^R(g(t))]_{ac} [D^R(h)]_{cd}\\
& \hspace{0.5cm} \times [D^R(g(t)^{-1})]_{de}(R,e,b)\\
&= \sum_e [D^R(g(t)hg(t)^{-1})]_{ae} (R,e,b), 
\end{align*}
from which we see that the charge experiences the same transform as if it travelled around a flux of $g(t)hg(t)^{-1}$, as expected from the rules for changing the start-points of fluxes.

We can also obtain the braiding relations of the magnetic fluxes using this picture. Consider the case where we have two flux tubes, which we define with the same start-point. We want to keep track of the measurement paths and the flux labels, as we move the fluxes around. We consider exchanging the two flux loops by pushing one through the other, as shown in Figure \ref{braiding_fluxes_1}. When we move a flux loop, the measurement path (which is associated with the flux label) moves with it so that the measurement path and flux tube remain linked (we can imagine the flux tube dragging the measurement path with it). For example, in the top-right part of Figure \ref{braiding_fluxes_1}, which shows the situation after we perform the braiding move, we see that the measurement path for the initially lower flux tube (the blue tube) is pulled through the other (now lower) tube (the red tube). This new deformed path carries the original flux label ($h$ in Figure \ref{braiding_fluxes_1}) and so this flux label is now associated to a process where we pull a charge through the lower loop, then around the upper one and then back through the lower loop, rather than a process where we simply braid the charge around the upper loop. We want to define our fluxes with respect to our original measurement paths, in order to find the labels associated to the original measurement processes and so to find the change to the system under braiding. That is, we want to find the labels associated to the original measurement paths ($\alpha$ and $\beta$ in Figure \ref{braiding_fluxes_1}). To do this, we need to write the original paths in terms of the new deformed ones, for which we know the path labels. This will allow us to obtain the labels of the original paths and so tell us the result of braiding our fluxes.

\begin{widetext}
	\begin{minipage}{\linewidth}
		\begin{figure}[H]
			\begin{center}
				\includegraphics{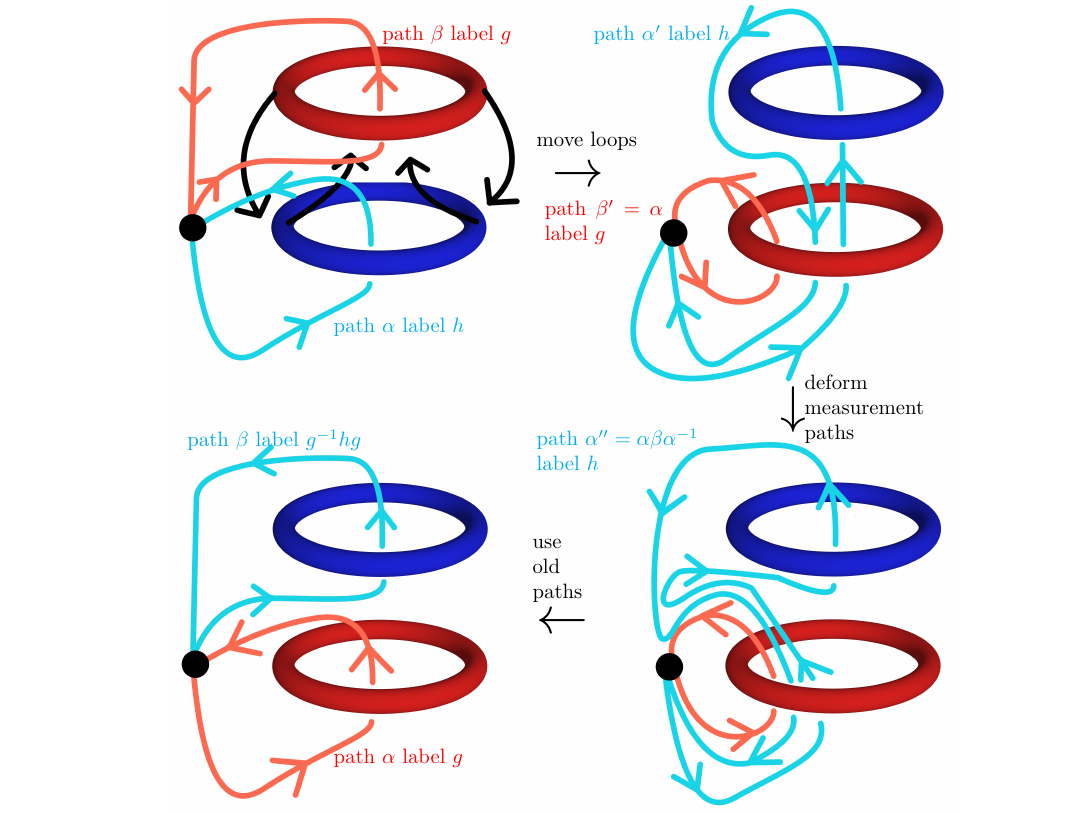}
				\caption{Starting with two fluxes (top-left) we can move the lower one through the upper one, swapping their positions. When we do so, we must also move the measurement paths associated with each flux (top-right). Then we can deform these paths (bottom-right) to write them as products of the original paths. Knowing the labels for these new paths (which are just the original flux labels of the two loops) allows us to find new labels for the fluxes, when measured along the original paths (bottom-left)}
				\label{braiding_fluxes_1}
			\end{center}
		\end{figure}
		
	\end{minipage}
\end{widetext}

Looking at Figure \ref{braiding_fluxes_1}, we see that $\beta$ is the path originally associated to the upper (red) flux, with label $g$. When we deform space to push the lower (blue) flux tube through the upper (red) flux tube, this path $\beta$ is deformed to $\beta'$. This means that the label of this new path, $\beta'$, is equal to the original label $g$ of path $\beta$. However, this path is equivalent to (i.e., can be smoothly deformed into) the original path $\alpha$ around the old lower flux. So we have $\beta' = \alpha$ and so the new label of the path $\alpha$ is $g$ (which is now associated with the red flux tube). On the other hand consider the path $\alpha$ originally associated with the blue (initially lower) flux. When we move the fluxes, this path is deformed into $\alpha'$ and it keeps its label of $h$, as indicated in the upper right diagram. We want to write $\alpha'$ in terms of our old paths $\alpha$ and $\beta$. To do this we note that $\alpha'$ can be smoothly deformed into another path $\alpha''$, which is equal to the path $\alpha \beta \alpha^{-1}$ obtained by traversing $\alpha$ then $\beta$ and then $\alpha$ in reverse, as shown in the bottom-right figure. Therefore we have $\alpha' = \alpha \beta \alpha^{-1}$ and so $\beta = \alpha^{-1} \alpha' \alpha $. Using the fact that $\alpha$ now has the label $g$ and $\alpha'$ has the label $h$, we see that $\beta$ has the label $g^{-1}hg$. We can write this braiding relation in the following way. We start with $(h,g)$, where the first symbol in brackets is the label given to path $\alpha$ and the second is the one given to path $\beta$. Then under braiding we have $(h,g) \rightarrow (g,g^{-1}hg)$, where on both sides the first symbol refers to the value of path $\alpha$ and the second to the value of path $\beta$, rather than giving the label of a particular one of the excitations (blue or red). If we instead keep track of the labels of each tube, we see that $h \rightarrow g^{-1}hg$ for the blue tube and $g \rightarrow g$ for the red tube. Therefore we see that the label of one of our flux tubes is conjugated by the label of the other one under our braiding.

 \subsection{Higher gauge theory}

We have so far described the excitations for ordinary lattice gauge theory, but we can use very similar ideas to explore higher lattice gauge theory. Our vertex terms still have the same algebra as in ordinary gauge theory. Namely, we have a group isomorphic to $G$ at each vertex. Therefore we expect to find electric charges labelled by irreps of $G$, just as with lattice gauge theory. Again, under parallel transport the charges will transform according to this irrep. This suggests that our electric excitations will be largely unchanged when compared to those from the Quantum Double model. Similarly, we expect to find magnetic flux tubes that are similar to those from lattice gauge theory. However, there is some subtlety in considering the braiding between these two types of excitation. This is because the lattice does not satisfy flatness in the ground state, but instead fake-flatness. This means that deforming a path over an unexcited region causes the path label to pick up a factor of the form $\partial(e)$. As we discussed earlier, moving an electric excitation through space, such as when we braid it around a magnetic flux tube, causes it to transform according to the label of the path traversed. This suggests that the result of braiding an electric charge around such a tube depends on the precise path chosen for the braiding, not just its homotopy class, implying that the braiding relation is not topological. As we shall discuss further in Section \ref{Section_Part_One_Confinement}, the resolution to this is that any electric excitation that is sensitive to such factors of $\partial(e)$ must be confined (i.e., cost energy to separate from its antiparticle), and so is not topological. In addition, the fact that we have fake-flatness rather than flatness indicates that a closed path may have a non-trivial label $\partial(e)$ even in the ground state. This implies that fluxes with label in $\partial(E)$ cannot be distinguished from trivial fluxes just by measuring the closed path. Furthermore, deforming the measurement path for a magnetic flux tube will change the label measured by an element of $\partial(E)$. Therefore, when we talk about the flux of a magnetic excitation, we should only define it up to elements in $\partial(E)$. This leads to magnetic excitations with label in $\partial(E)$ becoming topologically trivial, in a sense that we explain in Section \ref{Section_Part_One_Condensation}.

In addition to the vertex gauge transforms we also have the 2-gauge (or edge) transforms, which again have a group structure. These operators form a group isomorphic to $E$ for each edge: $\mathcal{A}_i^e \cdot \mathcal{A}_i^f = \mathcal{A}_i^{ef}$. Therefore we expect to find edge excitations that are labelled by irreps of the group $E$. Recall that the 2-gauge transform is equivalent to parallel transport of an extended object (a line object) over a surface. Therefore an object which transforms as a particular irrep under the 2-gauge transform should also transform as that irrep under parallel transport over a surface. Because of the fact that this transport is over a surface rather than a path, we expect our ``2-charges" to be extended objects. In fact, we will find that these 2-charges are loop-like objects. Then when we transport a loop, labelled by an irrep $\mu$ and matrix indices $a$ and $b$, over a surface labelled by $e$, we should obtain the transformation $(\mu,a,b) \rightarrow \sum_c [D^\mu(e^{-1})]_{ac} (\mu,c,b)$. There is some subtlety to this, however. Whenever we define a surface element we must give it a base-point. If we change the base-point we change the label of that surface from $e$ to some $g \rhd e$ for some $g \in G$. Then how do we know if we should have $[D^\mu(e)]$ or $[D^\mu(g \rhd e)]$ in our transformation when we transport the loop? That is, where should we take the base-point of our surfaces? The answer is that we must define the 2-charges with respect some start-point, just like the flux excitations. When we move the loop over a surface, we always take the label of that surface with respect to the start-point of our loop excitation. This start-point is particularly important for these loop excitations, because the action of the group $G$ on the start-point, which changes the surface label by some $g \rhd$ map, enables $G$ to affect the loop excitations. This action can even change the irrep labelling a 2-charge loop, suggesting that the irrep is not a conserved quantity. Instead there is some mixing within certain classes of irreps, which we term $\rhd$-Rep classes of irreps of $E$, with the irreps in a particular class being related by the action of $\rhd$. When $E$ is Abelian we define the classes with the equivalence relation
\begin{equation}
\mu_1 \sim \mu_2 \iff \exists g \in G \ \text{s.t} \ \mu_1(e) = \mu_2(g \rhd e) \  \forall e \in E
\end{equation} 
and when $E$ is non-Abelian we must generalize this to account for irreps related by conjugation. We therefore see that $\rhd$ plays a significant role in determining how the excitations behave.

Just as we have magnetic fluxes that are associated with non-trivial loops, there are also ``2-fluxes" associated to non-trivial closed surfaces, labelled by elements in $E$. We expect excitations corresponding to the 2-flux of a sphere to be point excitations, because we can shrink a sphere to enclose just a single point. Because these 2-fluxes are measured by closed surfaces, and every surface must be defined with a base-point, we must choose a base-point for our 2-flux excitation, which we call the start-point of the excitation. Note that we use the term \textit{start-point} to refer to privileged vertices related to the excitations, while \textit{base-point} is used to refer to the base-point of a surface. For the 2-flux excitations the base-point of our measurement surface is the start-point of the excitation. Moving the base-point of a surface along a path labelled by $g$ changes the surface label from $e$ to $g^{-1} \: \rhd e$, so similarly changing the start-point of our 2-flux changes its label by this $g^{-1} \: \rhd$ action. We also expect the 2-flux excitation to transform in this way as it moves along a path. This tells us that the 2-flux label is not conserved, but rather each group element belongs in a class of elements related by the $\rhd$ action. The equivalence relation defining such a class is that two elements $e,f \in E$ satisfy 
\begin{equation}
e \sim f \iff \exists \: g \in G \text{ such that }e= g\rhd f.
\end{equation}
These ``$\rhd$-classes" are then conserved under motion.

As this picture gives us the transport properties of these excitations, we can obtain their braiding relations as well. We know how our $E$-valued loops, the 2-charges, transform under transport over a surface, which tells us how they transform when pulled over the surface assigned to a 2-flux. For a loop excitation $(\mu,a,b)$ and a 2-flux $e$, defined with the same start-point, the loop excitation becomes $\sum_c [D^\mu(e^{-1})]_{ac} (\mu,c,b)$ when it is pulled over the 2-flux. We can also work out how the 2-fluxes braid with ordinary fluxes. When a 2-flux moves along a path $t$, the 2-flux label changes from $e$ to $g(t) \rhd e$. Therefore when moving a 2-flux labelled by $e$ around an ordinary magnetic flux labelled by $h$, the 2-flux becomes $h \rhd e$ (or $h^{-1} \rhd e$, depending on the orientation of the magnetic flux).

This picture therefore tidily describes several types of simple excitation that we expect to find. However, as explained in Ref. \cite{Bullivant2018}, there may be more complicated excitations as well. We may expect to find loop particles that generate both a 2-flux and a 1-flux. The non-trivial magnetic 1-flux is associated to a closed path that links with the excitation, while the 2-flux corresponds to a spherical surface enclosing that excitation. In Ref. \cite{Bullivant2018}, the braiding relation between two such loops is established, using geometric arguments and study of the loop braid group. The authors look at the situation where they braid two such excitations labelled by $(g,e)$ and $(h,f)$, where the first label of each pair gives the magnetic flux and the second the 2-flux. When the excitation labelled by $(g,e)$ is pushed through the one labelled by $(h,f)$, the excitations should transform under braiding to become $(h^{-1}gh, h^{-1} \rhd e)$ and $(h, e f [h^{-1} \rhd e^{-1}])$. As will be explained in Section \ref{Section_Part_One_Braiding}, we do indeed find such excitations with these braiding statistics in the lattice model.

\section{Excitations}
\label{Section_Part_One_Excitations}

The aim of this study is to find the excitations of the higher lattice gauge theory model and their properties. In the previous section, we gave brief arguments about these characteristics from geometric arguments. However, we wish to show that the lattice model does indeed support such excitations, and give a fuller description of their properties. A significant feature of this model is that we can explicitly find the operators to produce the excitations in various broad cases. Here we will explicitly construct these operators for the higher lattice gauge theory model and use the operators to study the excitations directly.

 Point-like excitations are produced by \textit{ribbon operators}. These ribbon operators act on a linearly extended region (often with some finite width), called a ribbon, and produce excitations at the two ends of the ribbon. One of the defining properties of the ribbon operators that produce the topological excitations is that they commute with the Hamiltonian everywhere except at the start and end of the ribbon and so act to produce a pair of anyons \cite{Kitaev2003, Levin2005}. Because the bulk of the ribbon does not produce any excitations, the ribbon itself is largely invisible apart from its end-points. Indeed, ribbon operators are \textit{topological}, in the sense that they can be smoothly deformed through the ground state (or any unexcited region of the lattice) without affecting the action of the operator. Important exceptions to this are the ribbon operators which produce confined excitations. These excitations cost energy to separate from their anti-particle, and so the corresponding ribbon operators have an energy cost associated to the length of the ribbon and are therefore not topological (the location of the ribbon can be detected by the energy terms along the length of the ribbon).

\hfill

The fact that the topological excitations must be produced in pairs at the ends of ribbon operators, rather than locally, is no accident. These excitations carry a conserved charge, known as topological charge. In 3+1d, there are multiple types of topological charge, corresponding to different measurement surfaces, as we explain in Section \ref{Section_Part_One_Topological_Sectors} (and in more detail in Ref. \cite{HuxfordPaper3}). In this work we will consider the charge measured by a sphere, which captures the point-like character of an excitation or set of excitations, and the charge measured by a torus, which captures the loop-like character. Because this charge is conserved, it can only be produced in one region by moving it out of another region. Ribbon operators do exactly this, and the charge carried by the excitation produced at one end of the ribbon operator must be balanced by the charge carried by the excitation at the other end.

\hfill

Instead of being produced at the ends of ribbon operators, the loop-like excitations are created at the boundary of so-called \textit{membrane operators}. As the name suggests, membrane operators act across some extended surface in the 3d spatial lattice, often with some finite thickness. These membrane operators must be applied on unexcited regions of the lattice, otherwise they may produce additional excitations or in some cases become ill-defined. The membrane operators, just like the ribbon operators, are topological, meaning that the membrane is largely invisible and deforming it through an unexcited region without changing its boundary leaves the action of the corresponding operator unchanged.

\hfill

These ribbon and membrane operators carry significant information about topological phases. They can be used to obtain the fusion rules for the associated particles, which describe how two anyons can be combined into a single particle \cite{Levin2005}. Furthermore, ribbon and membrane operators allow us to find the braiding statistics of the topological excitations, because these operators encode the creation and motion of the excitations. 

\hfill

In the following sections we construct the ribbon and membrane operators for the higher lattice gauge theory model and use the operators to study the properties of the excitations directly. We find four types of excitation in our model, with a rough correspondence to the four energy terms of the Hamiltonian. In 3+1d, two of these types of excitation are point-like and two types are loop-like.

\subsection{Electric excitations}
\label{Section_Part_One_Electric}

\begin{figure}[h]
	\begin{center}
	\includegraphics{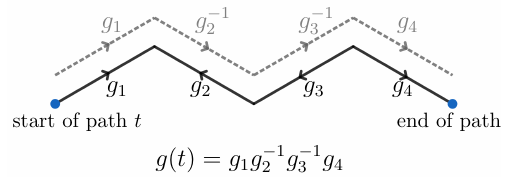}
		\caption{An electric ribbon operator measures the value of a path and assigns a weight to each possibility, creating excitations at the two ends of the path. In this example, the edges along the path are shown in black. Some of the edges are anti-aligned with the path and so we must invert the elements associated to these edges to find their contribution to the path element. This is represented by the grey dashed lines, which are labelled with the contribution of each edge to the path.}
		\label{path_image}
	\end{center}
\end{figure}

The first type of excitation we construct is called the electric excitation and is primarily associated to the vertex terms of our Hamiltonian. These electric excitations therefore correspond to the ``electric charges" that we described at the start of Section \ref{Section_Properties_From_Gauge_Theory}. In order to create these electric excitations, we measure the group element associated to some path on our lattice and apply weights depending on the result. In order to measure a path element, we take the product of edge elements along the path, with inverses if the orientation of the edge is against the orientation of the path. An example of this is shown in Figure \ref{path_image}. In order to measure each possible value of the path label, we apply a ribbon operator of the form
\begin{equation}
S^{\vec{\alpha}}(t) = \sum_{g \in G} \alpha_g \delta(\hat{g}(t),g)
\label{Equation_electric_ribbon_1}
\end{equation}
where $\hat{g}(t)$ is the path element for path $t$ and $\alpha_g$ is a coefficient (or weight) for the element $g$. This operator can excite the two vertex terms at the ends of the path $t$. We call the start of path $t$ the start-point of the operator, and it can be thought of as the position where the pair of excitations is created before the excitations are moved.

These electric excitations are equivalent to the electric excitations found in Kitaev's Quantum Double model \cite{Kitaev2003}, even up to the precise form of the ribbon operator that creates them. As we go on we will find that several features of the Quantum Double model (which is based on lattice gauge theory) carry over to the higher lattice gauge theory model. However, we will also see important distinctions between the two models. For instance, as we describe in Section \ref{Section_Part_One_Confinement} (and prove in Section S-I of Ref. \cite{HuxfordPaper2}, with the proof holding in both 2+1d and 3+1d), some of our electric excitations are confined, with the ribbon operator having an energy cost that scales with its length.

Any set of weights $\alpha_g$ that we choose in Equation \ref{Equation_electric_ribbon_1} will give a valid ribbon operator. Varying the weights therefore takes us through a space of these electric ribbon operators. A particularly useful basis for this space has the weights described using representations of $G$, as we anticipated in Section \ref{Section_Properties_From_Gauge_Theory}. Each basis element is labelled by an irreducible representation (irrep) of the group $G$, along with two matrix indices. Then for the irrep $R$ and the indices $a$ and $b$, the corresponding basis ribbon operator is given by
\begin{equation}
\hat{S}^{R,a,b}(t)= \sum_{g \in G} [D^{R}(g)]_{ab}\delta( \hat{g}(t), g), \label{Equation_electric_irrep_basis_1}
\end{equation}
where $D^{R}(g)$ is the matrix representation of the element $g$ in the irrep $R$. The operator labelled by the identity irrep is then the identity operator 
$$1 = \sum_{g \in G} \delta(g, \hat{g}(t))$$
and so does not produce any excitations. Any other basis operator (i.e., an operator labelled by a non-trivial irrep) does produce excitations at the two ends of the ribbon.

In addition to determining which operators excite the vertices, this basis is a good choice for examining the topological charge of the excitations. As we show in Ref. \cite{HuxfordPaper3} (in Section IX A 1), the point-like topological charge for (non-confined) pure electric excitations is labelled by the irreps of the group $G/ \partial(E)$ (the quotient removes the confined excitations) and the basis operators given above transport definite values of topological charge. That is, in the operator above $R$ labels a conserved charge, while the matrix indices $a$ and $b$ describe some internal space to the sector (as we expect from our discussion of gauge theory in Section \ref{Section_Properties_From_Gauge_Theory}). In particular, when $R$ is the trivial irrep the ribbon operator transports the vacuum charge, as we require from the fact that the ribbon operator is just the identity. On the other hand, if $R$ is non-trivial in the subgroup $\partial(E)$ then the excitation is confined.

\subsection{Blob excitations}
\label{Section_Blob_Excitations_1}
\begin{figure}[h]
	\begin{center}
	\includegraphics{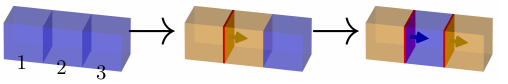}
		\caption{We consider a series of blobs in the ground state (leftmost image). In the ground state, all of the blob terms are satisfied, which we represent here by colouring the blobs blue (dark gray in grayscale). Changing the label of the plaquette between blobs 1 and 2 excites both adjacent blobs, as can be seen in the middle image (we represent excited blobs by colouring them orange, or lighter gray in grayscale). Multiplying another plaquette label on blob 2 to try to correct it just moves the right-hand excitation from blob 2 to blob 3 (rightmost image). In each step, the plaquettes whose labels we changed are indicated by the (red) squares and their orientations are indicated by an arrow.}
		\label{blob_ribbon_operator_sequential}
	\end{center}
\end{figure}

The next point-like excitations that we find are called blob excitations, because they primarily correspond to violations of the blob terms of the Hamiltonian. These excitations are therefore associated to the non-trivial closed surfaces we discussed in Section \ref{Section_Gauge_Invariants_HLGT} of the introduction. That is, the blob excitations correspond to non-trivial ``2-fluxes" on a sphere. In order to excite a blob term, which enforces that the surface element of that blob is trivial, we can multiply one of the plaquette labels on that blob by an element $e$ of $E$. However, each plaquette is shared by two blobs, one on either side of the plaquette. Changing a plaquette's label will therefore excite both of the adjacent blobs. We can try to correct this by changing another plaquette on this second blob. However, that plaquette will in turn be connected to a third blob, which will become excited, as illustrated in Figure \ref{blob_ribbon_operator_sequential}. This means that we simply move the second excitation from the second to the third blob. We therefore see that we produce these blob excitations in pairs, just as with the electric excitations. The series of plaquettes that we have to change to produce the excitations forms a string that passes through the centres of the blobs (3-cells). The operator that changes the plaquette labels is therefore another of our ribbon operators.

The precise action of our blob ribbon operator depends on which special case from Section \ref{Section_Special_Cases} we consider. In the simplest case, where $\rhd$ is trivial (Case 1 in Table \ref{Table_Cases_intro}), the action is fairly simple. We choose an element of $E$ to label the operator. We also choose a path on the dual lattice, which passes between the centres of blobs just as a path on the direct lattice passes between vertices. This path cuts through the plaquettes that separate the blobs (analogous to direct paths passing along edges), such as the red plaquettes in the example in Figure \ref{blob_ribbon_operator_sequential}. The choice of element $e \in E$ and path $r$ gives us a blob ribbon operator $B^e(r)$. The action of this operator is just to multiply the labels of all of the plaquettes pierced by this dual path by $e$ or $e^{-1}$, depending on the orientation of these plaquettes relative to the ribbon (where the orientation of the plaquette is obtained from its circulation using the right-hand rule). This results in the two blobs at the end of the path being excited, as we discussed for the example in Figure \ref{blob_ribbon_operator_sequential}. As we explain in Section \ref{Section_Part_One_Confinement}, some of the blob ribbon operators (those labelled by an element $e \in E$ for which $\partial(e)$ is not $1_G$) also excite the plaquettes pierced by the ribbon operator, resulting in the corresponding blob excitations being confined.

 We must modify the action of the blob ribbon operators slightly if $\rhd$ is non-trivial, as in Cases 2 and 3 of Table \ref{Table_Cases_intro}. When $\rhd$ is non-trivial, we must keep track of the base-points of the plaquettes that we want to change. We first move all of their base-points to a common location at the start of our operator (for example, the base-point of the first plaquette that we want to change), which we call the start-point. Then we multiply the label of each plaquette by $e$ or $e^{-1}$ before moving their base-point back to its original location. Recall from the introduction that moving the base-point of a plaquette along path $t$ changes its label from $e_p$ to $g(t)^{-1} \rhd e_p$ (see Figure \ref{whiskering_algebraic} for a reminder). The total change to $e_p$ is therefore 
\begin{align*}
e_p &\rightarrow g(t)^{-1} \rhd e_p \text{ (move base-point)} \\
&\rightarrow [g(t)^{-1} \rhd e_p] \: e^{-1} (\text{ postmultiply by }e^{-1})\\
&\rightarrow g(t) \rhd ([g(t)^{-1} \rhd e_p] \: e^{-1}) \text{ (move base-point back)}\\
&= e_p [g(t) \rhd e^{-1}]
\end{align*}
where $t$ is the path from the base-point $v_0(p)$ of plaquette $p$ to the start-point of our operator. This gives us the action
\begin{align}
&B^e(r):e_p  \notag \\
&= \begin{cases}  e_p [g(s.p(r)-v_0(p))^{-1} \rhd e^{-1}] & \text{if $p$ is aligned} \\ & \text{ with $r$} \\ [g(s.p(r)-v_0(p))^{-1} \rhd e] e_p & \text{if $p$ is anti-aligned} \\ &\text{ with $r$,} \end{cases}
\end{align}
where $g(s.p(r)-v_0(p))$ is the path element from the start-point of our operator to the base-point of $p$ and the plaquette is aligned with the ribbon if the circulation of the plaquette can be obtained from the local direction of the ribbon by using the right-hand rule. The path from the start-point to the base-point of the plaquette can be deformed over a fake-flat region without affecting the action of the operator, so its precise position is not usually important. However, the start-point itself is important and the vertex term there may be excited by the ribbon operator. This is because the vertex transform affects the path element $g(s.p(r)-v_0(p))$ and so does not commute with the ribbon operator in general. While we leave a more detailed discussion and proof of this for Ref. \cite{HuxfordPaper3}, we can understand this somewhat intuitively using the geometric interpretation discussed in Section \ref{Section_Properties_From_Gauge_Theory}. The blob excitations correspond to non-trivial 2-fluxes, and these 2-fluxes must be defined with respect to a base-point, which is the start-point of the ribbon operator. Moving this base-point induces a $g \: \rhd$ action on the label of the 2-flux. Then, because applying a vertex transform at a vertex is analogous to parallel transport of that vertex (see Section \ref{Section_HLGT}), the vertex transform also has a similar $g \: \rhd$ action on the label of the 2-flux produced by the ribbon operator. The ribbon operators which are invariant under this action therefore commute with the vertex transforms and so with the vertex energy term at the start-point. However, generic ribbon operators are not invariant under this action, which leads to some of them exciting the start-point. Specifically, given a linear combination $\sum_e \alpha_e B^e(r)$ of group-labelled blob ribbon operators, the start-point is not excited if the coefficients $\alpha_e$ are a function of $\rhd$-class, so that $\alpha_f = \alpha_{g \rhd f}$ for all pairs $g \in G$ and $f \in E$. On the other hand, the start-point is definitely excited if the sum of the coefficients in each $\rhd$-class is zero (i.e., $\sum_{g \in G} \alpha_{g \rhd e}=0$ for all $e \in E$). Ribbon operators that do not satisfy either of these conditions do not produce eigenstates of the vertex term when acting on the ground state, but can be written as a linear combination of operators that excite the vertex and operators that do not.

\subsection{$E$-valued loops}
\label{Section_E_Valued_Loops_1}
\begin{figure}[h]
	\begin{center}
		\includegraphics[width=\linewidth]{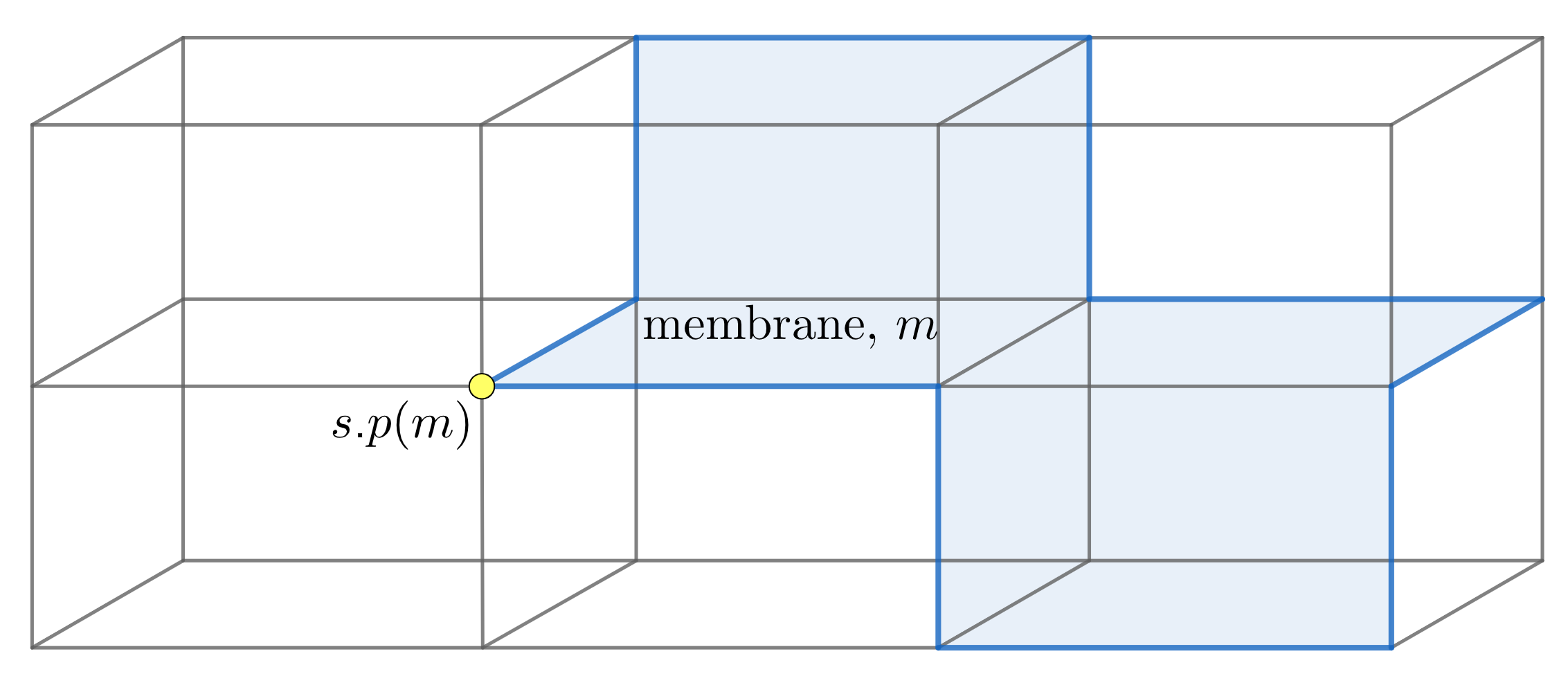}
		\caption{We consider applying an $E$-valued membrane on the shaded membrane $m$ in a fragment of the three-dimensional lattice. The membrane operator measures the surface label of the membrane, with a weight for each possible label. When measuring a surface, if $\rhd$ is non-trivial we must specify the base-point of that surface. The base-point of the surface measured by the membrane operator (the yellow dot) is called the start-point of the membrane. A non-trivial $E$-valued membrane operator excites the edges (solid blue lines) on the boundary of the membrane, and may also excite the start-point of the membrane.}
		\label{E_valued_membrane_example}
	\end{center}
\end{figure}

The first loop-like excitations that we find, which we call $E$-valued loop excitations, are produced by membrane operators which act primarily on the surface labels. These membrane operators measure the label of a surface in our lattice, using the rules for combining surface elements given in Sections \ref{Section_HLGT} and \ref{Section_Hamiltonian_Model} and apply a weight depending on the result. This is very similar to the electric ribbon operators except that the membrane operator measures a surface rather than a path. Indeed, if the electric excitations are charges for the 1-gauge field, then the $E$-valued loops are ``2-charges" corresponding to the higher gauge (2-gauge) field. In the same way as with the electric excitations, the weights describe a space of operators and an appropriate basis is given using irreps, this time of the group $E$. For an irrep $\mu$ and matrix indices $a$ and $b$, the operator acting on a membrane $m$ is given by
\begin{equation}
L^{\mu,a,b}(m)= \sum_{e \in E} [D^{\mu}(e)]_{ab} \delta(\hat{e}(m),e),
\label{E_membrane_definition}
\end{equation}
where $\hat{e}(m)$ is the total surface element of membrane $m$. This surface element can be written in terms of the labels of the plaquettes making up the membrane as
\begin{equation}
\hat{e}(m) = \prod_{\text{plaquettes }p \in m} g(s.p(m) -v_0(p)) \rhd e_p^{\sigma_p}, \label{Equation_surface_label_definition}
\end{equation}
where $v_0(p)$ is the base-point of the plaquette $p$, $s.p(m)$ is the base-point with respect to which we measure the surface label (and which we call the start-point of the membrane) and $\sigma_p$ is $\pm 1$ depending on the plaquette's orientation ($+1$ if it matches that of $m$ and $-1$ if it is anti-aligned with $m$). Note that when $E$ is non-Abelian, the order of the product in Equation \ref{Equation_surface_label_definition} is important and must be obtained by applying the rules for composing surfaces given in Sections \ref{Section_HLGT} and \ref{Section_Hamiltonian_Model}. There are generally many ways of composing the surfaces and each way is associated to different paths $(s.p(m)-v_0(p))$ and different orders of multiplication, which should give the same result as long as fake-flatness is satisfied on the membrane. Applying the membrane operator in Equation \ref{E_membrane_definition} causes the edges along the boundary of the surface to become excited (as long as $\mu$ is non-trivial), as indicated in Figure \ref{E_valued_membrane_example}. If $\mu$ is the trivial irrep, then the membrane operator is instead the identity operator, so the operator does not produce any excitations.

As with the blob excitations, there are some features that depend on which special case from Section \ref{Section_Special_Cases} that we look at. As we explained in Section \ref{Section_Special_Cases}, when $\rhd$ is trivial, the Peiffer conditions (Equations \ref{P1} and \ref{P2} in Section \ref{Section_HLGT}) imply that $E$ is Abelian. In this case, the irreps are all one dimensional, so we can drop the matrix indices $a$ and $b$ in Equation \ref{E_membrane_definition}. In addition, a trivial $\rhd$ means that we do not need to keep track of the start-point of the membrane.

On the other hand, when $\rhd$ is non-trivial, $E$ may be non-Abelian and so generally we must include the matrix indices. In addition, the start-point of the membrane becomes important and cannot generally be changed without affecting the action of the operator. Much as with the blob excitations, this start-point can be excited by the operator, which reflects the non-trivial transformation undergone by the operator when we move the start-point (due to the connection between vertex transforms and parallel transport). The start-point is not excited if the membrane operator is made of a linear combination $\sum_e \alpha_e \delta(\hat{e}(m),e)$ whose set of coefficients $\alpha_e$ is a function of $\rhd$-class (i.e., $\alpha_e = \alpha_{g \rhd e}$). If the start-point is not excited, then it can be moved without affecting the operator. On the other hand, the start-point may be excited if the coefficient $\alpha_e$ transforms non-trivially under the $\rhd$ action, as we describe in more detail in Refs. \cite{HuxfordPaper2} and \cite{HuxfordPaper3}. The start-point excitation is significant because it can carry a (point-like) topological charge, which must be balanced by a point-like charge on the loop itself, as we prove explicitly in Ref. \cite{HuxfordPaper3}, in Section IX A 1. Furthermore, as we describe in Ref. \cite{HuxfordPaper3} in Section V, this point-like charge can be confined for certain membrane operators, in which case the charge drags a line of excited edges between the start-point and the loop excitation. We note that this confinement can only occur when $E$ is non-Abelian, so it does not occur for Cases 1 and 2 in Table \ref{Table_Cases_intro}.

\subsection{Magnetic excitations}
\label{Section_Magnetic_Excitations_1}
\begin{figure}[h]
	
	\begin{center}
	\includegraphics{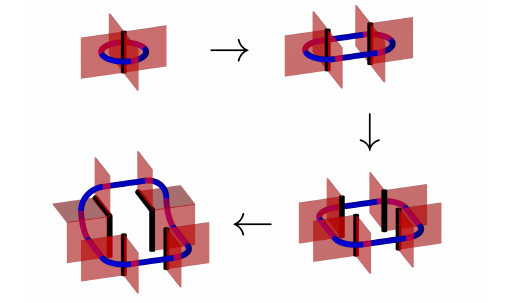}
		
		\caption{In order to excite one of the plaquettes in the lattice and produce a magnetic excitation, we change the label of one of the edges (black cylinders) on the boundary of the plaquette. However, this excites all of the plaquettes adjacent to that edge, as shown in the first image (the excited plaquettes are shown in red). Note that these plaquettes lie on a closed loop (blue tube) through their centres. If we change another edge label to try to prevent some of the plaquette excitations, we will excite the other plaquettes adjacent to that edge, as shown in the second image. Repeating the process, by changing the additional edges shown in black in each step, simply changes the shape of this loop (unless we change all of the edges bisected by a closed membrane and shrink the loop to nothing). }
		\label{magnetic_step_by_step}
	\end{center}
	
\end{figure}

\begin{figure}[h]
	\begin{center}
	\includegraphics{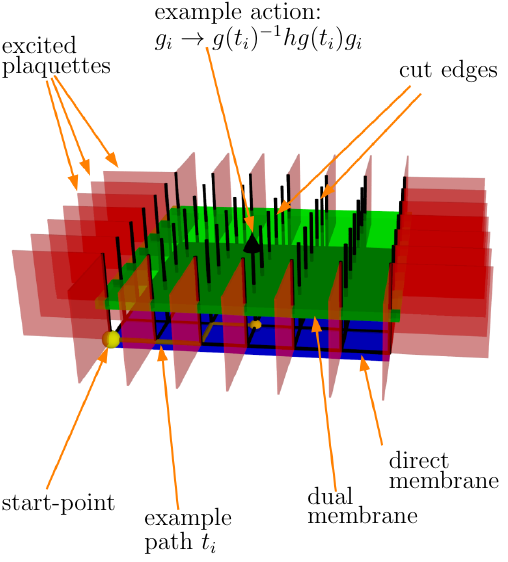}
		\caption{Here we give an example of the membranes for the flux creation operator (magnetic membrane operator). The dual membrane (green) cuts through the edges changed by the operator. The direct membrane (blue) contains a vertex at the end of each of these cut edges (such as the orange sphere). A path from a privileged start-point to the end of the edge (such as the example path, $t_i$) determines the action on the edge. This action leads to the plaquettes around the boundary of the membrane being excited.}
		\label{fluxmembrane1}
	\end{center}
\end{figure}

The final type of elementary excitation is the magnetic excitation, named so due to its correspondence to the magnetic excitations in Kitaev's Quantum Double model \cite{Kitaev2003} and its analogy to magnetic flux. The magnetic loop excitation is primarily associated with excitation of the plaquette energy terms. Recall from Section \ref{Section_Hamiltonian_Model} (see Figure \ref{plaquette_term_bigon_1}) that the plaquette term checks that the 1-flux of the plaquette is trivial, where the 1-flux is given by the product of edge elements around the boundary multiplied by the image under $\partial$ of the plaquette's surface element. In order to create an elementary flux excitation, we must excite the fewest number of plaquettes by changing edge labels (changing the plaquette label could also excite the plaquette, but this results in blob excitations as we saw in Section \ref{Section_Blob_Excitations_1}). We therefore consider trying to excite a single plaquette by changing one of the edges on that plaquette. However, in three spatial dimensions the edge will generally be shared by multiple plaquettes. Therefore changing the edge label will excite all of the plaquettes surrounding this edge. We can try to fix one of these additional excited plaquettes by changing the label of another edge on that plaquette, but this will in turn excite the other plaquettes surrounding that edge. We can repeat this process, but always get a closed string of excited plaquettes, as shown in Figure \ref{magnetic_step_by_step} (unless we collapse the loop to nothing). This means that the magnetic excitation is indeed loop-like. The edges that we have to change to produce the magnetic excitation are bisected by a membrane bounded by the excited loop, as shown in Figure \ref{fluxmembrane1}. We call the membrane cutting these edges the dual membrane, because the membrane bisects the edges of the lattice rather than lying on the lattice itself. The fact that we must change degrees of freedom across a membrane in order to produce a general magnetic excitation means that the creation operator is a membrane operator, which we call the magnetic membrane operator.

Next we will explicitly describe the action of the membrane operator. The features of the magnetic excitation depend strongly on the special case that we take. We first consider the $\rhd$ trivial case (Case 1 in Table \ref{Table_Cases_intro}). In this case the operator to produce the excitation is analogous to the 2+1d magnetic ribbon operator from Kitaev's Quantum Double model \cite{Kitaev2003}. We denote the magnetic membrane operator labelled by an element $h\in G$ and acting on a membrane $m$ by $C^h(m)$. When the group $G$ is Abelian, the action of this membrane operator is simple. We just multiply the labels of each of the affected edges (those cut by the dual membrane) by the element $h$ or its inverse, depending on the orientation of the edge. On the other hand, when the group is non-Abelian, we must multiply each edge by some element in the same conjugacy class as $h$ (or the inverse). To determine which element this is, we must first endow our operator with a privileged point, called the start-point. Furthermore, we must specify a path from this start-point to each edge cut by the membrane. Denoting the path to edge $i$ by $t_i$, the action of the membrane operator on edge $i$ is

\begin{equation}
	C^h(m): g_i = \begin{cases} g(t_i)^{-1}hg(t_i)g_i  &\text{if $i$ points away from} \\ & \text{ the direct membrane} \\g_i g(t_i)^{-1}h^{-1}g(t_i)  &\text{if $i$ points towards} \\ &\text{ the direct membrane} \end{cases}
\end{equation}

This action is shown in Figure \ref{fluxmembrane1}. The paths involved in this action lie on a second membrane, which we call the direct membrane, so the support of the membrane operator actually lies on both the direct and dual membranes, which we sometimes refer to together as the thickened membrane. This is analogous to how the ribbon operators in Kitaev's Quantum Double model act on a ribbon \cite{Kitaev2003}, which is a thickened string.

 The precise choice of the paths on the direct membrane is not usually significant, as we can deform them over a fake-flat region without affecting the action of the membrane operator. This is because the group elements assigned to two paths differing by such a deformation only differ by an element of $\partial(E)$, due to the fake-flatness condition. When $\rhd$ is trivial, elements of $\partial(E)$ are in the centre of $G$ (see Section \ref{Section_Special_Cases}) and so do not affect the expression $g(t_i)^{-1}hg(t_i)$. Because we usually apply membrane operators on regions without any other excitations, this means that we do not generally need to specify the precise positions of the paths. An important exception that we describe in Ref. \cite{HuxfordPaper3} is when we produce two linked magnetic excitations by applying intersecting membrane operators.

As with the blob and $E$-valued loop excitations, the start-point of the magnetic membrane operator may be excited. As mentioned previously, the magnetic excitations in this model are analogous to the magnetic excitations from Kitaev's Quantum Double model \cite{Kitaev2003}, and the potential start-point excitation is also present for the magnetic ribbon operators in that model (see for example Ref. \cite{Komar2017}). We can interpret the start-point of the magnetic membrane operators in this model, as well as the start-points of the magnetic ribbon operators in Kitaev's Quantum Double model, in terms of gauge theory. Recall from Section \ref{Section_Properties_From_Gauge_Theory} that whenever we measure a flux, we must do so with respect to a certain start-point. The flux created by $C^h(m)$ is only $h$ when we measure with respect to the start-point of the membrane operator (or ribbon operator for the Quantum Double model). Measuring the flux from a different point gives us a result of $\hat{g}(t)h\hat{g}(t)^{-1}$, where $\hat{g}(t)$ is a path element operator for which we are not generally in an eigenstate (even if there are no excitations present other than the magnetic flux). Note that the element $\hat{g}(t)h\hat{g}(t)^{-1}$ is still in the conjugacy class of $h$, indicating that this conjugacy class is independent of the start-point even if the flux element is not. This interpretation of the start-point is then connected to whether the start-point of a magnetic membrane operator is excited. The vertex transform at a vertex acts like parallel transport of that vertex, and so we can think of the vertex transform at the start-point as moving that start-point, which conjugates the flux label. In order to diagonalize the vertex term at the start-point, we must therefore take a linear combination of magnetic membrane operators with different labels in the conjugacy class of $h$ (so that we are considering a state in a superposition of different flux labels), as we prove in Ref. \cite{HuxfordPaper3}. If the start-point is unexcited, it means that the membrane operator is not sensitive to changes to the start-point from which we measure the flux, which occurs when the membrane operator produces an equal combination of fluxes in the conjugacy class (i.e., the coefficients of the linear combination are the same for each element in the conjugacy class). This equal combination of elements in the conjugacy class produces a flux tube with a trivial point-like charge (or in the 2+1d case such as the Quantum Double model, a pair of excitations that can be annihilated, as described in Ref. \cite{Preskill2004}).

\begin{figure*}[t!]
	\begin{center}
	\includegraphics{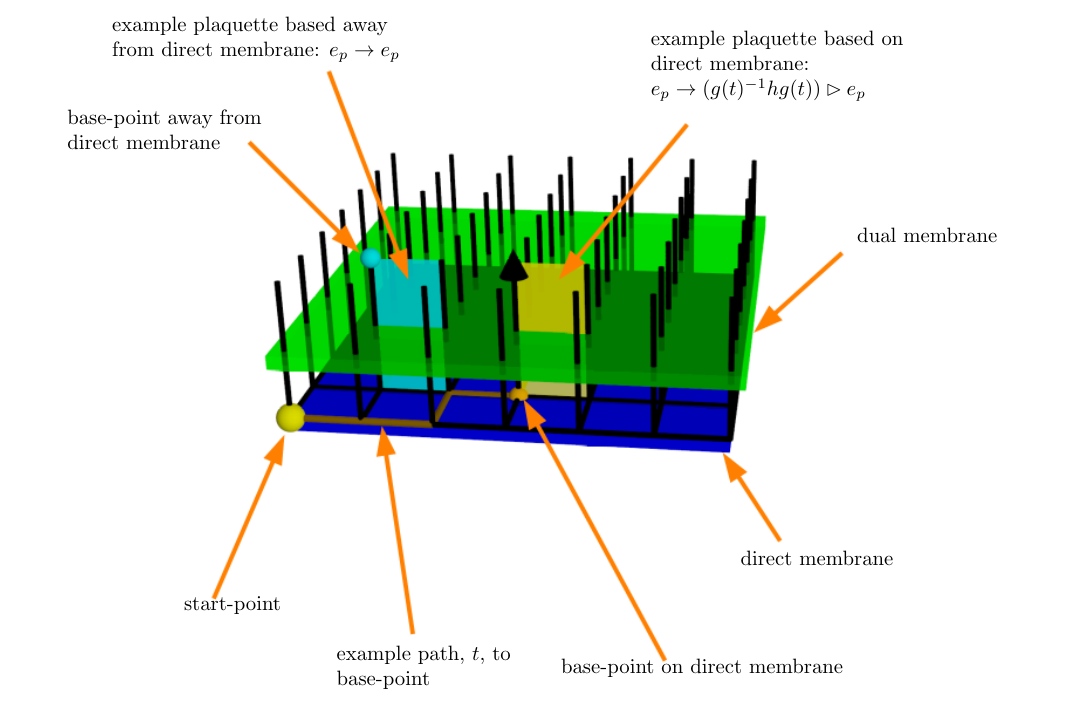}
		\caption{In addition to changing the edges cut by the dual membrane, when $\rhd$ is non-trivial the magnetic membrane operator affects the plaquettes cut by the dual membrane if their base-points lie on the direct membrane}
		\label{modmembranecutplaquettespart1}
	\end{center}
\end{figure*}

So far this consideration of the magnetic excitations has all been in the case where $\rhd$ is trivial. In the case where $\rhd$ is non-trivial but we restrict to fake-flat configurations (Case 3 in Table \ref{Table_Cases_intro}), we cannot include the magnetic excitations at all, because the magnetic excitations violate the plaquette terms and hence break fake-flatness. The most interesting case is Case 2 from Table \ref{Table_Cases_intro}, where we loosen the restrictions on the crossed module without throwing out the non-fake-flat configurations. Specifically, we require that $\partial$ maps to the centre of $G$ and that $E$ is Abelian. In this case, we are allowed to keep the magnetic excitations, though their operators must be modified. We briefly describe this modification here, but give a full description in Ref. \cite{HuxfordPaper3}. The new membrane operators act on edge elements in the same way as described above in the $\rhd$ trivial case, but they also act on the plaquette elements around the membrane in two ways. Firstly, the membrane operator directly affects the plaquettes that are cut by the dual membrane. If a cut plaquette $p$ has its base-point on the direct membrane, then its label $e_p$ is changed to $(g(t)^{-1}hg(t)) \rhd e_p$, where $g(t)$ is the path from the start-point of the operator to the base-point of the plaquette. Any plaquette whose base-point is away from the direct membrane is left unaffected. An example of this action is shown in Figure \ref{modmembranecutplaquettespart1}. This action on the plaquettes is analogous to how the vertex transform affects plaquettes based at that vertex, but not plaquettes that are not based at the vertex. Indeed, the vertex transform is like a closed magnetic membrane operator, whose dual membrane encloses that vertex and whose direct membrane is just the vertex itself (it is this equivalence that leads to the membrane operators being topological, as we explain in Ref. \cite{HuxfordPaper3}).

In addition to this $\rhd$ action, we have to multiply the membrane operator by blob ribbon operators, with these ribbons running from a special blob defined in the membrane operator, which we call blob 0, to the plaquettes of the direct membrane. In the $\rhd$ trivial case (Case 1 of Table \ref{Table_Cases_intro}), the plaquettes around the membrane were excited and the start-point could also be excited. In this more general case (Case 2 of Table \ref{Table_Cases_intro}) the special blob, blob 0, may also be excited, as may the edges and blobs surrounding the membrane. The word ``may" is important for the edges and blobs at the boundary of the membrane, because near an excited plaquette, the edge and blob terms cease to commute and also become inconsistent with changes to the branching structure of the lattice. This effect would only become worse if we lifted the condition on $\partial$. In that case, the blob ribbon operators that we add would generally be confined, which could lead to plaquette excitations away from the boundary, which is the reason that we do not consider the fully general case. In Case 2 the problematic plaquette excitations are restricted to the boundary of the membrane, and any topological quantities can be measured far from this boundary (if the membrane is sufficiently large), so the inconsistencies from the plaquette excitations are not important.

Due to the extra features of the magnetic membrane operator in this case, the magnetic excitation may carry both an ordinary 1-flux and a 2-flux, as we discussed in Section \ref{Section_Properties_From_Gauge_Theory}. Recall that a non-trivial 2-flux indicates a closed surface with a non-trivial label. We can see that the loop excitation must be associated to a non-trivial surface when blob 0 of the membrane operator is excited. This is because blob 0 being excited indicates that this blob carries a non-trivial 2-flux, which must be balanced by a 2-flux belonging to the loop-like excitation itself.

\section[Condensation and confinement]{Condensation and confinement}

\label{Section_Part_One_Condensation_Confinement}
\subsection{Confinement}
\label{Section_Part_One_Confinement}

As we alluded to in Section \ref{Section_Part_One_Excitations}, we found that some of the excitations in the higher lattice gauge theory model are confined, meaning that there is an energetic cost to separating particles (or growing and moving loop-like excitations) which grows at least linearly with separation (or with the area swept by the loop). Specifically, some of the point-like (electric and blob) excitations are confined. 

For the blob excitations, the mechanism for this confinement is the plaquette terms. Recall from Section \ref{Section_Blob_Excitations_1} that the blob ribbon operator $B^e(t)$ multiplies each plaquette $p$ pierced by the ribbon by an element $[g(s.p(t)-v_0(p))^{-1} \rhd e^{\pm 1}]$ of $E$. In addition to exciting the blobs at the ends of the ribbon, this action may excite the plaquettes that are pierced by the ribbon. This is because the plaquette term projects onto states for which the plaquette label $e_p$ and the path label $g_p$ of the boundary of the plaquette are related by $\partial(e_p)g_p=1_G$. If the ribbon operator changes $\partial(e_p)$ for the plaquettes that it pierces, it will excite those plaquettes. This occurs when $\partial(g(s.p(t)-v_0(p))^{-1} \rhd e)$ is non-trivial. The Peiffer condition Equation \ref{P1} states that $\partial(g \rhd e)= g \partial(e)g^{-1}$, and so $\partial(g(s.p(t)-v_0(p))^{-1} \rhd e)$ is the identity element when $\partial(e)=1_G$ regardless of the value of $g(s.p(t)-v_0(p))$. That is, the blob ribbon operator $B^e(t)$ excites every plaquette it pierces when $e$ is outside the kernel of $\partial$. In this case, the ribbon operator has an energetic cost that grows linearly with the length of the ribbon and so the associated blob excitations are confined. Note that for Case 3 in Table \ref{Table_Cases_intro}, where we exclude states that excite the plaquette terms from the Hilbert space, we must not allow the confined blob excitations.

We also find that some of the electric excitations are confined. The mechanism of this confinement is the edge terms along the ribbon. This is because the edge transforms change the path element measured by the electric ribbon operator by an element in $\partial(E)$ (the image of $\partial$) and so may fail to commute with the ribbon operator. To determine which electric ribbon operators are confined, we use the irrep basis for the ribbon operators. Given an electric ribbon operator labelled by irrep $R$ of $G$, we can determine whether the corresponding excitations are confined by evaluating the irrep $R$ on elements of the normal subgroup $\partial(E)$ and treating this as a representation of the subgroup. Restricting the irrep $R$ to the subgroup in this way produces a generally reducible representation of the subgroup. When we decompose this representation into irreps of $\partial(E)$, by Clifford's theorem \cite{Clifford1937} the constituent irreps will all be related by conjugation. This means that if one of the irreps of $\partial(E)$ found by restricting $R$ is the trivial irrep, then they all are (this result can also be obtained by Schur's Lemma in the case where $\partial(E)$ is in the centre of $G$). If $R$ branches to the trivial irrep in this way, then the ribbon operator will transform trivially when we alter the path element by an element in $\partial(E)$. This means that the ribbon operator commutes with the edge transforms and the electric excitations are not confined. On the other hand if $R$ does not branch to the trivial irrep then the ribbon operator does not commute with the edge transforms. In particular, it transforms as a non-trivial irrep of $\partial(E)$ under them and so gives zero when we act with the edge energy term, which is an average over all labels of the edge transform, due to the Grand Orthogonality Theorem. This means that all of the edges along the ribbon are excited and the corresponding excitations are confined. We note that this is equivalent to the confinement for the field theory discussed in Ref. \cite{Gukov2013}, where the electric operators are confined if they can detect factors in a subgroup $\pi_1(H)$ (equivalent to $\partial(E)$ here).

Some of the $E$-valued loop-like excitations can also be confined in a certain sense when $E$ is non-Abelian, as described in Section \ref{Section_E_Valued_Loops_1}. This confinement does not give an energy cost to growing the loop-like excitation, but instead it costs energy to move the excitation away from the start-point of the membrane. As we discussed in Section \ref{Section_E_Valued_Loops_1}, it seems like the point-like charge carried by the loop excitation is confined, rather than the loop-like charge. This is also reflected in the topological properties of the membrane operator. Normally, the creation operator (ribbon or membrane operator) for a confined excitation is not topological, because the position of the operator can be detected by the energy terms it excites. In the case of the membrane operators producing the particular confined loop-like excitations in this model, however, the membrane operator is still partially topological: we can deform the membrane without affecting the action of the operator, but we must keep the location of the excited edges fixed (and so the location of the confining string is fixed). This is again because it is the point-like charge that is confined (and so the motion of the point-like charge is not topological).

\subsection{Condensation}
\label{Section_Part_One_Condensation}

The phenomenon of confinement is closely related to a process known as condensation. Consider a topological model with some set of topological charges. By deforming the Hamiltonian, we may find that some of the non-trivial topological charges from the old Hamiltonian are present in the ground state of the deformed Hamiltonian. Because the ground state corresponds to the topological vacuum, this means that the excitations which carried the previously non-trivial charges now carry the trivial charge, although they may still remain energetic. We say that those excitations \textit{condense} \cite{Bais2009, Burnell2018, Neupert2016, Bais2003, Eliens2014}. When this occurs, any excitations that braided non-trivially with those excitations in the original become confined in the deformed model \cite{Bais2009, Burnell2018} (this is a bit of a simplification, but is sufficient to describe this model). This process is known as a condensation-confinement transition. While this is fairly well understood in 2+1d, there has been comparatively little study of such transitions in the 3+1d case (examples of work in this area include Refs. \cite{Burnell2013} and \cite{Ye2016}), and so it is interesting to see how condensation and confinement arise in this 3+1d model.

In the higher lattice gauge theory models, we can consider the process of condensation by constructing two models related by such a transition. That is, we consider cases where we can turn the confinement on and off by changing a parameter in the Hamiltonian. We cannot change the groups $G$ and $E$, which also fix the Hilbert space, but we can change the map $\partial$. When $E$ is Abelian we can construct a model with no confinement, described by the crossed module $(G,E,\partial \rightarrow 1_G,\rhd)$, and called the uncondensed model. Here $\partial \rightarrow 1_G$ indicates that $\partial$ maps to the identity of $G$. The condition that $E$ be Abelian is required from the second Peiffer condition, Equation \ref{P2}, which enforces that $\partial(e) \rhd f =efe^{-1}$ for every pair of elements $e$ and $f$ in $E$. Because $\partial(e)=1_G$ for the uncondensed model and $1_G \: \rhd$ is always the trivial map, the second Peiffer condition ensures that conjugation is also trivial: $efe^{-1} =1_G \rhd f =f$. Starting from this uncondensed model we can ``turn on" $\partial$ (while keeping the groups fixed), by moving into a model described by a crossed module where $\partial$ maps to some non-trivial subgroup of $G$. This can be done by interpolating between the two Hamiltonians, because the two models have the same Hilbert space. During this process, some of the excitations condense. When this occurs, the condensing excitations may still cost energy in the new model, but they carry trivial charge in this condensed phase. We therefore refer to such excitations in the condensed phase as condensed excitations. These condensed excitations can be produced by an operator local to the excitation. For loop-like excitations this means that a condensed loop-like excitation can be produced with an operator that only acts near the loop, rather than on an entire membrane. Indeed, finding such an operator to produce the loop excitation is one way to show that the excitation is condensed.

In this model, we find that some of the loop-like excitations are indeed condensed. Recall from Section \ref{Section_Magnetic_Excitations_1} that the magnetic excitations are labelled by elements of the group $G$. If that element is in the image of $\partial$, then the excitation is condensed. This can be seen from considering the magnetic excitations in light of the plaquette term. We discussed in Section \ref{Section_Properties_From_Gauge_Theory} that the magnetic excitations are associated with closed loops that have non-trivial label. However, the plaquette energy term enforces that the label of closed loops in our lattice match the image under $\partial$ of the surface element bounded by the loop, rather than just being the identity element. Therefore the ground state contains closed loops with all values in the image of $\partial$. A magnetic membrane operator with label in the image of $\partial$ modifies the labels of closed loops that link with the excitation only by multiplication by another element of $\partial(E)$, and so results in closed loop values already found in the ground state. Therefore the topological charges of the corresponding magnetic excitations (which we measure with closed paths) belong in the ground state, and so these charges have been condensed (in the uncondensed model, only $1_G$ is in the image of $\partial$). However, note that this does not mean that these magnetic excitations are not excitations at all. Changing the path label by an element of $\partial(E)$ still leads to plaquette excitations, because the path labels do not match the surface enclosed by the loop. In Ref. \cite{HuxfordPaper3} (in Section S-III) we explicitly show that the action of a condensed membrane operator on the ground state is equivalent to the action of a (confined) blob ribbon operator around the boundary of the membrane operator, which is local to the excitation, confirming that the corresponding loop-like excitation is condensed. Like the confinement of the electric excitations, this condensation is equivalent to that for the field theory discussed in Ref. \cite{Gukov2013}, where fluxes in the subgroup $\pi_1(H)$ (equivalent to $\partial(E)$) are condensed.

Some of the $E$-valued loops are also condensed. This is because some of the membrane operators that produce them are equivalent (when acting on the ground state) to an electric ribbon operator acting around the boundary of that membrane. The membrane operators for the $E$-valued loops measure the surface element of the membrane that they are placed on. However, in the ground state the group element $e_m$ assigned to a surface $m$ is related to the path around the boundary of the surface (labelled by $g(\text{boundary})$) by $\partial(e_m)g(\text{boundary})=1_G$, due to the plaquette terms. This suggests that we can measure the surface element just by examining the boundary, but because this expression involves only $\partial(e_m)$, this correspondence between the surface and boundary does not fully fix the value $e_m$. The surface element can be split into a part that describes the image under $\partial$ of that element and a part in the kernel of $\partial$, with the former part fixed by the boundary label. Therefore, if a membrane operator is only sensitive to the former part of the surface element, and not to the part in the kernel, then it is equivalent to an operator that simply measures the boundary path element. In this case, the corresponding loop excitation can be produced by an electric ribbon operator that only acts near the loop itself, indicating that the loop cannot carry non-trivial loop topological charge and is condensed.

 We can be more precise about this notion of sensitivity to the kernel by using the irrep basis for the membrane operators. Recall from Section \ref{Section_E_Valued_Loops_1} that the $E$-valued loops are labelled by irreps of $E$, as seen in Equation \ref{E_membrane_definition}. We can construct an irrep of the kernel of $\partial$ (which is a subgroup of $E$) by restricting the irrep of $E$ to the kernel. This results in a (generally reducible) representation of ker($\partial$). The kernel is always central in $E$ due to the second Peiffer condition, Equation \ref{P2}, so Schur's Lemma applies. This means that the matrix representation of any group element in the kernel must be a scalar multiple of the identity, with the scalar being an irrep of the kernel. If this irrep of the kernel is trivial then the excitation is condensed, otherwise it is not condensed. That is, for an excitation produced by a membrane operator labelled by an irrep $\mu$ of $E$, if the matrix representation satisfies $D^{\mu}(e_K)=\mathbf{1}$ for all $e_K$ in the kernel then the excitation is condensed.

\section{Braiding}
\label{Section_Part_One_Braiding}

\begin{figure}[h]
	\begin{center}
		\includegraphics{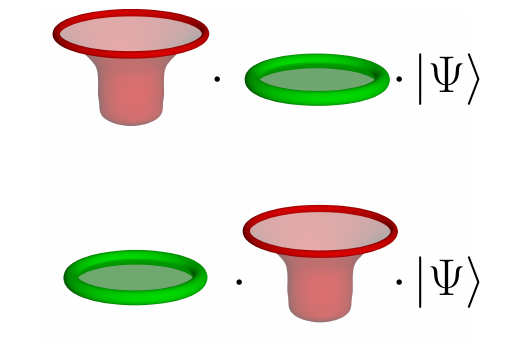}
		\caption{The commutation of operators used to calculate the braiding. The partially transparent surfaces indicate the membranes for the operators, while the opaque loops indicate the excited regions, which are the boundaries of the membranes.}
		\label{LoopLoopCommutation}
	\end{center}
\end{figure}

In this section we will first explain how braiding relations can be obtained from the ribbon and membrane operators, then describe our results for this model. Ribbon operators can be thought of as creating a pair of excitations and then separating them along the ribbon. The ribbon operator therefore encodes the result of moving an excitation. Similarly a membrane operator can be thought of as nucleating a small loop and moving it across the membrane. Therefore we should be able to find the result of braiding by applying successive membrane operators. In particular the braiding is related to the commutation relations of the membrane operators. For instance, consider the commutation relation shown in Figure \ref{LoopLoopCommutation}, which we will see relates to loop-loop braiding. The images represent membrane operators, which are displaced horizontally to indicate an order of operators (although the membranes intersect in space). In the first line, we first act with an operator that produces a loop excitation (the green loop) and then act with an operator that creates another loop (shown in red) and moves it through that first loop. This performs a braiding move, specifically the one from the left side of Figure \ref{LoopMovesIntro}. In the second line of Figure \ref{LoopLoopCommutation}, we first act with the operator that creates and moves the red loop through empty space, before creating the other loop excitation. In this case no braiding move occurs. Comparing these two lines (that is, working out the commutation relation for the two membrane operators) therefore lets us compare the situation with braiding to the one without, but where the excitations have the same final positions in each case. This latter point is important because ensuring that the excitations have the same final positions in the two cases isolates the effect of braiding from any other effects.

There are some subtleties when determining the braiding relations, which mean that we have to take care when interpreting the commutation relations between membrane operators. The formulae given in this section will refer to ``same-site" braiding. This refers to the case where the start-points of the operators (as defined when discussing the excitations in Section \ref{Section_Part_One_Excitations}) are in the same location. In this case the excitations involved have a definite fusion product (i.e., well defined combined topological charge). In a non-Abelian anyon theory, where two anyons may fuse to multiple different types of anyon, the braiding of two particles depends not only on the charges of the two particles, but also which charge they fuse to \cite[Preliminaries]{Pachos2012}. To find the braiding relations we therefore want to consider the case where they have definite total charge, otherwise the braiding relations are not well-defined (which in practice is reflected in this model by the presence of operator labels in the braiding relations). In simple cases, where the fusion is Abelian (i.e., where two charges only have one fusion channel), this requirement for same-site braiding is lifted (and this is usually reflected by the start-point of the corresponding membrane or ribbon operators being irrelevant to the commutation relation). While we only discuss the same-site case here, we will give more general results in Ref. \cite{HuxfordPaper3}.

Having discussed the method for finding the braiding, we now discuss our results. Firstly, we note that any braiding of two point-like excitations (which we term permutation) is trivial. This is because in this model the permutation is implemented by ribbon operators that connect the initial and final positions of the excitations. In 3+1d, the ribbon operators that permute the two excitations do not need to intersect and so commute, which leads to trivial (bosonic) exchange statistics. That is, the result of moving a point-like excitation past another one is always the same as moving that excitation through empty space (and then producing the other point-like particle), as long as the excitations stay well separated. Even if the ribbons do intersect, they can be deformed away from one another using the topological property without affecting their action. This contrasts with the 2+1d case where crossings cannot always be removed without pulling a ribbon over an excitation. For the same reason, exchange involving loop-like excitations moving past each-other (which we refer to as permutation), rather than through each-other (which we will refer to as loop-braiding or just braiding) is trivial.

Next we consider the braiding where we move a point-like or loop-like excitation through a loop-like excitation, as shown in Figure \ref{chargethroughloop2} for the point-like case. First consider the case where $\rhd$ is trivial. In this case, the excitations split into two separate sets, with non-trivial braiding only within each set. Firstly we have the excitations that are labelled by objects related to $G$, namely the electric and magnetic excitations. We have non-trivial point-loop braiding between the electric and magnetic excitations, where the magnetic excitation labelled by a group element $h$ acts on the electric excitation, labelled by an irrep $R$ and indices $a$ and $b$, by multiplication by the matrix $D^{R}(h)$. That is, given an electric ribbon operator $S^{R,a,b}(t)=\sum_g [D^{R}(g)]_{ab} \delta(\hat{g}(t), g)$ and a magnetic membrane operator $C^h(m)$, commuting the operators gives the following braiding relation:

\begin{center}
	\noindent\fcolorbox{black}{myblue1}{%
		\parbox{0.9\linewidth}{%
			\centering  \textbf{Flux-Charge Braiding Relation:}
\vspace{-5pt}
			\begin{align}
			S^{R,a,b}(t) C^h(&m) \ket{GS} \notag \\& = C^h(m) \sum_{c=1}^{|R|} [D^{R}(h)]_{ac} S^{R,c,b}(t) \ket{GS}. \label{Equation_electric_magnetic_irrep_braiding}
			\end{align}
		}}
\end{center}

This braiding mixes ribbon operators labelled by different indices, but not operators labelled by different irreps. This is because $R$ labels a conserved charge, whereas the matrix indices only represent some internal space to the topological sector, as we mentioned in Section \ref{Section_Part_One_Electric}. When the irrep is one dimensional, such as when the group $G$ is Abelian, the braiding relation results in the accumulation of a phase of $R(h)$, because the matrices belonging to a 1d unitary representation are just phases. This braiding relation holds for a specific orientation of the loop-like excitation and direction with which the point-like particle is moved through the loop. If either orientation was reversed, then we would replace $h$ with its inverse in the braiding relation. This braiding relation is natural from the gauge theory perspective and indeed is equivalent to the one that we predicted in Section \ref{Section_Properties_From_Gauge_Theory}.

We also find non-trivial loop-loop braiding between the magnetic excitations when the group $G$ is non-Abelian. When one loop, labelled by $g$, is passed through another, labelled by $h$, the first loop has its label conjugated, while the other label is left unchanged, as indicated by Equation \ref{Equation_flux_flux_tri_trivial}. This braiding is just as we would expect for the braiding of the magnetic excitations in ordinary lattice gauge theory, as explained in Section \ref{Section_Properties_From_Gauge_Theory}. 

 \begin{center}
	\noindent\fcolorbox{black}{myblue1}{%
		\parbox{0.9\linewidth}{%
			\centering  \textbf{Flux-Flux Braiding Relation ($\rhd$ trivial):}
			\vspace{-5pt}
			\begin{equation}
				C^g(m_1)C^h(m_2)\ket{GS}= C^h(m_2)C^{h^{-1}gh}(m_1)\ket{GS}. \label{Equation_flux_flux_tri_trivial}
			\end{equation}
		}
	}
\end{center}

Next we consider the other set of excitations, those labelled by objects corresponding to the group $E$. In this set the only non-trivial braiding, whether point-loop or loop-loop, is the braiding between the point-like blob excitations and the $E$-valued loop excitations. When a blob excitation labelled by an element $e \in E$ passes through a loop excitation labelled by the 1D irrep $\alpha$ of $E$, a phase of $\alpha(e)$ (or the inverse phase, depending on the orientation of the loop and direction of braiding) is accumulated. When $\rhd$ is trivial, the group $E$ is Abelian, so all of the irreps are 1D and so the braiding transformation is only a phase, as shown in Equation \ref{Equation_loop_blob_braiding}.

 \begin{center}
	\noindent\fcolorbox{black}{myblue1}{%
		\parbox{0.9\linewidth}{%
			\centering  \textbf{Loop-Blob Braiding Relation:}
			\vspace{-5pt}
			\begin{align}
				B^e(t) & \sum_{f \in E} \alpha({f}) \delta(f,\hat{e}(m)) \ket{GS} \notag\\ &=\alpha(e) \sum_{f \in E} \alpha(f) \delta(f,\hat{e}(m)) B^e(t) \ket{GS}. \label{Equation_loop_blob_braiding}
			\end{align}
		}
	}
\end{center}

The braiding relations are a little different when $\rhd$ is non-trivial. If we restrict to fake-flat configurations (Case 3 from Table \ref{Table_Cases_intro}), we have to throw out the magnetic excitations (and the blob excitations labelled by elements outside the kernel of $\partial$) and the only non-trivial braiding is between the blob excitations and the $E$-valued loops. When we pass a blob excitation labelled by $e$ through a loop labelled by the irrep $\mu$ and indices $a$ and $b$, the loop transforms by multiplication by the matrix $D^{\mu}(e)$ or the inverse. On the other hand, if we take our other special case (Case 2 in Table \ref{Table_Cases_intro}), where $\partial$ maps from an Abelian $E$ to the centre of $G$, then the braiding is richer than in the $\rhd$ trivial case. While braiding not involving the magnetic excitations is the same as in the $\rhd$ trivial case, the magnetic excitations now braid non-trivially with all of the types of excitation. To obtain the braiding relations we have to combine the magnetic excitation with the $E$-valued loop, giving us an excitation we call a higher-flux loop excitation. This is because the magnetic excitation now carries a 2-flux, but this flux is not well-defined unless we also apply an $E$-valued membrane operator $\delta(e, \hat{e}(m))$ on the same region of space (which fixes the 2-flux). Then the combined membrane operator produces an excitation labelled by a pair of group elements $(g,\tilde{e})$, where $g$ is in $G$ and gives the 1-flux, while $\tilde{e}$ is in $E$ and gives the 2-flux. Here $g$ is simply the label of the magnetic membrane operator, but $\tilde{e}$ is related to the label $e$ of the additional $E$-valued membrane operator by $\tilde{e}=e [g^{-1} \rhd e^{-1}]$, as we explain in more detail in Ref. \cite{HuxfordPaper3}. When a pair of these higher-flux excitations braid, we have
 \begin{center}
	\noindent\fcolorbox{black}{myblue1}{%
		\parbox{0.95\linewidth}{%
			\centering  \textbf{Higher-Flux--Higher-Flux Braiding Relation:}
			\vspace{-5pt}
			\begin{equation}
				((g,\tilde{e}_2),(h,\tilde{e}_1))\rightarrow ((h,\tilde{e}_1\tilde{e}_2 [h\rhd \tilde{e}_2^{-1}]), (hgh^{-1}, h \rhd \tilde{e}_2)).
			\end{equation}
		
	}}
\end{center}
 This braiding then matches the prediction of Ref. \cite{Bullivant2018}, where the result is argued from geometric grounds. Note that the conjugation of the 1-flux label is slightly different from Equation \ref{Equation_flux_flux_tri_trivial} due to a different convention we use for the orientation of the higher-flux excitations.

Finally, when a blob excitation labelled by $e$ is passed through a higher-flux excitation labelled by $(h,\tilde{f})$, the braiding relation is given by

 \begin{center}
	\noindent\fcolorbox{black}{myblue1}{%
		\parbox{0.9\linewidth}{%
			\centering  \textbf{Higher-Flux--Blob Braiding Relation:}
			\vspace{-5pt}
			\begin{align}
				B^e&(t)C^{h,f}_T(m)\ket{GS} \notag\\
				& = C^{h,f e }_T(m) B^e(t_1') B^{ h^{-1} \rhd e}(t_2') \ket{GS}, \label{higher_flux_blob_commutation_same_sp}
			\end{align}
			}}
\end{center}
where $t_1'$ and $t_2'$ are the parts of the ribbon before and after the intersection with the membrane. Note that the labels of the excitations change in two ways. Firstly the label $f$ of the pinned $E$-valued loop is changed by multiplication by $e$ or by $h \rhd e^{-1}$ (depending on orientation), which induces a change in the 2-flux $\tilde{f}=f [h^{-1} \rhd e^{-1}]$ of the excitation by multiplication by $e [h^{-1} \rhd e^{-1}]$ or $e [h \rhd e^{-1}]$. Secondly, the blob ribbon operator labelled by $e$ is acted on by the magnetic operator, so that $e \rightarrow h^{-1} \rhd e$ or $h \rhd e$ after the intersection (again depending on orientation). From this we can see that the product $ \tilde{f}e$ of the 2-fluxes of the two excitations is preserved by the braiding.

\section{Topological charge}
\label{Section_Part_One_Topological_Sectors}

\begin{figure}[h]
 	\begin{center}
 	\includegraphics[width=0.5\linewidth]{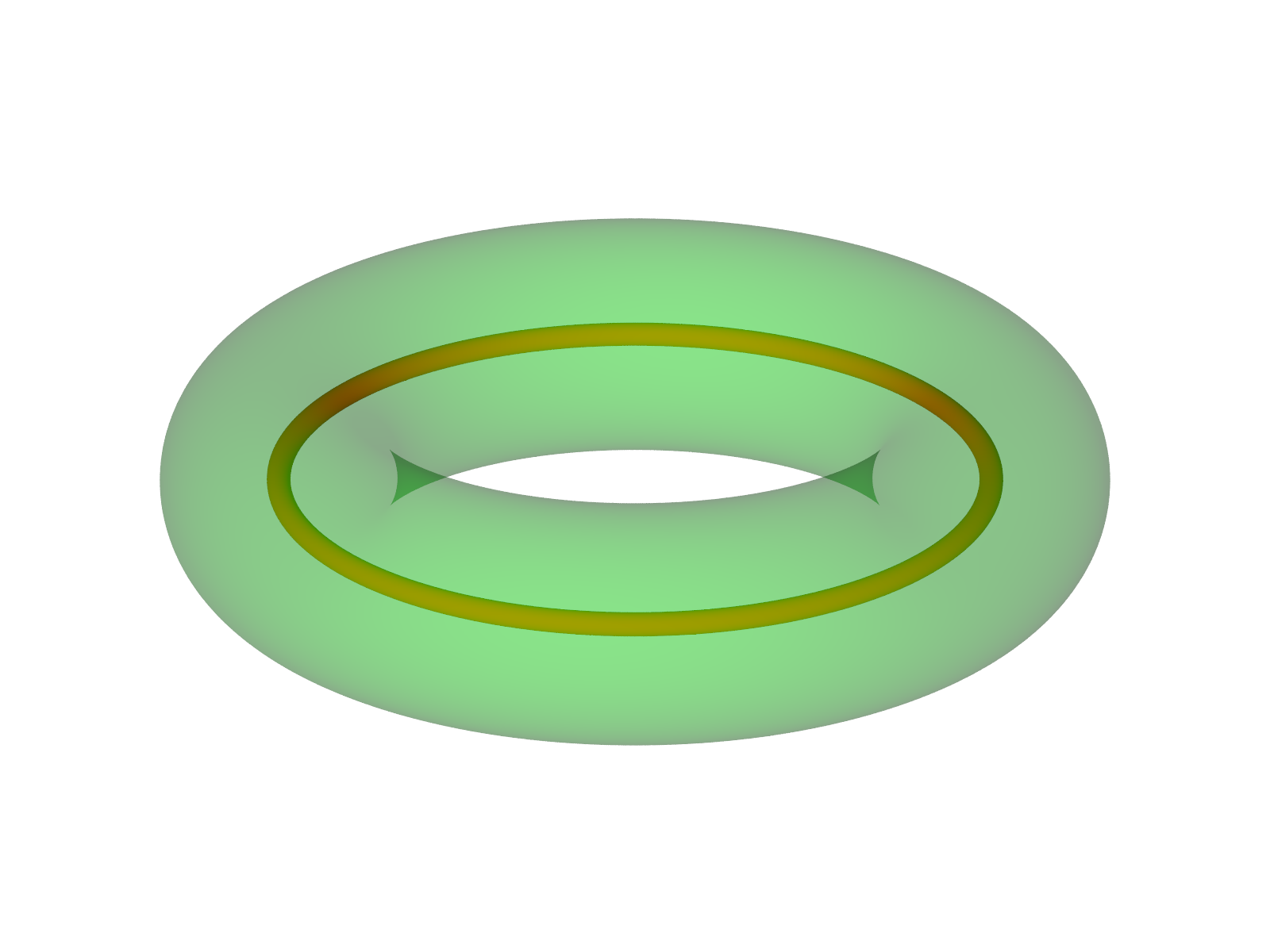}

 		\caption{Given a loop excitation (thin red torus), we can measure its topological charge with a toroidal surface (larger green torus) enclosing it}
 		\label{loop_measurement}
 	\end{center}
 \end{figure}

In this section we will explain in more detail what we mean by topological charge and how we can measure it. Topological charge is a conserved quantity associated with the excitations of the model (the anyons), while the ground state carries the trivial charge, also called the vacuum charge. The charge held within a region can only be changed by moving that charge from inside the region to outside or moving charge from the outside in. This means that the charge in a region can only be changed by operators that connect the inside of the region to the outside. We note that there is no need for a symmetry to enforce this conservation. Topological charge is conserved on the level of the Hilbert space and can be defined without reference to any Hamiltonian (although in this case we must define the vacuum charge in another way). We can measure the topological charge associated to a region using operators on the boundary of that region, which is reminiscent of the way that we can determine the electric charge in a region by measuring the flux of the electric field through the boundary of that region. For example, we can measure the charge associated to a loop excitation using a torus that encloses that loop, as shown in Figure \ref{loop_measurement}. Any operator that would move topological charge from inside a region to outside it must cross the boundary of that region, and so can be detected by the measurement operator on the surface. 

The choice of measurement surface is important, not just because it determines where we want to measure the charge, but also because the set of charges to be measured depends on the topology of the surface. A spherical measurement surface measures a different set of topological charges from a toroidal surface for example, because a spherical measurement surface cannot distinguish a loop-like excitation from a point-like one. For this reason we say that a spherical surface measures the point-like charge of an excitation (or set of excitations). In order to determine the loop-like charge of an excitation, we must use a toroidal surface, such as the one shown in Figure \ref{loop_measurement} (or a surface of higher genus, although we will not consider these in this work).
 
In order to identify the operators that measure the topological charge, we first consider the characteristics that we require such operators to have. While the topological charges are properties of the Hilbert space, the Hamiltonian picks out a certain set of charges, such that the ground state has the trivial charge. Then, because a measurement operator should not change the charge in any region, we require that the measurement operators do not create any excitations and so must commute with the Hamiltonian. In addition, smoothly deforming the measurement operator without crossing any excitations should preserve the measured charge, because the ground state has trivial charge. Following the method of Bombin and Martin-Delgado \cite{Bombin2008}, we construct such measurement operators using closed ribbon and closed membrane operators (the latter because we consider the 3+1d case, whereas Ref. \cite{Bombin2008} considered a 2+1d model). This may seem restrictive, but all operators in this model can be expressed in terms of the ribbon and membrane operators, and the ribbon operators that commute with the energy terms must be closed. However, we may need to take additional steps to guarantee that the closed operators commute with all of the energy terms, because there may be some obstruction to closing the operator without producing some excitations.

\begin{figure}[h]
	\begin{center}
		\includegraphics[width=0.9\linewidth]{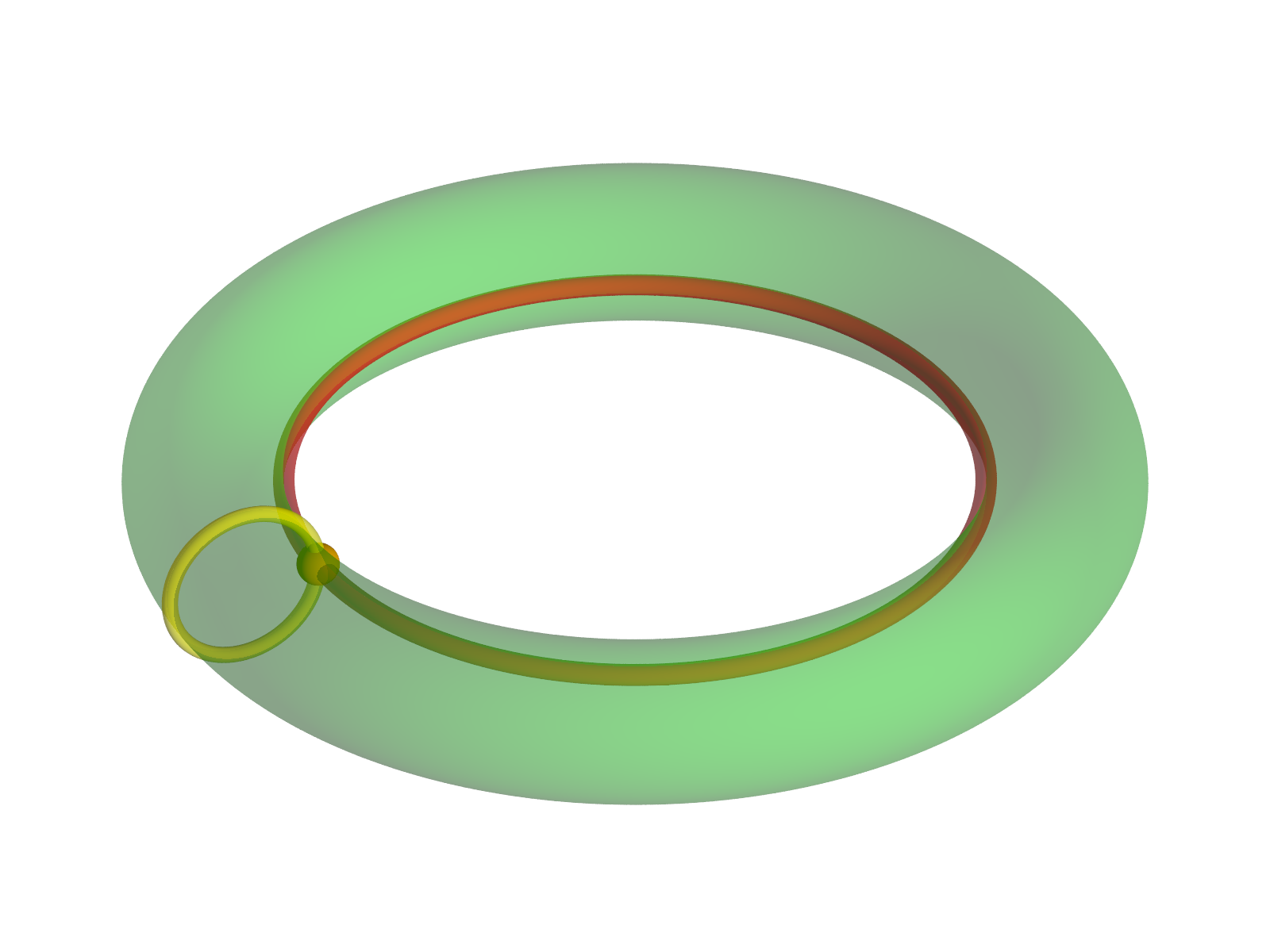}

		\caption{We apply closed ribbon operators on the two cycles of the torus (the thin red and yellow loops) and a membrane operator over the surface itself (the thicker green torus)}
		\label{torus_measurement}
	\end{center}
\end{figure}

As an example, consider the case of a torus, shown in Figure \ref{torus_measurement}. We apply ribbon operators around the two non-contractible cycles of the torus, and membrane operators on the torus itself. Any other closed ribbon operators that we could apply would not be independent (i.e., could be reduced to operators of the types already considered by deformation or other means), or would leave excitations on the surface. This means that the only operators we can apply are an electric and a blob ribbon operator running around each independent non-contractible cycle and a magnetic and $E$-valued membrane operator over the torus itself. This gives us six labels, one for each operator. We then construct a linear combination over all possible labels, with coefficients chosen so that the sum commutes with the Hamiltonian:
\begin{align}
	&\sum_{e_{c_1}, e_{c_2}, e_m \in E} \sum_{g_{c_1},g_{c_2},h \in G} \alpha_{	(g_{c_1},g_{c_2},h,e_{c_1},e_{c_2},e_m)} B^{e_{c_1}}(c_1) \notag \\& \ B^{e_{c_2}}(c_2)C^h_T(m)
	\delta(\hat{e}(m),e_m) \delta(\hat{g}(c_1),g_{c_1}) \delta(\hat{g}(c_2),g_{c_2}),
\end{align}
where $c_1$ and $c_2$ are the two cycles of the torus, while $m$ is its surface. Unfortunately, not being able to construct the magnetic excitations for a general crossed module (Case 3 in Table \ref{Table_Cases_intro}) prevents us from constructing all of our charge measurement operators in every case. However, when $E$ is Abelian and $\partial$ maps from an Abelian group to the centre of $G$ (Case 2 in Table \ref{Table_Cases_intro}, of which Case 1 is a subset), we are able to explicitly construct the measurement operators. As explained in Ref. \cite{HuxfordPaper3}, we find that the following restrictions on the are necessary for the operator to commute with the Hamiltonian.

\begin{center}
	\noindent\fcolorbox{black}{myblue1}{%
		\parbox{0.9\linewidth}{%
			{\centering  \textbf{Conditions for the measurement operators:} \par}

 Firstly, the coefficients $\alpha_{	(g_{c_1},g_{c_2},h,e_{c_1},e_{c_2},e_m)}$ are only non-zero when

\begin{align}
	&\partial(e_m)=[g_{c_2},g_{c_1}], \label{Equation_torus_charge_flatness}\\
	&\partial(e_{c_2})=[g_{c_1},h], \label{Equation_torus_charge_plaquette_seam_1}\\
&	\partial(e_{c_1})=[h,g_{c_2}], \label{Equation_torus_charge_plaquette_seam_2}\\
	&1_E =[h \rhd e_m^{-1}] \: e_m e_{c_1}^{-1} [g_{c_1}^{-1} \rhd e_{c_1}] e_{c_2}^{-1} [g_{c_2}^{-1} \rhd e_{c_2}] \label{Equation_torus_charge_blob_0}, 
\end{align}	
	
	Secondly, the coefficients must satisfy

	\begin{align}
	&\alpha_{(g_{c_1},g_{c_2},h,e_{c_1},e_{c_2},e_m) } \notag \\	 & \ = \alpha_{(gg_{c_1}g^{-1} ,gg_{c_2}g^{-1},ghg^{-1},g \rhd e_{c_1},g \rhd e_{c_2},g \rhd e_m)} \ \forall g \in G, \label{Equation_torus_charge_vertex}\\
	&\alpha_{(g_{c_1},g_{c_2},h,e_{c_1}, e_{c_2},e_m)} \notag \\	& \  = \alpha_{ (\partial(e)^{-1}g_{c_1},g_{c_2}, h, e_{c_1}, e_{c_2} [h \rhd e] \: e^{-1}, e_m e^{-1} [g_{c_2}^{-1} \rhd e])} \ \forall e \in E, \label{Equation_torus_charge_edge_seam_1}\\
	&\alpha_{(g_{c_1}, g_{c_2},h, e_{c_1},e_{c_2}, e_m)} \notag \\	& \ = \alpha_{ (g_{c_1}, \partial(r)g_{c_2}, h, e_{c_1} [h \rhd r] \: r^{-1}, e_{c_2}, e_m r^{-1} [g_{c_1}^{-1} \rhd r])} \ \forall r \in E, \label{Equation_torus_charge_edge_seam_2} \\
	&\alpha_{(g_{c_1}, g_{c_2}, h, e_{c_1},e_{c_2}, e_m)} \notag \\ & \ = \alpha_{ (g_{c_1}, g_{c_2}, \partial(e)h, e_{c_1} [g_{c_2}^{-1} \rhd e] \: e^{-1}, e_{c_2} [g_{c_1}^{-1}\rhd e^{-1}] \: e, e_m )} \ \forall e \in E \label{Equation_torus_charge_perpendicular_edge}.
\end{align}}}
\end{center}

The number of linearly independent operators satisfying these conditions is then the number of topological charges measured by the torus. In Ref. \cite{HuxfordPaper3} we explicitly construct a basis for this space, consisting of operators that project to definite charge. In this case we find that the number of charges matches the ground-state degeneracy of the 3-torus. This is perhaps to be expected for a topological phase; in topological quantum field theories (TQFTs) there is a correspondence between the partition function associated to a closed (four-dimensional) manifold $M \times S^1$ and the dimension of the Hilbert space associated to the open manifold $M \times I$ (where $I$ is the interval) \cite{Atiyah1988}. The ground state degeneracy of the higher lattice gauge theory model on a (three-dimensional) manifold $M$ is equal to the partition function of the Yetter TQFT \cite{Bullivant2017, Yetter1993} on a manifold $M \times S^1$. Taking $M$ to be the 3-torus $T^3 = S^1 \times T^2$, we may therefore perhaps expect a relationship between the ground-state degeneracy of the 3-torus (which matches the partition function of the TQFT on $T^3 \times S^1 = S^1 \times T^2 \times S^1 =S^1 \times T^3$) and the dimension of the space for the degrees of freedom on thickened 2-torus $I \times T^2$ (which should match the size of the space associated to $I \times T^2 \times S^1= I \times T^3$ in the TQFT). This in turn should match the number of independent measurement operators that we can apply on the toroidal measurement surface (which is really a thickened torus). This correspondence between ground-state degeneracy and the charges for this model was also discovered in Ref. \cite{Bullivant2020}, using a different method (employing tube algebra). Indeed, the projection operators we construct in Ref. \cite{HuxfordPaper3} are labelled by the same objects labelling the simple modules of the tube algebra found in Ref \cite{Bullivant2020}.

\begin{figure}[h]
	\begin{center}
		\includegraphics[width=0.7\linewidth]{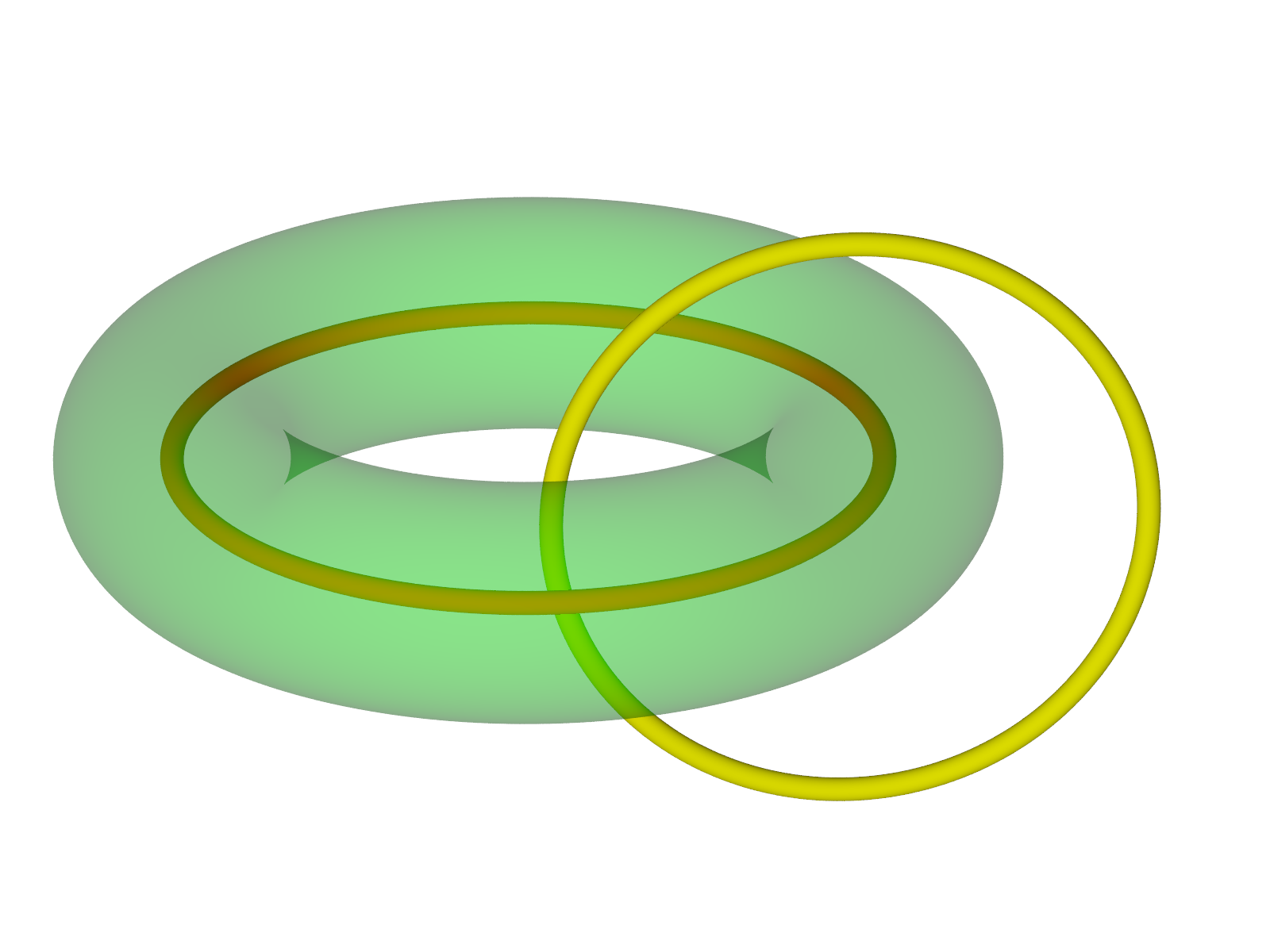}

		\caption{The torus can measure loops that are held within it (e.g., the red loop shown) or that link with it but are outside (e.g., the yellow loop shown here). More generally the surface can measure two linked loops, or collections of loops that link together in this way, together with point-particles within the torus.}
		\label{torus_two_loop_measurement}
	\end{center}
\end{figure}

One important thing to note is that the torus surface can measure loops that link with either of the cycles of the torus, as shown in Figure \ref{torus_two_loop_measurement}. This means that the general object measured by the surface is a pair of linked loops (or a set of objects that can be fused into such a pair). In particular, we note that the number of charges measured by the torus (and so the ground state degeneracy) is not equal to the number of distinct loop-like excitations, but instead the number of link-like excitations (counting those obtained by fusion with point-like excitations as well). This suggests that it is important to consider the linking of loop-like excitations when studying 3+1d topological phases. This reinforces work by other authors \cite{Wang2014, Jiang2014} that shows that so-called three-loop (or necklace) braiding, which is braiding of two loops while both are linked to a third loop, is important for characterising 3+1d topological phases. 

\section{Conclusion}

In this paper we constructed the membrane and ribbon operators that produce the excitations of the higher lattice gauge theory model in three broad cases and used these operators to obtain various properties of the excitations. The model has two classes of excitations: those present in (the 3d version of) the Quantum Double model and those that involve the surface holonomy, which only appear when we consider higher lattice gauge theory. The character of the excitations depends on which of the special cases we consider. When the map $\rhd$ from our crossed module is trivial, the first class of excitation, those familiar from the Quantum Double model, are largely unchanged by the move to higher lattice gauge theory. The excitations related to the surface holonomy are also simple, with Abelian fusion rules. Furthermore the braiding between the two sectors is trivial, with non-trivial braiding only within the sectors (i.e., between the $G$-valued excitations and between the $E$-valued excitations). However, unlike in the Quantum Double model, some of the point-like excitations are confined, costing additional energy to separate pairs of excitations after producing them, while some of the loop-like excitations are condensed and carry trivial topological charge. We discussed how this arises from a condensation-confinement transition between higher lattice gauge theory models described by crossed modules with the same groups $G$ and $E$ but different maps $\partial$ between the groups.

We saw that the excitations are more interesting when $\rhd$ is non-trivial. In particular, we looked at the case where the group $E$ labelling the 2-flux is Abelian and where the map $\partial$ maps to the centre of the group $G$. In this case, even though the group $E$ is Abelian, the excitations related to 2-flux have an internal space controlled by $\rhd$. In addition, there is non-trivial braiding between excitations from the two different sectors. In particular, it is sensible to consider loop excitations, which we called higher-flux loops, built from a combination of excitations from the two sectors. These excitations carry both an ordinary magnetic flux along paths that link with the loop and a 2-flux associated to a surface enclosing the loop. The higher-flux loops have potentially non-trivial braiding with excitations from both sectors. We found the braiding relations of these higher-flux loops and discovered that the braiding between two higher-flux loops matches a braiding scheme for loop-like excitations described in Ref. \cite{Bullivant2018}.

In addition to using the membrane operators to find the braiding properties of the various excitations, we have discussed the explicit construction of the membrane operators to construct operators to measure the topological charge in a region. The charge measurement operators are operators with support on or near the surface bounding a region, with different surfaces having different potential topological charges. A related idea is the fact that in 3+1d there are both point-like and loop-like excitations, with associated point-like and loop-like charge. While loop-like excitations may possess a point-like charge, which may be measured by enclosing the excitation in a sphere, they also possess loop-like charge, which can only be measured by surfaces with non-contractible handles such as the torus. This approach of explicitly constructing the charge measurement operators (which we present more fully in Ref. \cite{HuxfordPaper3}) could allow us to measure the charge of collections of excitations and so provide a tool to consider fusion of topological charge.

There are many interesting avenues of research that would build on the results we've obtained in this work, either directly or indirectly. Firstly, and most obviously, there are some results for the higher lattice gauge theory model that we have so far not been able to obtain. We have considered certain special cases (described by Table \ref{Table_Cases_intro}) and while there are some conceptual issues associated with taking the most general case (as described in Section \ref{Section_Special_Cases}), there may still be a way to work around these issues. It is possible that there are some features exhibited by the general case that we have not been able to study in our special cases.

It would also be interesting to further study the different types of topological charge possible in 3+1d topological phases, perhaps in a more general and conceptual setting than this specific model. In this work we have mostly considered the simple excitations, and only considered measurement operators for the topological charge within a sphere and torus. We believe that it would be useful to do more work with the charges themselves, either in this model or more generally. There are several questions that could be explored in this direction. For example, which surfaces do we need to consider to obtain all unique charges? How do we consider the fusion of charges that are measured by surfaces other than simple spheres? For instance, considering two tori, we could fuse the enclosed charges by bringing the tori together and stacking them on top of each-other, so that a single torus can enclose both. After some preliminary calculations, we found that, in the higher lattice gauge theory model at least, this leads to consistency conditions for the ``threading flux" passing up through the two tori, which must be satisfied in order to be able to fuse them, but a complete calculation is left for future work. Another way to combine two torus charges would be to bring the two tori side-by-side, so that a 2-handled torus would be needed to enclose both. For more general surfaces, there could be even more ways of fusing charges.

Another sensible direction for future study would be to extend our approach of utilizing explicit construction of membrane operators, to more general models for topological phases. In particular, it is known that twisted gauge theory models can have non-trivial three-loop braiding statistics \cite{Wang2014, Wang2015}. It has been claimed that these models cover all phases that can be produced from bosonic degrees of freedom and that result in bosonic point excitations (i.e., phases with trivial braiding between point-like particles) \cite{Lan2018}. Because these models have a similar structure to the model considered in this paper, the application of our methods to the twisted gauge theory model seems feasible. By looking at other models, we may be able to see if the relation between the ground state degeneracy of the 3-torus and the number of charges measured by a torus surface holds more generally than just in the higher lattice gauge theory model, as we may expect.

Finally, it would be interesting to study the condensation-confinement transitions in 3+1d topological phases in more detail. We have seen some examples in the higher lattice gauge theory model, but in the cases where $E$ is Abelian (which we studied in more detail) the pattern of condensation is rather simple. The current understanding of condensation in 3+1d is perhaps incomplete, particularly when it comes to loop-like excitations condensing, and the examples considered in this work may be useful in studying this process.

\begin{acknowledgments}
	The authors thank Jo\~{a}o Faria Martins and Alex Bullivant for helpful discussions about the higher lattice gauge theory model. We also thank Paul Fendley for advice on preparing our manuscript. We acknowledge support from EPSRC Grant EP/S020527/1. Statement of compliance with EPSRC policy framework on research data: This publication is theoretical work that does not require supporting research data.
	
\end{acknowledgments}

\onecolumngrid

\appendix

\section{Consistency of the higher lattice gauge theory model under changes to the branching structure}
\label{Section_rebranching}

As we discussed in Section \ref{Section_Hamiltonian_Model}, when we define a lattice for the higher-lattice gauge theory model, we must give the lattice a \textit{branching structure}. This determines the orientations of the edges of the lattice as well as the orientation and base-point of each plaquette, which we call the ``dressing" of the lattice. In order for the model to be topological, it must somehow be resilient to changes to these details. In this section we will show that there is a certain sense in which the energy terms are invariant under changes to the branching structure (to be more precise, a branching structure forbids local oriented loops, and has an ordering of vertices, but we will not make these restrictions for the altered dressing). We note that this issue of consistency under changes to the branching structure is also addressed by Ref. \cite{Koppen2021}, but we will need the more explicit treatment used here for future results. To understand what we mean by consistency under changes to the branching structure, let us first consider an example of how we can change the dressing of the lattice. In Section \ref{Section_Lattice_Gauge_Theory}, we described the notion of parallel transport across an edge of the lattice. The label of the edge describes the result of parallel transport and the orientation of the edge describes the direction of the parallel transport that gives this result. If we were to perform parallel transport in the opposite direction, we would expect the inverse transformation. This suggests that there is a natural operation in which we can take an edge with a given orientation and a label $g$ and reverse the orientation of the edge, while simultaneously changing the label to $g^{-1}$, without changing the physical meaning of the label. In a similar way, in Section \ref{Section_Hamiltonian_Model} we described how reversing the orientation of a surface should change its label from $e$ to $e^{-1}$. We can also move the base-point of a plaquette from $v_1$ to $v_2$, which should change its label from $e$ to $g(v_1-v_2)^{-1} \rhd e$, where $g(v_1-v_2)$ is the group element assigned to the path along which we move the base-point. We can treat these transformations as maps between two copies of our Hilbert space, where each Hilbert space corresponds to different dressings of the lattice. We extend the action of these transformations to states where the edges or plaquettes are labelled by linear combinations of the group elements (and to states where the different degrees of freedom are entangled) in the sensible way, to make these transformations linear. Then the transformations, which involve changes to the dressing of a lattice (orientations and base-points) as well as to the labels attached to the different cells of the lattice, are maps that connect states from different Hilbert spaces that should have the same physical content, but have different descriptions. Each of the Hilbert spaces is equipped with a Hamiltonian of the form given by Equation \ref{Equation_Hamiltonian} from Section \ref{Section_Hamiltonian_Model}. If the states related by the transformations do indeed contain the same physical content, then the energy of a state should be preserved under these transformations. That is, if we first apply the Hamiltonian on a state in one of the Hilbert spaces and then apply a transformation to change the dressing of the lattice, this should give us the same result as applying the transformation then the Hamiltonian on the new Hilbert space. Equivalently, given a transformation $\hat{T}$ that changes the dressing, then the Hamiltonian $H_1$ in the original space should be related to the Hamiltonian $H_2$ in the second space by $\hat{T}^{-1}H_2\hat{T}=H_1$. If this is true then we have a way to move between the different dressed lattices, while preserving the structure endowed on the space by the Hamiltonian. This demonstrates the desired resilience of the higher lattice gauge theory model to changes to the dressing of the lattice. In this section we will show that the Hamiltonian does indeed have this property, subject to certain caveats related to fake-flatness and the map $\rhd$.

\subsection{Reversing the orientation of an edge}
\label{Section_rebranching_flip_edge}
Let us start by considering the procedure where we reverse the orientation of an edge and simultaneously invert the label of that edge. We denote this edge flipping transformation on an edge $i$ by $\hat{P}_i$. We wish to show that this map preserves each energy term individually (which is a slightly stronger condition than just preserving the Hamiltonian). Suppose that we start in one copy of our Hilbert space $\mathcal{H}_1$, and $P_i$ maps to another copy $\mathcal{H}_2$ of the Hilbert space. Then for any vertex, edge, plaquette or blob in our lattice we have an energy term $\hat{O}_1$ acting on the first Hilbert space, and a corresponding energy term $\hat{O}_2$ acting on the second Hilbert space, where these energy terms are defined in Section \ref{Section_Hamiltonian_Model}. We wish to show that the energy is preserved under the transformation $\hat{P}_i$, by which we mean that $\hat{P}_i \hat{O}_1= \hat{O}_2 \hat{P}_i$. Thinking in terms of the eigenstates, if this is true then for a state $\ket{\psi}$ which is an eigenstate of the energy term $\hat{O}_1$, with eigenvalue $\lambda$, the state $\hat{P}_i \ket{\psi}$ reached by flipping the edge $i$ is an eigenstate of the equivalent energy term $\hat{O}_2$ with the same eigenvalue, $\lambda$. Because the map $\hat{P}_i$ is invertible (we can undo $P_i$ by flipping the edge back and inverting the edge element again, so $P_i$ is in fact its own inverse), we can also write the condition as $\hat{O}_1=\hat{P}_i^{-1}\hat{O}_2 \hat{P}_i$. Therefore, when we say that the energy is preserved by the transformation, we mean that changing the orientation of the edge, applying an energy term and then changing the orientation back has the same net effect as simply applying the corresponding energy term without flipping the edge.

We first wish to show that the edge flipping procedure is consistent with the vertex terms $A_v$. Each vertex term is an equal sum of the vertex transforms: $A_v = \frac{1}{|G|}\sum_{g \in G} A_v^g$. Therefore, if we can show that the vertex transforms $A_v^g$ are invariant under $P_i$, this will also be true for the vertex terms $A_v$. The vertex transform on a vertex $v$ only has support on the edges that are adjacent to that vertex (and neighbouring plaquettes, although the action on plaquettes is independent of edge orientation). Therefore the vertex transform is only sensitive to the edge-flipping procedure on adjacent edges. Recall that the action of the vertex transform on an adjacent edge $i$, labelled by $g_i$, is
$$A_v^x :g_i = \begin{cases} xg_i & i \text{ points away from } v\\ g_i x^{-1} & i \text{ points towards } v.\end{cases}$$

The action of the vertex transform on edge $i$ is only sensitive to the orientation of edge $i$ itself and not to the orientation of any other edges. Therefore we only need to consider the effect of flipping edge $i$ itself. We have
\begin{align*}
\hat{P}_i^{-1}A_v^x \hat{P}_i :g_i &= \hat{P}_i^{-1} A_v^x: g_i^{-1}\\
&= \hat{P}_i^{-1} : \begin{cases} g_i^{-1}x^{-1} & i \text{ originally pointed away from (now towards) } v\\ xg_i^{-1} & i \text{ originally pointed towards } v\end{cases}\\
&= \begin{cases} xg_i & i \text{ points away from } v\\ g_i x^{-1} & i \text{ points towards } v,\end{cases}
\end{align*}
from which we see that the action of the vertex transform is preserved under the edge-flipping procedure $P_i$.

 Next we consider the plaquette and blob energy terms. These terms only involve the edges through path elements (the path around the boundary of the plaquette for the plaquette term, and paths between the base-points of surfaces for the blob term). When we construct a path element from the edge elements, each edge contributes the group element it would have if it pointed along the path. That is, if the edge is aligned with the path the edge contributes its group element, but if the edge is anti-aligned with the path then it instead contributes the inverse of its group element. This means that path elements are invariant under the flipping procedure. If an edge with label $g_i$ already points along the path then it contributes $g_i$ to the path. If we flip the edge, then we change its label to $g_i^{-1}$. However, the flipped edge then points against the path, so it contributes $(g_i^{-1})^{-1} =g_i$ to the path due to the way that we calculate path elements. Similarly, if the edge originally points against the path then it contributes $g_i^{-1}$. If we flip it, then we change its label to $g_i^{-1}$. However, the edge then points along the path, so it still contributes $g_i^{-1}$. The path elements being invariant under the edge flipping procedure in this way means that both the plaquette and blob energy terms are similarly invariant under this procedure.

This leaves us to consider the edge energy terms. The energy term for an edge $i$ is $\mathcal{A}_i$, which is made of a sum of edge transforms: $\mathcal{A}_i = \frac{1}{|E|}\sum_{e \in E} \mathcal{A}_i^e$. Let us consider how an edge transform is affected by the edge flipping procedures. Recall that the edge transform $\mathcal{A}_i^e$ acts on the edge $i$ and on neighbouring plaquettes. The support of $\mathcal{A}_i^e$ also includes some of the other edges along the boundaries of the adjacent plaquettes, because the action of $\mathcal{A}_i^e$ on an adjacent plaquette $p$ is
\begin{equation}
\mathcal{A}_i^e :e_p = \begin{cases} e_p \: [g(v_0(p)-v_i) \rhd e^{-1}] & i \text{ points along the boundary of } p \\ (g(\overline{v_0(p)-v_{i+1}}) \rhd e) \: e_p & i \text{ points against the boundary,}\end{cases} \label{Edge_transform_plaquette_appendix}
\end{equation}
which depends on the path element $g(v_0(p)-v_i)$ or $g(\overline{v_0(p)-v_{i+1}})$. Flipping the orientation of the edges along these paths will not affect the action on the edge transform, because the path elements are invariant under the flipping operation, as we just saw. However, the action of $\mathcal{A}_i^e$ directly depends on the orientation of $i$ itself. Therefore we need to consider whether $\mathcal{A}_i^e$ is invariant under the edge-flipping operation on edge $i$. First consider the action of $\mathcal{A}_i^e$ on the edge $i$ itself. If the edge initially has label $g_i$, then the edge transform acts as $\mathcal{A}_i^e:g_i = \partial(e)g_i$. Therefore
\begin{align*}
P_i^{-1} \mathcal{A}_i^e P_i: g_i &= P_i^{-1} \mathcal{A}_i^e: g_i^{-1}\\
&= P_i^{-1}: \partial(e)g_i^{-1}\\
&= g_i \partial(e)^{-1}\\
&= g_i \partial(e)^{-1} g_i^{-1} g_i\\
&=\partial(g_i \rhd e^{-1})g_i\\
&=\mathcal{A}_i^{g_i \rhd e^{-1}}:g_i,
\end{align*}
where we used the Peiffer condition Equation \ref{P1} to write $\partial(g_i \rhd e^{-1}) = g_i \partial(e)g_i^{-1}$. We therefore see that, unlike the vertex transforms, the individual edge transforms are not generally invariant under the edge-flipping procedure. However, this does not mean that the edge energy term itself is not invariant. We have, at least for the action on the edge
\begin{align*}
P_i^{-1} \mathcal{A}_i P_i:g_i &= \frac{1}{|E|}\sum_{e \in E} P_i^{-1} \mathcal{A}_i^e P_i:g_i\\
&= \frac{1}{|E|}\sum_{e \in E} \mathcal{A}_i^{g_i \rhd e^{-1}}:g_i\\
&= \frac{1}{|E|}\sum_{e' = g_i \rhd e^{-1}} \hspace{-0.2cm} \mathcal{A}_i^{e'} :g_i\\
&=\mathcal{A}_i:g_i,
\end{align*}
which suggests that the energy term is invariant under the procedure, even if the individual transforms are not. However, we have only shown this for the action on the edges, and it must also hold for the plaquettes in order to obtain the operator relation $P_i^{-1} \mathcal{A}_i^e P_i=\mathcal{A}_i^{g_i \rhd e^{-1}}$ (i.e., we must also satisfy the relationship $P_i^{-1} \mathcal{A}_i^e P_i:e_p=\mathcal{A}_i^{g_i \rhd e^{-1}}:e_p$ for all plaquettes $p$).

\begin{figure}[h]
	\begin{center}
		\includegraphics{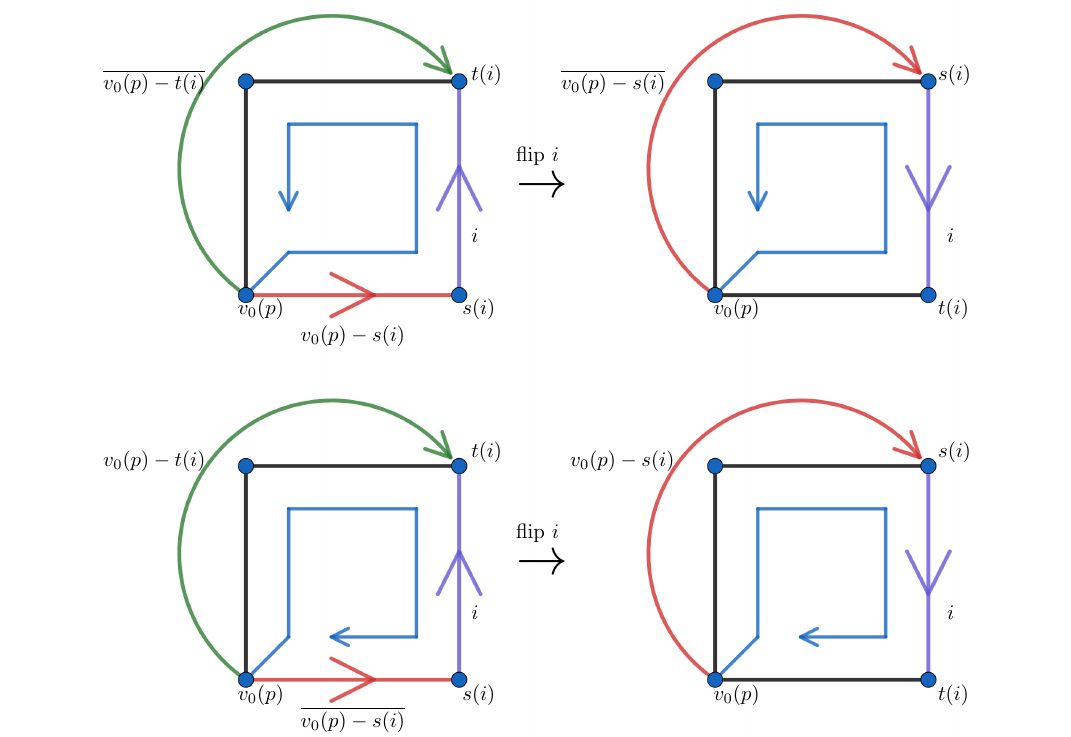}
		\caption{The action of an edge transform applied on edge $i$ on an adjacent plaquette $p$ depends on the orientation of that edge. If we flip the edge, then the paths which appear in the action of the edge transform change. In the top line, we consider the case where the edge $i$ initially points along the plaquette's orientation. The path which appears in the action of the edge transform is then $v_0(p)-s(i)$. If we flip the edge and then apply the edge transform, then instead the path $\overline{v_0(p)-s(i)}$ from the right side appears in the edge transform. Because we have flipped the edge, the source of the edge after it is flipped is the target of the edge before the flip. Therefore the path $\overline{v_0(p)-s(i)}$ that appears in the edge transform after the flip is the same path as $\overline{v_0(p)-t(i)}$ (green path on the left side) before the flip. In the lower line, we show the analogous situation when the edge initially points against the plaquette's orientation.}
		
		\label{flip_edge_edge_transform}
	\end{center}
\end{figure}

In order to determine whether the action on the plaquettes also satisfies this relationship, consider Equation \ref{Edge_transform_plaquette_appendix} which defines this action on the adjacent plaquettes. When we flip the edge, we reverse the relative orientation of $i$ and the plaquette, and so change which path element $g(v_0(p)-v_i)$ or $g(\overline{v_0(p)-v_i})$ appears in the expression for the action on the plaquette (i.e., we change whether the path travels along the circulation of the plaquette or against it). Furthermore, note that the vertex $v_i$ or $v_{i+1}$ which is the end-point of the path is always the source of the edge $i$, the vertex which the edge points away from. When we flip the edge, we exchange the source and target of the edge (where the target is the vertex the edge points towards), which therefore changes the end-point of the path that appears in the edge transform. These changes are shown in Figure \ref{flip_edge_edge_transform}. Denoting the original source of the edge $i$ by $s(i)$ and the original target of $i$ by $t(i)$, the action of the edge transform on an adjacent plaquette (without flipping the edge) is
$$\mathcal{A}_i^e :e_p = \begin{cases} e_p \: [g(v_0(p)-s(i)) \rhd e^{-1}] & i \text{ points along the boundary of } p \\ [g(\overline{v_0(p)-s(i)}) \rhd e] \: e_p & i \text{ points against the boundary.}\end{cases} $$

On the other hand if we act with $P_i^{-1} \mathcal{A}_i^e P_i$, we have
\begin{align*}
P_i^{-1} \mathcal{A}_i^e P_i:e_p &= \begin{cases} [g(\overline{v_0(p)-t(i)}) \rhd e] \: e_p & i \text{ originally points along the boundary of } p \\ e_p \: [g(v_0(p)-t(i)) \rhd e^{-1}] & i \text{ originally points against the boundary.}\end{cases} 
\end{align*}

Note that if $\rhd$ is trivial, we can forget the path elements and so we get $P_i^{-1}\mathcal{A}_i^eP_i=\mathcal{A}_i^{e^{-1}}$ and the overall edge term will commute with the flipping procedure. On the other hand, if $\rhd$ is non-trivial we need to consider the path elements more closely. Looking at Figure \ref{flip_edge_edge_transform}, we see that when $i$ points along the boundary of $p$, we have $$g(\overline{v_0(p)-t(i)})g_i^{-1}g(v_0(p)-s(i))^{-1}=g(\text{boundary}(p))^{-1}.$$ Provided that the plaquette satisfies fake-flatness, so that $g(\text{boundary}(p))^{-1}=\partial(e_p)$, this means that 
\begin{align*}
g(\overline{v_0(p)-t(i)}) &= g(\text{boundary}(p))^{-1} g(v_0(p)-s(i)) g_i \\
&= \partial(e_p)g(v_0(p)-s(i)) g_i 
\end{align*}
and so
\begin{align*}
P_i^{-1}\mathcal{A}_i^eP_i: e_p &=[g(\overline{v_0(p)-t(i)}) \rhd e] e_p\\
& =\big[(\partial(e_p) g(v_0(p)-s(i)) g_i ) \rhd e \big] e_p.
\end{align*}
Using the Peiffer condition Equation \ref{P2}, this becomes
\begin{align*}
P_i^{-1}\mathcal{A}_i^eP_i: e_p &= e_p \big[(g(v_0(p)-s(i)) g_i ) \rhd e \big]  e_p^{-1} e_p\\
&= e_p [g(v_0(p)-s(i)) \rhd (g_i \rhd e^{-1})^{-1}]\\
&=\mathcal{A}_i^{g_i \rhd e^{-1}}:e_p.
\end{align*}

In a similar way, if $i$ points against the boundary of the plaquette, we see from Figure \ref{flip_edge_edge_transform} that $$g(v_0(p)-t(i))g_i^{-1}g(\overline{v_0(p)-s_i})^{-1}= g(\text{boundary}(p)).$$ Again, if the plaquette satisfies fake-flatness, so that $g(\text{boundary}(p))=\partial(e_p)^{-1}$, then (following the same steps as before)
\begin{align*}
P_i^{-1}\mathcal{A}_i^eP_i: e_p &= e_p g(v_0(p)-t(i)) \rhd e^{-1}\\
&=e_p \big([\partial(e_p)^{-1}g(\overline{v_0(p)-s_i})g_i] \rhd e^{-1}\big)\\
&= e_p e_p^{-1} \big([g(\overline{v_0(p)-s_i})g_i] \rhd e^{-1}\big) e_p\\
&= \big(g(\overline{v_0(p)-s_i})\rhd (g_i \rhd e)^{-1}\big) e_p\\
&=\mathcal{A}_i^{g_i \rhd e^{-1}}:e_p.
\end{align*}

We can therefore see that, for the action on every degree of freedom (and so for the operators themselves), $P_i^{-1}\mathcal{A}_i^eP_i=\mathcal{A}_i^{g_i \rhd e^{-1}}$ and so $P_i^{-1}\mathcal{A}_iP_i=\mathcal{A}_i$. However, note that (unless $\rhd$ is trivial) this relies on the plaquettes around the edge satisfying fake-flatness. If fake-flatness is not satisfied near the edge and $\rhd$ is not trivial then we cannot say that the edge energy term is invariant under the edge flipping procedure. This indicates that the energy of the edge energy term is not independent of the supposedly arbitrary dressing of the lattice and so we do not consider the energy of the edge to be well-defined (although note that if we fix a branching structure and never consider changing it, the energy can still be defined). However, in the ground state, or more generally in regions where fake-flatness is satisfied, the energy of the edge is well-defined. We are typically interested in the ground state, and states with a few excitations. For such states, the edge energy is only ill-defined near these excitations. This does not affect the topological quantities of the theory, such as the braiding relations or topological charges, because these can be measured far away from the affected regions, where the theory is still fully consistent.

\subsection{Reversing the orientation of a plaquette}

Having considered the procedure of flipping an edge and inverting the corresponding edge label, and having shown that this is consistent with the various energy terms, we now consider the analogous procedure where we reverse the orientation of a plaquette and invert its label. We denote this operation by $Q_p$ for a plaquette $p$. First consider the vertex transform, $A_v^g$. Apart from the edges, which are unaffected by the orientation of plaquettes, this transform acts only on plaquettes with base-point at $v$. We have
$$A_v^g: e_p = \begin{cases} g \rhd e_p & \text{ if } v_0(p)=v\\ e_p & \text{ otherwise,}\end{cases}$$
where $v_0(p)$ is the base-point of plaquette $p$. The action of the vertex transform on plaquette $p$ can only be affected by flipping the plaquette $p$ itself (it is not affected by flipping other plaquettes). We have
\begin{align*}
Q_p^{-1}A_v^g Q_p: e_p &= Q_p^{-1}A_v^g: e_p^{-1}\\
&= Q_p^{-1}:\begin{cases} g \rhd e_p^{-1} & \text{ if } v_0(p)=v\\ e_p^{-1} & \text{ otherwise}\end{cases}\\
&=\begin{cases} g \rhd e_p & \text{ if } v_0(p)=v\\ e_p & \text{ otherwise}\end{cases}\\
&=A_v^g:e_p,
\end{align*}
from which we see that the vertex transforms (and so the vertex energy term) are invariant under the plaquette flipping procedure.

Next consider the plaquette term. Recall that the plaquette term for a plaquette $p$ is
$$B_p = \delta\big(\partial(e_p)g(\text{boundary}(p)),1_G\big).$$
If we flip the orientation of the plaquette, then we also reverse the orientation of its boundary, so that $g(\text{boundary}(p))$ is inverted. Therefore
\begin{align*}
Q_p^{-1}B_pQ_p &= \delta\big(\partial(e_p^{-1})g(\text{boundary}(p))^{-1},1_G\big)\\
&=\delta\big(g(\text{boundary}(p))^{-1},\partial(e_p)\big)\\
&=\delta\big(1_G, \partial(e_p)g(\text{boundary}(p))\big)\\
&=B_p,
\end{align*}
so that the plaquette energy term is preserved by the plaquette flipping procedure.

The blob energy term is similarly preserved by $Q_p$. Recall that the blob energy term checks that the blob 2-holonomy is equal to the identity of group $E$, where the blob 2-holonomy is a product of the plaquette elements around the boundary of the blob. For a blob $B$, the 2-holonomy is
$$H_2(B) =\prod_{p \in \text{Bd}(B)} g(v_0(B)-v_0(p)) \rhd e_p^{ \sigma_p},$$
where $\sigma_p$ is $1$ or $-1$ depending on the orientation of the plaquette $p$, $v_0(p)$ is the base-point of plaquette $p$ and $v_0(B)$ is the base-point of the blob $B$. Flipping the orientation of plaquette $p$ swaps $\sigma_p$ between the two, which cancels with the inverse from the flipping procedure itself (in a similar way to how the contribution of edge elements to a path was invariant under the edge flipping procedure), so that the blob 2-holonomy, and thus the blob energy term, is preserved by the plaquette flipping orientation.

\begin{figure}[h]
	\begin{center}
		\includegraphics{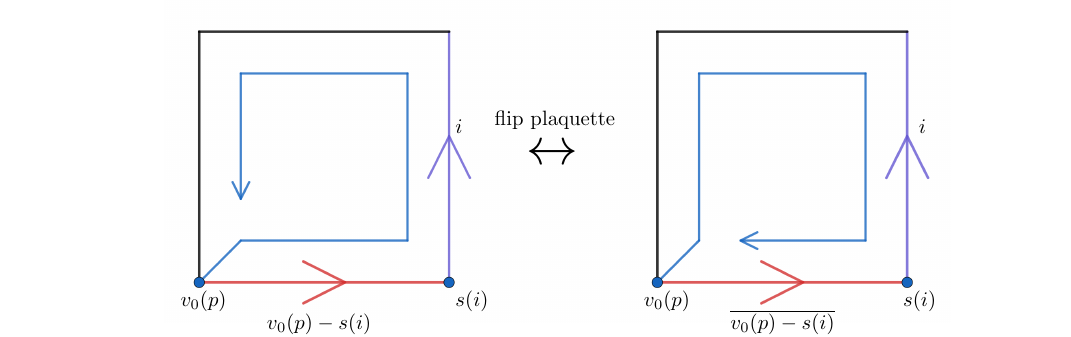}
		\caption{The paths used for the action of an edge transform on a plaquette depend on the orientation of the plaquette with respect to the edge. We may therefore think that flipping the orientation of the plaquette will change the path used for the action of the edge transform. However, we can see that while we do swap whether the path $v_0(p)-s(i)$ or $\overline{v_0(p)-s(i)}$ is used, we also swap the definition of these paths. If the plaquette is initially aligned with the edge and we swap the orientation of the plaquette, then the path $v_0(p)-s(i)$ before the flip is the same as the path $\overline{v_0(p)-s(i)}$ after the flip (and vice-versa).}
		
		\label{flip_plaquette_edge_transform}
	\end{center}
\end{figure}

This just leaves the edge energy terms to consider. Recall that the action of the edge transform $\mathcal{A}_i^e$ on an adjacent plaquette $p$ depends on the relative orientation of the edge $i$ and the plaquette $p$. We have
$$\mathcal{A}_i^e :e_p = \begin{cases} e_p \: [g(v_0(p)-s(i)) \rhd e^{-1}] & i \text{ points along the boundary of } p \\ [g(\overline{v_0(p)-s(i)}) \rhd e] \: e_p & i \text{ points against the boundary.}\end{cases} $$
Swapping the orientation of the plaquette does not change the source of $i$, unlike the edge flipping procedure. However, it does change the relative orientation of the plaquette and the edge. The orientation of the plaquette determines which path between the base-point of the plaquette and source of the edge we use. When the edge aligns with the orientation of the plaquette we use the path $(v_0(p)-s(i))$ that also aligns with the orientation of the plaquette, and when the edge is anti-aligned with the plaquette we use the path $(\overline{v_0(p)-s(i)})$ that travels against the orientation of the plaquette. Therefore flipping the plaquette changes which of these paths we use. However, when we flip the plaquette, the path that previously was aligned with the plaquette is now anti-aligned with the plaquette. This means that $(v_0(p)-s(i))$ before the flip is equal to $(\overline{v_0(p)-s(i)})$ afterwards (and vice-versa), as we show in Figure \ref{flip_plaquette_edge_transform}. These two effects cancel, and therefore flipping the orientation of the plaquette has no net effect on the path that appears in the edge transform (we swap whether we should use the aligned or anti-aligned path, but also swap which one is which). In addition, reversing the orientation of the plaquette changes whether we use pre-multiplication (by $g(\overline{v_0(p)-s(i)})\rhd e$) or post-multiplication (by $g(v_0(p)-s(i)) \rhd e^{-1}$). Using this, we see that
\begin{align*}
Q_p^{-1}\mathcal{A}_i^eQ_p: e_p &= Q_p^{-1}\mathcal{A}_i^e: e_p^{-1}\\
&=Q_p^{-1}: \begin{cases} [g(v_0(p)-s(i)) \rhd e] \: e_p^{-1} & i \text{ originally points along the boundary of } p \\ e_p^{-1} \: [g(\overline{v_0(p)-s(i)}) \rhd e^{-1} ]& i \text{ originally points against the boundary of } p \end{cases}\\
&= \begin{cases}e_p \: [g(v_0(p)-s(i)) \rhd e^{-1}] & i \text{ points along the boundary of } p \\ [g(\overline{v_0(p)-s(i)}) \rhd e] \: e_p & i \text{ points against the boundary}\end{cases}\\
&= \mathcal{A}_i^e: e_p,
\end{align*}
so the edge transforms are invariant under the plaquette flipping procedure. This means that every energy term is invariant under this procedure.

\subsection{Moving the base-point of a plaquette}
\label{Section_consistency_move_base_point}
The final procedure to consider is changing the base-point of a plaquette. We denote the procedure that moves the base-point of plaquette $p$ from a vertex $v_1$ to a vertex $v_2$ by $E_p({v_1 \rightarrow v_2})$. In cases where the precise path by which we move the base-point (rather than just the end-points) is important, we will state that this is the case. Under the operation $E_p({v_1 \rightarrow v_2})$, as well as changing the base-point of the plaquette, we must change the plaquette's label from $e_p$ to $g(v_1 - v_2)^{-1} \rhd e_p$, where $g(v_1-v_2)$ is the group element assigned to the path along which we move the base-point. In the case where $\rhd$ is trivial, the base-point of a plaquette is irrelevant, but in the more general cases the base-point of a plaquette affects the action of all of the energy terms. We first consider a vertex transform. The vertex transform at a vertex $v$ affects any plaquette whose base-point is at the vertex $v$. Furthermore, it affects path elements which start or end at the vertex $v$. This is relevant because the transformation of the plaquette label under $E_p({v_1 \rightarrow v_2})$ depends on the path element $g(v_1 - v_2)$, which is affected by vertex transforms at $v_1$ and $v_2$. This means that the vertex transforms at $v_1$ and $v_2$ might not commute with the procedure for moving the base-point. First consider the vertex transform $A_{v_1}^x$. We have $A_{v_1}^x: e_p = x \rhd e_p$. On the other hand, we have
\begin{align*}
E_p^{-1}({v_1 \rightarrow v_2})A_{v_1}^x E_p({v_1 - v_2}): e_p &= E_p^{-1}({v_1 \rightarrow v_2})A_{v_1}^x: g(v_1 - v_2)^{-1} \rhd e_p\\
&= E_p^{-1}({v_1 \rightarrow v_2}): g(v_1 - v_2)^{-1} \rhd e_p, 
\end{align*}
where in the last line the vertex transform leaves the plaquette element unchanged because the base-point of $p$ is no longer at $v_1$. However, the path element for $v_1 - v_2$ has been changed by the action of the vertex transform from its original value of $g(v_1 - v_2)$ to $xg(v_1 - v_2)$. This means that when we move the base-point back along the path, we pick up a factor of $xg(v_1 - v_2)$ acting on the plaquette, rather than just a factor of $g(v_1 - v_2)$. This gives us
\begin{align*}
E_p^{-1}({v_1 \rightarrow v_2}): g(v_1 - v_2)^{-1} \rhd e_p&= (xg(v_1 - v_2)) \rhd (g(v_1 - v_2)^{-1} \rhd e_p)\\
&= x \rhd e_p, 
\end{align*}
so that
\begin{align*}
E_p^{-1}({v_1 \rightarrow v_2})A_{v_1}^x E_p({v_1 \rightarrow v_2}): e_p &= A_{v_1}^x:e_p.
\end{align*} 

We therefore see that the vertex transforms (and so the vertex energy term) at $v_1$ are invariant under changing the base-point. Now consider the vertex transforms at $v_2$, the position of the base-point after we move it. If we do not move the base-point, we have $A_{v_2}^x: e_p =e_p$. On the other hand, if we do move the base-point we have:
\begin{align*}
E_p^{-1}({v_1 \rightarrow v_2})A_{v_2}^x E_p({v_1 \rightarrow v_2}): e_p &= E_p^{-1}({v_1 \rightarrow v_2})A_{v_2}^x: g(v_1 \rightarrow v_2)^{-1} \rhd e_p\\
&=E_p^{-1}({v_1 \rightarrow v_2}): x \rhd [g(v_1 \rightarrow v_2)^{-1} \rhd e_p],
\end{align*}
where the vertex transform affects the plaquette label because the base-point is at $v_2$ after the action of $E_p({v_1 \rightarrow v_2})$. However, the vertex transform also changes the path element $g(v_1 \rightarrow v_2)$ to $g(v_1 \rightarrow v_2)x^{-1}$. Therefore
\begin{align*}
E_p^{-1}({v_1 \rightarrow v_2}): x \rhd [g(v_1 \rightarrow v_2)^{-1} \rhd e_p]&= (g(v_1 \rightarrow v_2)x^{-1}) \rhd \big(x \rhd [g(v_1 \rightarrow v_2)^{-1} \rhd e_p]\big)\\
&= e_p, 
\end{align*}
from which we see that
\begin{align*}
E_p^{-1}({v_1 \rightarrow v_2})A_{v_2}^x E_p({v_1 \rightarrow v_2}): e_p &= A_{v_2}^x :e_p.
\end{align*}
This means that the vertex energy term at $v_2$ is also unaffected by our procedure for changing the base-points of plaquettes, and so all of the vertex transforms are preserved by this procedure.

\begin{figure}[h]
	\begin{center}
	\includegraphics[width=0.9\linewidth]{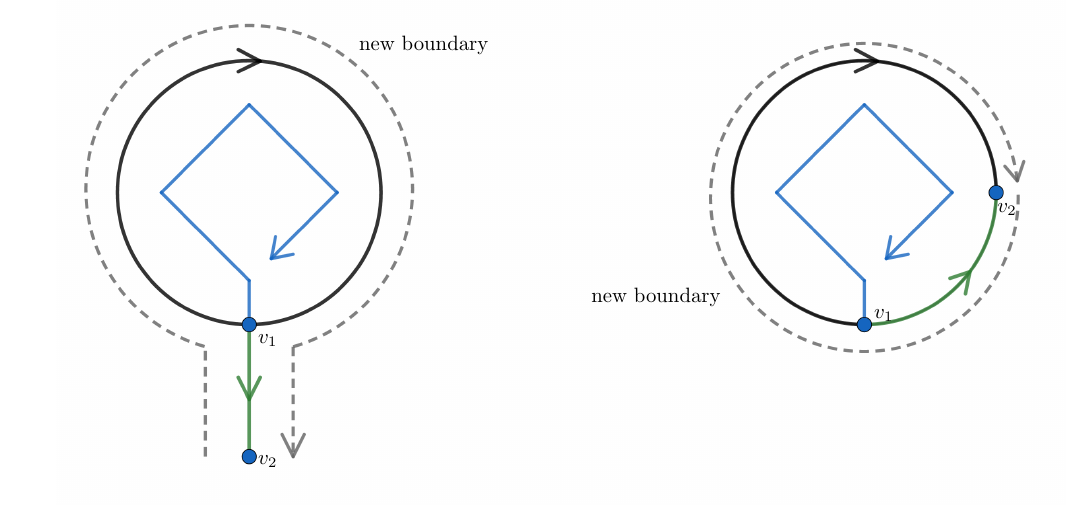}
		\caption{When we change the base-point of a plaquette, we also change the boundary of that plaquette. In the left image we show a case where the base-point is moved from a vertex $v_1$ to $v_2$, along a path $v_1-v_2$ (green) that leaves the plaquette. That is, we whisker the plaquette. We can see that the new boundary (grey dotted line) is $(v_1-v_2)^{-1}\cdot(\text{boundary})\cdot(v_1-v_2)$, where (boundary) is the original boundary of the plaquette (black). In the right image, we instead show a case where the base-point of the plaquette is moved along the boundary of the plaquette. In this case, it is still true that the boundary after moving the base-point is $(v_1-v_2)^{-1}\cdot(\text{boundary})\cdot(v_1-v_2)$, although this time there is some cancellation between sections of the original boundary and $(v_1-v_2)$.}
		\label{branching_change_base_point_plaquette}
	\end{center}
\end{figure}

Next consider the plaquette energy terms. Moving the base-point of a plaquette also affects the boundary of that plaquette, as shown in Figure \ref{branching_change_base_point_plaquette}. We see that under $E_p({v_1 \rightarrow v_2})$, the boundary of plaquette $p$ transforms as 
$$g(\text{boundary}(p)) \rightarrow g(v_1 \rightarrow v_2)^{-1}g(\text{boundary}(p))g(v_1 \rightarrow v_2).$$
Therefore the plaquette holonomy $H_1(p)= \partial(e_p)g(\text{boundary}(p))$ becomes
\begin{align*}
\partial(g(v_1 \rightarrow v_2)^{-1} \rhd e_p ) & g(v_1 \rightarrow v_2)^{-1}g(\text{boundary}(p))g(v_1 \rightarrow v_2)\\
&= g(v_1 \rightarrow v_2)^{-1} \partial(e_p) g(v_1 \rightarrow v_2) g(v_1 \rightarrow v_2)^{-1}g(\text{boundary}(p))g(v_1 \rightarrow v_2)\\
&=g(v_1 \rightarrow v_2)^{-1} \partial(e_p)g(\text{boundary}(p))g(v_1 \rightarrow v_2),
\end{align*}
where we used the Peiffer condition Equation \ref{P1} to write $\partial(g(v_1 \rightarrow v_2)^{-1} \rhd e_p )= g(v_1 \rightarrow v_2)^{-1} \partial(e_p) g(v_1 \rightarrow v_2)$. We see that the plaquette holonomy is merely conjugated by a path element, which preserves the identity element. Therefore the energy term (which checks if the plaquette holonomy is equal to the identity) is unaffected by the base-point changing procedure.

The next energy term to consider is the blob term. This checks that the blob 2-holonomy is the identity, where the blob 2-holonomy is
$$H_2(B) =\prod_{p \in \text{Bd}(B)} g(v_0(B)-v_0(p)) \rhd e_p^{ \sigma_p},$$
where $\sigma_p$ depends on the orientation of the plaquette. The contribution of a particular plaquette $p$ to the blob 2-holonomy is $g(v_0(B)-v_0(p)) \rhd e_p^{ \sigma_p}$. If we change the base-point of this plaquette from $v_1$ to $v_2$ then we must change the label $e_p$ to $g(v_1 \rightarrow v_2)^{-1} \rhd e_p$ due to the effect of $E_p$, but we must also change the base-point $v_0(p)$ that appears in the expression $g(v_0(B)-v_0(p)) \rhd e_p^{ \sigma_p}$ along the same path (the same path so that the resulting surface is the same as the original). This means that when we change the base-point of plaquette $p$, its contribution to the blob 2-holonomy transforms as
\begin{align*}
g(v_0(B)-v_0(p)) \rhd e_p^{ \sigma_p} &\rightarrow g(v_0(B)-v_2) \rhd (g(v_1 \rightarrow v_2)^{-1} \rhd e_p^{\sigma_p})\\
&=\big(g(v_0(B)-v_1)g(v_1 \rightarrow v_2) \big) \rhd (g(v_1 \rightarrow v_2)^{-1} \rhd e_p^{\sigma_p})\\
&= g(v_0(B)-v_1) \rhd e_p^{\sigma_p},
\end{align*}
so that the contribution of the plaquette to the blob 2-holonomy is unchanged by moving the base-point of the plaquette.

The final energy terms to consider are the edge energy terms. There are several cases to consider. We need to consider edge terms that directly affect the plaquette whose base-point is being moved (i.e., the edges on the boundary of that plaquette), but also any edge transforms along the path on which we move the plaquette's base-point. First consider the latter kind of edge, which is not on the original boundary of the plaquette, but is on the path along which we move the base-point. If we move the base-point of the plaquette before acting with the edge transform, then the path becomes part of the boundary of the plaquette. That is, the plaquette is whiskered along the path through which we move the base-point. The action of the edge transform, as defined in Equation \ref{Equation_edge_transform_definition_2}, does not account for the possibility of a whiskered plaquette. We give a more general definition of the edge transform in Ref. \cite{HuxfordPaper2} (in Section S-I C), which describes the action on whiskered plaquettes and other special cases. We define this generalized edge transform in a way that is consistent with the procedure for changing the base-point by whiskering (at least if fake-flatness is satisfied), as we show in Ref. \cite{HuxfordPaper2} when we define the transform. We will therefore not prove this consistency here.

Next we consider the case where we move the base-point along the boundary of the plaquette and $i$ is one of the edges on the plaquette. Recall that the action of the edge transform $\mathcal{A}_i^e$ on an adjacent plaquette is
$$\mathcal{A}_i^e :e_p = \begin{cases} e_p [g(v_0(p)-s(i)) \rhd e^{-1}] & i \text{ points along the boundary of } p \\ [g(\overline{v_0(p)-s(i)}) \rhd e] e_p & i \text{ points against the boundary.}\end{cases} $$
Now consider $E_p(v_1 \rightarrow v_2)^{-1}\mathcal{A}_i^eE_p(v_1 \rightarrow v_2)$, where $v_1$ is the initial base-point of the plaquette $p$, $v_0(p)$. Then
\begin{align*}
E_p(v_1 \rightarrow v_2)^{-1}&\mathcal{A}_i^eE_p(v_1 \rightarrow v_2):e_p\\
&= E_p(v_1 \rightarrow v_2)^{-1}\mathcal{A}_i^e: g(v_1 -v_2)^{-1} \rhd e_p\\
&=E_p(v_1 \rightarrow v_2)^{-1}: \begin{cases} (g(v_1 -v_2)^{-1} \rhd e_p) \: [g(v_2-s(i)) \rhd e^{-1}] & i \text{ points along the boundary of } p \\ [g(\overline{v_2-s(i)}) \rhd e] (g(v_1 -v_2)^{-1} \rhd e_p) & i \text{ points against the boundary.}\end{cases} 
\end{align*}

Then, if the edge $i$ is not on the path $v_1-v_2$, so that the label of the path is unaffected by the edge transform, moving the base-point of the plaquette back gives
\begin{align*}
E_p(v_1 \rightarrow v_2)^{-1}&\mathcal{A}_i^eE_p(v_1 \rightarrow v_2):e_p\\
&=\begin{cases}g(v_1-v_2) \rhd \big((g(v_1 -v_2)^{-1} \rhd e_p) \: [g(v_2-s(i)) \rhd e^{-1}]\big) & i \text{ points along the boundary of } p \\ g(v_1-v_2) \rhd \big([g(\overline{v_2-s(i)}) \rhd e]\: (g(v_1 -v_2)^{-1} \rhd e_p)\big) & i \text{ points against the boundary}\end{cases}\\
&= \begin{cases} e_p \: \big[\big(g(v_1-v_2)g(v_2-s(i))\big) \rhd e^{-1} \big]& i \text{ points along the boundary of } p \\ \big[\big(g(v_1-v_2)g(\overline{v_2-s(i)})\big) \rhd e\big] e_p & i \text{ points against the boundary.}\end{cases}
\end{align*}

\begin{figure}[h]
	\begin{center}
		\includegraphics[width=0.8\linewidth]{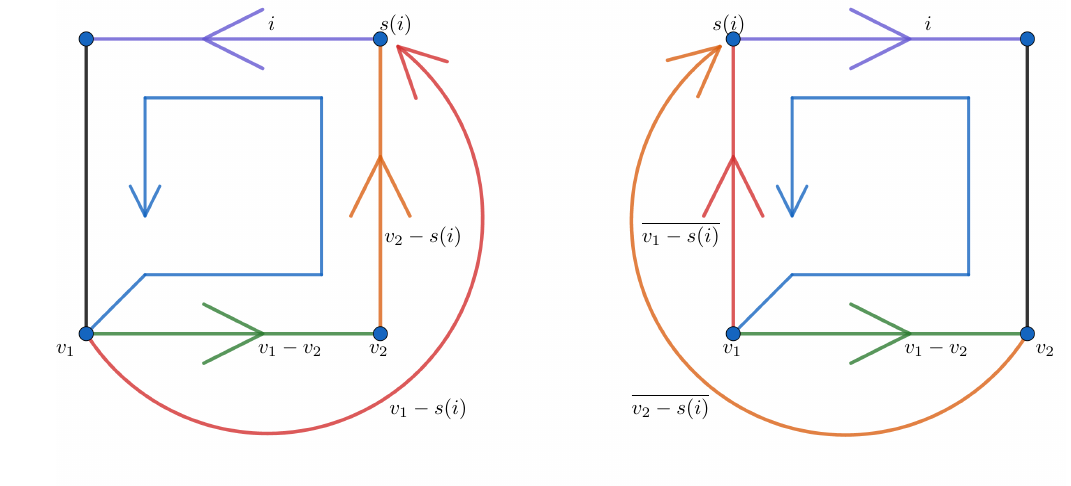}
		\caption{The action of the edge transform on a plaquette $p$ depends on various paths, which change when we change the base-point of the plaquette. In this figure, we show two examples. In the left case, the orientation of the edge on which we apply the transform ($i$) matches the orientation of the plaquette (shown in blue). Because these orientations match, the path which appears in the edge transform is $v_0(p)-s(i)$, which has the same orientation as the plaquette. We see that if the original base-point is $v_1$ and we move it to $v_2$ along the path $v_1-v_2$, where the edge $i$ is not on this path, then $v_1-s(i)=(v_1-v_2) \cdot (v_2-s(i))$. Similarly, in the right of the figure we show a case where the edge points against the orientation of the plaquette, so that the path $\overline{v_0(p)-s(i)}$, which has the opposite orientation to the plaquette, appears in the edge transform. In this case $\overline{v_1-s(i)}=(v_1-v_2) \cdot(\overline{v_0(p)-s(i)})$. Note that there are two other cases, where we flip the orientation of the plaquette in each of these images, where the paths obey similar relations.}
		\label{change_base_point_edge_1}
	\end{center}
\end{figure}

We then need to consider the path elements involved in the expression above. There are various cases, as shown in Figure \ref{change_base_point_edge_1}. We see that $g(v_1-v_2)g(v_2-s(i))=g(v_1-s(i))$ if the edge is aligned with the boundary of $p$ and $g(v_1-v_2)g(\overline{v_2-s(i)}))=g(\overline{v_1-s(i)})$ if it is anti-aligned. Therefore
\begin{align*}
E_p(v_1 \rightarrow v_2)^{-1}\mathcal{A}_i^eE_p(v_1 \rightarrow v_2):e_p &= \begin{cases} e_p \: [(g(v_1-v_2)g(v_2-s(i))) \rhd e^{-1}] & i \text{ points along the boundary of } p \\ [(g(v_1-v_2)g(\overline{v_2-s(i)})) \rhd e]\: e_p & i \text{ points against the boundary}\end{cases}\\
&=\begin{cases} e_p \:[g(v_0(p)-s(i)) \rhd e^{-1}] & i \text{ points along the boundary of } p \\ [g(\overline{v_0(p)-s(i)}) \rhd e] \:e_p & i \text{ points against the boundary}\end{cases}\\
&= \mathcal{A}_i^e:e_p,
\end{align*}
and so the action of the edge transform is preserved.

\begin{figure}[h]
	\begin{center}
		\includegraphics[width=0.9\linewidth]{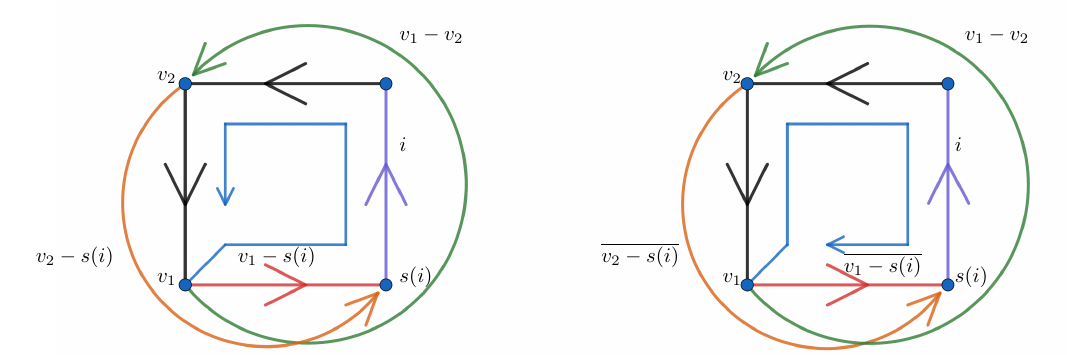}
		\caption{In this figure, we consider the case where the path on which we move the base-point includes the edge $i$ on which we apply the edge transform. In the left case, the edge $i$ has the same orientation as the plaquette. In this case, the path $(v_1-v_2) \cdot(v_2-s(i))$ includes the entire boundary of the plaquette, so $(v_1-v_2) \cdot(v_2-s(i))= (\text{boundary})\cdot (v_1-s(i))$. In the right image, we consider the case where the edge and plaquette have opposite orientations. In this case $(v_1-v_2) \cdot(\overline{v_2-s(i)})= (\text{boundary})^{-1}\cdot (v_1-s(i))$. }
		\label{change_base_point_edge_2}
	\end{center}
\end{figure}

Next we consider the case where $i$ is not only on the plaquette, but is also one of the edges on the path along which we move the base-point, as shown in Figure \ref{change_base_point_edge_2}. In this case, the edge transform affects the path element before we return it to its original position. If the edge $i$ points along $v_1-v_2$, then the edge transform acts on the path element $g(v_1-v_2) = g(v_1-s(i))g_i g(t(i)-v_2)$ as
\begin{align*}
\mathcal{A}_i^e: g(v_1-v_2)&=\mathcal{A}_i^e:\begin{cases} g(v_1-s(i))g_i g(t(i)-v_2) & i \text{ points along the boundary of $p$}\\ g(\overline{v_1-s(i)})g_i g(t(i)-v_2) & i \text{ points against the boundary of $p$} \end{cases}\\ 
&= \begin{cases} g(v_1-s(i))\partial(e)g_i g(t(i)-v_2) & i \text{ points along the boundary of $p$}\\ g(\overline{v_1-s(i)})\partial(e)g_i g(t(i)-v_2) & i \text{ points against the boundary of $p$,} \end{cases}\\ 
\end{align*}
where we split the path $g(v_1-v_2)$ into parts and then used the action of the edge transform on the edge $i$. Then we want to write this again in terms of $g(v_1-v_2)$, so we introduce a factor of the identity in the form of $ g(v_1-s(i))^{-1} g(v_1-s(i))$ (or the equivalent with $g(\overline{v_1-s(i)})$) to obtain

\begin{align}
\mathcal{A}_i^e: g(v_1-v_2)&=\begin{cases} g(v_1-s(i))\partial(e)g(v_1-s(i))^{-1} g(v_1-s(i))g_i g(t(i)-v_2) & i \text{ points along the boundary of $p$} \notag \\ g(\overline{v_1-s(i)})\partial(e)g(\overline{v_1-s(i)})^{-1} g(\overline{v_1-s(i)})g_i g(t(i)-v_2) & i \text{ points against the boundary of $p$} \end{cases} \notag \\ 
&=\begin{cases} \partial \big(g(v_1-s(i)) \rhd e\big) g(v_1-v_2) & i \text{ points along the boundary of $p$}\\ \partial \big(g(\overline{v_1-s(i)}) \rhd e\big)g(v_1-v_2) & i \text{ points against the boundary of $p$.} \end{cases} \label{Equation_edge_transform_path_move_plaquette}
\end{align}

In addition, when $i$ is on the path $v_1-v_2$, then the path $v_1-s(i)$ (or $\overline{v_1-s(i)}$) along the boundary is not $(v_1-v_2) \cdot (v_2-s(i))$ (or $(v_1-v_2) \cdot (\overline{v_2-s(i)})$), as we can see from Figure \ref{change_base_point_edge_2}. Instead $(v_1-v_2) \cdot (v_2-s(i))= (\text{boundary}(p)) \cdot (v_1-s(i))$ if $i$ points along the boundary and $(v_1-v_2) \cdot (\overline{v_2-s(i)})=(\text{boundary}(p))^{-1} \cdot (\overline{v_1-v_i})$ if the edge $i$ points against the boundary. If the plaquette satisfies fake-flatness, then $g(\text{boundary}(p))^{-1}=\partial(e_p)$. We can then use this to find the action of $E_p(v_1 \rightarrow v_2)^{-1}\mathcal{A}_i^eE_p(v_1 \rightarrow v_2)$ on the plaquette label. We have
\begin{align*}
E_p&(v_1 \rightarrow v_2)^{-1}\mathcal{A}_i^eE_p(v_1 \rightarrow v_2):e_p\\
&= E_p(v_1 \rightarrow v_2)^{-1}\mathcal{A}_i^e: g(v_1 -v_2)^{-1} \rhd e_p\\
&=E_p(v_1 \rightarrow v_2)^{-1}: \begin{cases} [g(v_1 -v_2)^{-1} \rhd e_p] \: [g(v_2-s(i)) \rhd e^{-1}] & i \text{ points along the boundary of } p \\ [g(\overline{v_2-s(i)}) \rhd e] \: [g(v_1 -v_2)^{-1} \rhd e_p] & i \text{ points against the boundary of p.}\end{cases} 
\end{align*}

Then because the edge $i$ is on the path $v_1-v_2$, so that the label of this path is altered by the edge transform according to Equation \ref{Equation_edge_transform_path_move_plaquette}, moving the base-point of the plaquette back gives
\begin{align*}
E_p&(v_1 \rightarrow v_2)^{-1}\mathcal{A}_i^eE_p(v_1 \rightarrow v_2):e_p\\ &=\begin{cases}\big(\partial(g(v_1-s(i)) \rhd e)g(v_1-v_2)\big) \rhd \big( [g(v_1 -v_2)^{-1} \rhd e_p] \: [g(v_2-s(i)) \rhd e^{-1}]\big) & i \text{ points along the boundary of } p \\ \big(\partial(g(\overline{v_1-s(i)}) \rhd e)g(v_1-v_2)\big) \rhd \big([g(\overline{v_2-s(i)}) \rhd e] \:(g(v_1 -v_2)^{-1} \rhd e_p)\big) & i \text{ points against the boundary of $p$.}\end{cases}
\end{align*}

We can then use the Peiffer condition Equation \ref{P2}, to remove the factor of $\partial(g(v_1-s(i)) \rhd e)$ in $[\partial(g(v_1-s(i)) \rhd e)g(v_1-v_2)]$ in favour of conjugation of the entire expression by $g(v_1-s(i)) \rhd e$ (and similar for the other orientation):
\begin{align*}
E_p&(v_1 \rightarrow v_2)^{-1}\mathcal{A}_i^eE_p(v_1 \rightarrow v_2):e_p\\
&= \begin{cases}[g(v_1-s(i)) \rhd e]  \big[g(v_1-v_2) \rhd \big([g(v_1 -v_2)^{-1} \rhd e_p] [g(v_2-s(i)) \rhd e^{-1}]\big)\big] [g(v_1-s(i)) \rhd e^{-1}] & i \text{ points along} \\ & \text{ the boundary of } p \\ [g(\overline{v_1-s(i)}) \rhd e] \big[g(v_1-v_2) \rhd \big([g(\overline{v_2-s(i)}) \rhd e] [g(v_1 -v_2)^{-1} \rhd e_p]\big)\big]  [g(\overline{v_1-s(i)}) \rhd e^{-1}] & i \text{ points against} \\& \text{ the boundary of $p$}\end{cases}\\
&= \begin{cases}[g(v_1-s(i)) \rhd e] \big[e_p [(g(v_1-v_2)g(v_2-s(i))) \rhd e^{-1}]\big] \: [g(v_1-s(i)) \rhd e^{-1}] & i \text{ points along the boundary of } p \\ [g(\overline{v_1-s(i)}) \rhd e] \big[[(g(v_1-v_2)g(\overline{v_2-s(i)})) \rhd e] e_p\big] [g(\overline{v_1-s(i)}) \rhd e^{-1}]& i \text{ points against the boundary of $p$,}\end{cases}
\end{align*}
where in the last line we used the group homomorphism property $g \rhd (e_1 e_2)= (g \rhd e_1) (g \rhd e_2)$ to distribute the action $g(v_1-v_2) \: \rhd$ across the terms in the curved brackets. Next we use the relationships 
$$g(v_1-s(i))=g(\text{boundary}(p))^{-1}g(v_1-v_2)g(v_2-s(i)) = \partial(e_p)g(v_1-v_2)g(v_2-s(i)) $$
and
$$g(\overline{v_1-s(i)})=g(\text{boundary}(p))g(v_1-v_2)g(\overline{v_2-s(i)})=\partial(e_p^{-1})g(v_1-v_2)g(\overline{v_2-s(i)})$$
(where in each case the latter equality is obtained from fake-flatness of the plaquette) to write
\begin{align*}
E_p&(v_1 \rightarrow v_2)^{-1}\mathcal{A}_i^eE_p(v_1 \rightarrow v_2):e_p\\
&= \begin{cases} [g(v_1-s(i)) \rhd e]  \big[ e_p [(\partial(e_p^{-1})g(v_1-s(i))) \rhd e^{-1}]\big] [g(v_1-s(i)) \rhd e^{-1}] & i \text{ points along the boundary of } p \\ [g(\overline{v_1-s(i)}) \rhd e]  \big[[(\partial(e_p)g(\overline{v_1-s(i)})) \rhd e] e_p\big] [g(\overline{v_1-s(i)}) \rhd e^{-1}]& i \text{ points against the boundary}\end{cases}\\
&= \begin{cases}[g(v_1-s(i)) \rhd e] \big[ e_p e_p^{-1}[g(v_1-s(i)) \rhd e^{-1}] \: e_p\big] [g(v_1-s(i)) \rhd e^{-1}] & i \text{ points along the boundary of } p \\ [g(\overline{v_1-s(i)}) \rhd e] \: \big[e_p[g(\overline{v_1-s(i)}) \rhd e]e_p^{-1} e_p\big] [g(\overline{v_1-s(i)}) \rhd e^{-1}]& i \text{ points against the boundary,}\end{cases}
\end{align*}
where we again used the Peiffer condition Equation \ref{P2}, this time on the factor $\partial(e_p)$. Then we can simplify the expression to give
\begin{align*}
E_p&(v_1 \rightarrow v_2)^{-1}\mathcal{A}_i^eE_p(v_1 \rightarrow v_2):e_p\\
&= \begin{cases}[(v_1-s(i)) \rhd e] [g(v_1-s(i)) \rhd e^{-1}]e_p  [g(v_1-s(i)) \rhd e^{-1}] & i \text{ points along the boundary of } p \\ [g(\overline{v_1-s(i)}) \rhd e] e_p[g(\overline{v_1-s(i)}) \rhd e] [g(\overline{v_1-s(i)}) \rhd e^{-1}]& i \text{ points against the boundary}\end{cases}\\
&= \begin{cases}e_p \: [g(v_1-s(i)) \rhd e^{-1}] & i \text{ points along the boundary of } p \\ [g(\overline{v_1-s(i)}) \rhd e] \: e_p& i \text{ points against the boundary}\end{cases}\\
&=\mathcal{A}_i^e:e_p.
\end{align*}

This means that again the action of the edge transform, and therefore the edge energy term, is invariant under moving the base-point. We also need to consider the case where $i$ points against the path $v_1-v_2$. The calculation for this case is very similar. Rather than go through it, we will justify the fact that this case works by noting that the edge energy term is invariant under flipping the edge (from previous calculations in this section). Similarly, the base-point moving procedure is invariant under such a flip. Therefore, we can always flip the edge so that it points along $v_1-v_2$ before we apply either the edge transform or move the base-point of the plaquette. This means that if $i$ points in the wrong way, we can apply $P_i$ first. Then, using the above argument for the case where $i$ points along the path, we have
$$\mathcal{A}_i^e P_i = E_p(v_1 \rightarrow v_2)^{-1}\mathcal{A}_i^eE_p(v_1 \rightarrow v_2)P_i$$
which implies that $\mathcal{A}_i^e= E_p(v_1 \rightarrow v_2)^{-1}\mathcal{A}_i^eE_p(v_1 \rightarrow v_2)$.

We have therefore shown that each of the procedures for changing the structure of the lattice (flipping the orientations of edges and plaquettes or moving the base-point of a plaquette) is consistent with each of the energy terms. It is important to note that to show this for some energy terms we had to require that fake-flatness was satisfied in the region where the energy operator has support, at least when $\rhd$ is non-trivial. This further shows the importance of fake-flatness in ensuring the consistency of the higher-lattice gauge theory model.

\subsection{Use of the re-branching procedures for other proofs}
\label{Section_rebranching_use}
In addition to demonstrating the consistency of the higher lattice gauge theory model, these procedures for changing the dressing of the lattice will be useful when considering the ribbon or membrane operators and their commutation relations with the energy terms, as we do in Refs. \cite{HuxfordPaper2}  and \cite{HuxfordPaper3}. When we consider such commutation relations, the dressing of the lattice determines the action of both the energy terms and the ribbon operators, which means we need to consider several different cases for each commutation relation, depending on the branching structure in the region of the membrane operator and energy term. However, we can avoid this by instead showing that the membrane and ribbon operators are invariant under changing the dressing of the lattice in the same way as the energy terms are. In this case, we can demonstrate the commutation relations for one choice of the branching structure. Then, because the operators are invariant under changes to the dressing, the commutation relation will hold for all choices of branching structure. Let $\hat{X}$ be a series of these re-branching operations, and suppose a commutation relation between two operators $\hat{O}_1$ and $\hat{O_2}$ holds when acting on states that are defined when the lattice has a particular structure. Then the relation also holds for the branching structure produced by acting with $\hat{X}$ on these states. For example, suppose the operators $\hat{O}_1$ and $\hat{O}_2$ commute for one choice of branching structure. Then for any state $\ket{\psi}$ defined on the original lattice, $\hat{O}_1 \hat{O}_2 \ket{\psi}=\hat{O}_2 \hat{O}_1\ket{\psi}$. If the operators are consistent with the re-branching then $X^{-1}\hat{O}_iX=\hat{O}_i$. Therefore
\begin{align*}
X^{-1}\hat{O}_1X X^{-1}\hat{O}_2X\ket{\psi}&=X^{-1}\hat{O}_2 X X^{-1}\hat{O}_1X\ket{\psi}\\
\implies X^{-1}\hat{O}_1 \hat{O}_2 X\ket{\psi}&= X^{-1} \hat{O}_2 \hat{O}_1 X\ket{\psi}\\
\implies \hat{O}_1 \hat{O}_2 X\ket{\psi}&= \hat{O}_2 \hat{O}_1 X\ket{\psi},
\end{align*}
so that the commutation relation also holds for the state $X \ket{\psi}$, that is for states defined on the lattice with the altered branching structure.

\twocolumngrid

\bibliography{references}{}

\begin{thebibliography}{96}%
\makeatletter
\providecommand \@ifxundefined [1]{%
 \@ifx{#1\undefined}
}%
\providecommand \@ifnum [1]{%
 \ifnum #1\expandafter \@firstoftwo
 \else \expandafter \@secondoftwo
 \fi
}%
\providecommand \@ifx [1]{%
 \ifx #1\expandafter \@firstoftwo
 \else \expandafter \@secondoftwo
 \fi
}%
\providecommand \natexlab [1]{#1}%
\providecommand \enquote  [1]{``#1''}%
\providecommand \bibnamefont  [1]{#1}%
\providecommand \bibfnamefont [1]{#1}%
\providecommand \citenamefont [1]{#1}%
\providecommand \href@noop [0]{\@secondoftwo}%
\providecommand \href [0]{\begingroup \@sanitize@url \@href}%
\providecommand \@href[1]{\@@startlink{#1}\@@href}%
\providecommand \@@href[1]{\endgroup#1\@@endlink}%
\providecommand \@sanitize@url [0]{\catcode `\\12\catcode `\$12\catcode
  `\&12\catcode `\#12\catcode `\^12\catcode `\_12\catcode `\%12\relax}%
\providecommand \@@startlink[1]{}%
\providecommand \@@endlink[0]{}%
\providecommand \url  [0]{\begingroup\@sanitize@url \@url }%
\providecommand \@url [1]{\endgroup\@href {#1}{\urlprefix }}%
\providecommand \urlprefix  [0]{URL }%
\providecommand \Eprint [0]{\href }%
\providecommand \doibase [0]{https://doi.org/}%
\providecommand \selectlanguage [0]{\@gobble}%
\providecommand \bibinfo  [0]{\@secondoftwo}%
\providecommand \bibfield  [0]{\@secondoftwo}%
\providecommand \translation [1]{[#1]}%
\providecommand \BibitemOpen [0]{}%
\providecommand \bibitemStop [0]{}%
\providecommand \bibitemNoStop [0]{.\EOS\space}%
\providecommand \EOS [0]{\spacefactor3000\relax}%
\providecommand \BibitemShut  [1]{\csname bibitem#1\endcsname}%
\let\auto@bib@innerbib\@empty
\bibitem [{\citenamefont {Landau}(1937)}]{Landau}%
  \BibitemOpen
  \bibfield  {author} {\bibinfo {author} {\bibfnamefont {L.~D.}\ \bibnamefont
  {Landau}},\ }\bibfield  {title} {\bibinfo {title} {On the theory of phase
  transitions. {I}.},\ }\href@noop {} {\bibfield  {journal} {\bibinfo
  {journal} {Phys. Z. Sowjetunion}\ }\textbf {\bibinfo {volume} {11}} (\bibinfo
  {year} {1937})}\BibitemShut {NoStop}%
\bibitem [{\citenamefont {Wen}(1989)}]{Wen1989}%
  \BibitemOpen
  \bibfield  {author} {\bibinfo {author} {\bibfnamefont {X.-G.}\ \bibnamefont
  {Wen}},\ }\bibfield  {title} {\bibinfo {title} {Vacuum degeneracy of chiral
  spin states in compactified space},\ }\href
  {https://doi.org/10.1103/physrevb.40.7387} {\bibfield  {journal} {\bibinfo
  {journal} {Phys. Rev. B}\ }\textbf {\bibinfo {volume} {40}},\ \bibinfo
  {pages} {7387} (\bibinfo {year} {1989})}\BibitemShut {NoStop}%
\bibitem [{\citenamefont {Wen}(1990)}]{Wen1990a}%
  \BibitemOpen
  \bibfield  {author} {\bibinfo {author} {\bibfnamefont {X.-G.}\ \bibnamefont
  {Wen}},\ }\bibfield  {title} {\bibinfo {title} {Topological orders in rigid
  states},\ }\href {https://doi.org/10.1142/s0217979290000139} {\bibfield
  {journal} {\bibinfo  {journal} {Int. J. Mod. Phys. B}\ }\textbf {\bibinfo
  {volume} {04}},\ \bibinfo {pages} {239 } (\bibinfo {year}
  {1990})}\BibitemShut {NoStop}%
\bibitem [{\citenamefont {Wen}(2013)}]{Wen2013}%
  \BibitemOpen
  \bibfield  {author} {\bibinfo {author} {\bibfnamefont {X.-G.}\ \bibnamefont
  {Wen}},\ }\bibfield  {title} {\bibinfo {title} {Topological order: From
  long-range entangled quantum matter to a unified origin of light and
  electrons},\ }\bibfield  {journal} {\bibinfo  {journal} {ISRN Condens. Matter
  Phys.}\ }\href {https://doi.org/10.1155/2013/198710} {10.1155/2013/198710}
  (\bibinfo {year} {2013})\BibitemShut {NoStop}%
\bibitem [{\citenamefont {Tsui}\ \emph {et~al.}(1982)\citenamefont {Tsui},
  \citenamefont {Stormer},\ and\ \citenamefont {Gossard}}]{Tsui1982}%
  \BibitemOpen
  \bibfield  {author} {\bibinfo {author} {\bibfnamefont {D.~C.}\ \bibnamefont
  {Tsui}}, \bibinfo {author} {\bibfnamefont {H.~L.}\ \bibnamefont {Stormer}},\
  and\ \bibinfo {author} {\bibfnamefont {A.~C.}\ \bibnamefont {Gossard}},\
  }\bibfield  {title} {\bibinfo {title} {Two-dimensional magnetotransport in
  the extreme quantum limit},\ }\href
  {https://doi.org/10.1103/physrevlett.48.1559} {\bibfield  {journal} {\bibinfo
   {journal} {Phys. Rev. Lett.}\ }\textbf {\bibinfo {volume} {48}},\ \bibinfo
  {pages} {1559} (\bibinfo {year} {1982})}\BibitemShut {NoStop}%
\bibitem [{\citenamefont {Wen}\ and\ \citenamefont {Niu}(1990)}]{Wen1990}%
  \BibitemOpen
  \bibfield  {author} {\bibinfo {author} {\bibfnamefont {X.-G.}\ \bibnamefont
  {Wen}}\ and\ \bibinfo {author} {\bibfnamefont {Q.}~\bibnamefont {Niu}},\
  }\bibfield  {title} {\bibinfo {title} {Ground state degeneracy of the {FQH}
  states in presence of random potential and on high genus {R}iemann
  surfaces},\ }\href {https://doi.org/https://doi.org/10.1103/PhysRevB.41.9377}
  {\bibfield  {journal} {\bibinfo  {journal} {Phys. Rev. B}\ }\textbf {\bibinfo
  {volume} {41}},\ \bibinfo {pages} {9377} (\bibinfo {year}
  {1990})}\BibitemShut {NoStop}%
\bibitem [{\citenamefont {Stern}(2008)}]{Stern2008}%
  \BibitemOpen
  \bibfield  {author} {\bibinfo {author} {\bibfnamefont {A.}~\bibnamefont
  {Stern}},\ }\bibfield  {title} {\bibinfo {title} {Anyons and the quantum
  {H}all effect - {A} pedagogical review},\ }\href
  {https://doi.org/10.1016/j.aop.2007.10.008} {\bibfield  {journal} {\bibinfo
  {journal} {Ann. Phys. (N. Y.)}\ }\textbf {\bibinfo {volume} {323}},\ \bibinfo
  {pages} {204–249} (\bibinfo {year} {2008})}\BibitemShut {NoStop}%
\bibitem [{\citenamefont {Chakraborty}\ and\ \citenamefont
  {Pietlinen}(1995)}]{Chakraborty1995}%
  \BibitemOpen
  \bibfield  {author} {\bibinfo {author} {\bibfnamefont {T.}~\bibnamefont
  {Chakraborty}}\ and\ \bibinfo {author} {\bibfnamefont {P.}~\bibnamefont
  {Pietlinen}},\ }\href@noop {} {\emph {\bibinfo {title} {The Quantum {H}all
  Effects: Integral and Fractional}}},\ \bibinfo {edition} {2nd}\ ed.,\
  Springer Ser. in Solid-State Sciences\ (\bibinfo  {publisher} {Springer},\
  \bibinfo {address} {New York},\ \bibinfo {year} {1995})\BibitemShut {NoStop}%
\bibitem [{\citenamefont {Sarma}\ and\ \citenamefont
  {Pinczuk}(1997)}]{Sarma1997}%
  \BibitemOpen
  \bibfield  {author} {\bibinfo {author} {\bibfnamefont {S.~D.}\ \bibnamefont
  {Sarma}}\ and\ \bibinfo {author} {\bibfnamefont {A.}~\bibnamefont
  {Pinczuk}},\ }\href@noop {} {\emph {\bibinfo {title} {Perspectives in Quantum
  {H}all Effects : Novel Quantum Liquids in Low-dimensional Semiconductor
  Structures}}}\ (\bibinfo  {publisher} {Wiley},\ \bibinfo {address} {New
  York},\ \bibinfo {year} {1997})\BibitemShut {NoStop}%
\bibitem [{\citenamefont {Chen}\ \emph {et~al.}(2013)\citenamefont {Chen},
  \citenamefont {Gu}, \citenamefont {Liu},\ and\ \citenamefont
  {Wen}}]{Chen2013}%
  \BibitemOpen
  \bibfield  {author} {\bibinfo {author} {\bibfnamefont {X.}~\bibnamefont
  {Chen}}, \bibinfo {author} {\bibfnamefont {Z.-C.}\ \bibnamefont {Gu}},
  \bibinfo {author} {\bibfnamefont {Z.-X.}\ \bibnamefont {Liu}},\ and\ \bibinfo
  {author} {\bibfnamefont {X.-G.}\ \bibnamefont {Wen}},\ }\bibfield  {title}
  {\bibinfo {title} {Symmetry protected topological orders and the group
  cohomology of their symmetry group},\ }\href
  {https://doi.org/10.1103/physrevb.87.155114} {\bibfield  {journal} {\bibinfo
  {journal} {Phys. Rev. B}\ }\textbf {\bibinfo {volume} {87}},\ \bibinfo
  {pages} {155114} (\bibinfo {year} {2013})}\BibitemShut {NoStop}%
\bibitem [{\citenamefont {Chen}\ \emph {et~al.}(2010)\citenamefont {Chen},
  \citenamefont {Gu},\ and\ \citenamefont {Wen}}]{Chen2010}%
  \BibitemOpen
  \bibfield  {author} {\bibinfo {author} {\bibfnamefont {X.}~\bibnamefont
  {Chen}}, \bibinfo {author} {\bibfnamefont {Z.-C.}\ \bibnamefont {Gu}},\ and\
  \bibinfo {author} {\bibfnamefont {X.-G.}\ \bibnamefont {Wen}},\ }\bibfield
  {title} {\bibinfo {title} {Local unitary transformation, long-range quantum
  entanglement, wave function renormalization, and topological order},\ }\href
  {https://doi.org/10.1103/physrevb.82.155138} {\bibfield  {journal} {\bibinfo
  {journal} {Phys. Rev. B}\ }\textbf {\bibinfo {volume} {82}},\ \bibinfo
  {pages} {155138} (\bibinfo {year} {2010})}\BibitemShut {NoStop}%
\bibitem [{\citenamefont {Mesaros}\ and\ \citenamefont
  {Ran}(2013)}]{Mesaros2013}%
  \BibitemOpen
  \bibfield  {author} {\bibinfo {author} {\bibfnamefont {A.}~\bibnamefont
  {Mesaros}}\ and\ \bibinfo {author} {\bibfnamefont {Y.}~\bibnamefont {Ran}},\
  }\bibfield  {title} {\bibinfo {title} {Classification of symmetry enriched
  topological phases with exactly solvable models},\ }\href
  {https://doi.org/10.1103/physrevb.87.155115} {\bibfield  {journal} {\bibinfo
  {journal} {Phys. Rev. B}\ }\textbf {\bibinfo {volume} {87}},\ \bibinfo
  {pages} {155115} (\bibinfo {year} {2013})}\BibitemShut {NoStop}%
\bibitem [{\citenamefont {Kitaev}(2003)}]{Kitaev2003}%
  \BibitemOpen
  \bibfield  {author} {\bibinfo {author} {\bibfnamefont {A.~Y.}\ \bibnamefont
  {Kitaev}},\ }\bibfield  {title} {\bibinfo {title} {Fault-tolerant quantum
  computation by anyons},\ }\href
  {https://doi.org/10.1016/s0003-4916(02)00018-0} {\bibfield  {journal}
  {\bibinfo  {journal} {Ann. Phys. (N. Y.)}\ }\textbf {\bibinfo {volume}
  {303}},\ \bibinfo {pages} {2} (\bibinfo {year} {2003})}\BibitemShut {NoStop}%
\bibitem [{\citenamefont {Dennis}\ \emph {et~al.}(2002)\citenamefont {Dennis},
  \citenamefont {Kitaev}, \citenamefont {Landahl},\ and\ \citenamefont
  {Preskill}}]{Dennis2002}%
  \BibitemOpen
  \bibfield  {author} {\bibinfo {author} {\bibfnamefont {E.}~\bibnamefont
  {Dennis}}, \bibinfo {author} {\bibfnamefont {A.}~\bibnamefont {Kitaev}},
  \bibinfo {author} {\bibfnamefont {A.}~\bibnamefont {Landahl}},\ and\ \bibinfo
  {author} {\bibfnamefont {J.}~\bibnamefont {Preskill}},\ }\bibfield  {title}
  {\bibinfo {title} {Topological quantum memory},\ }\href
  {https://doi.org/10.1063/1.1499754} {\bibfield  {journal} {\bibinfo
  {journal} {J. Math. Phys.}\ }\textbf {\bibinfo {volume} {43}},\ \bibinfo
  {pages} {4452} (\bibinfo {year} {2002})}\BibitemShut {NoStop}%
\bibitem [{\citenamefont {Terhal}(2015)}]{Terhal2015}%
  \BibitemOpen
  \bibfield  {author} {\bibinfo {author} {\bibfnamefont {B.~M.}\ \bibnamefont
  {Terhal}},\ }\bibfield  {title} {\bibinfo {title} {Quantum error correction
  for quantum memories},\ }\bibfield  {journal} {\bibinfo  {journal} {Rev. Mod.
  Phys.}\ }\textbf {\bibinfo {volume} {87}},\ \href
  {https://doi.org/10.1103/revmodphys.87.307} {10.1103/revmodphys.87.307}
  (\bibinfo {year} {2015})\BibitemShut {NoStop}%
\bibitem [{\citenamefont {Brown}\ \emph {et~al.}(2016)\citenamefont {Brown},
  \citenamefont {Loss}, \citenamefont {Pachos}, \citenamefont {Self},\ and\
  \citenamefont {Wootton}}]{Brown2016}%
  \BibitemOpen
  \bibfield  {author} {\bibinfo {author} {\bibfnamefont {B.~J.}\ \bibnamefont
  {Brown}}, \bibinfo {author} {\bibfnamefont {D.}~\bibnamefont {Loss}},
  \bibinfo {author} {\bibfnamefont {J.~K.}\ \bibnamefont {Pachos}}, \bibinfo
  {author} {\bibfnamefont {C.~N.}\ \bibnamefont {Self}},\ and\ \bibinfo
  {author} {\bibfnamefont {J.~R.}\ \bibnamefont {Wootton}},\ }\bibfield
  {title} {\bibinfo {title} {Quantum memories at finite temperature},\
  }\bibfield  {journal} {\bibinfo  {journal} {Rev. Mod. Phys.}\ }\textbf
  {\bibinfo {volume} {88}},\ \href
  {https://doi.org/10.1103/revmodphys.88.045005} {10.1103/revmodphys.88.045005}
  (\bibinfo {year} {2016})\BibitemShut {NoStop}%
\bibitem [{\citenamefont {Bridgeman}\ \emph {et~al.}(2016)\citenamefont
  {Bridgeman}, \citenamefont {Flammia},\ and\ \citenamefont
  {Poulin}}]{Bridgeman2016}%
  \BibitemOpen
  \bibfield  {author} {\bibinfo {author} {\bibfnamefont {J.~C.}\ \bibnamefont
  {Bridgeman}}, \bibinfo {author} {\bibfnamefont {S.~T.}\ \bibnamefont
  {Flammia}},\ and\ \bibinfo {author} {\bibfnamefont {D.}~\bibnamefont
  {Poulin}},\ }\bibfield  {title} {\bibinfo {title} {Detecting topological
  order with ribbon operators},\ }\href
  {https://doi.org/10.1103/physrevb.94.205123} {\bibfield  {journal} {\bibinfo
  {journal} {Phys. Rev. B}\ }\textbf {\bibinfo {volume} {94}},\ \bibinfo
  {pages} {205123} (\bibinfo {year} {2016})}\BibitemShut {NoStop}%
\bibitem [{\citenamefont {Leinaas}\ and\ \citenamefont
  {Myrheim}(1977)}]{Leinaas1977}%
  \BibitemOpen
  \bibfield  {author} {\bibinfo {author} {\bibfnamefont {J.~M.}\ \bibnamefont
  {Leinaas}}\ and\ \bibinfo {author} {\bibfnamefont {J.}~\bibnamefont
  {Myrheim}},\ }\bibfield  {title} {\bibinfo {title} {On the theory of
  identical particles},\ }\href {https://doi.org/10.1007/bf02727953} {\bibfield
   {journal} {\bibinfo  {journal} {Nuovo Cim. B}\ }\textbf {\bibinfo {volume}
  {37}},\ \bibinfo {pages} {1} (\bibinfo {year} {1977})}\BibitemShut {NoStop}%
\bibitem [{\citenamefont {Wilczek}(1982)}]{Wilczek1982}%
  \BibitemOpen
  \bibfield  {author} {\bibinfo {author} {\bibfnamefont {F.}~\bibnamefont
  {Wilczek}},\ }\bibfield  {title} {\bibinfo {title} {Magnetic flux, angular
  momentum, and statistics},\ }\href
  {https://doi.org/10.1103/physrevlett.48.1144} {\bibfield  {journal} {\bibinfo
   {journal} {Phys. Rev. Lett.}\ }\textbf {\bibinfo {volume} {48}},\ \bibinfo
  {pages} {1144} (\bibinfo {year} {1982})}\BibitemShut {NoStop}%
\bibitem [{\citenamefont {Arovas}\ \emph {et~al.}(1984)\citenamefont {Arovas},
  \citenamefont {Schrieffer},\ and\ \citenamefont {Wilczek}}]{Arovas1984}%
  \BibitemOpen
  \bibfield  {author} {\bibinfo {author} {\bibfnamefont {D.}~\bibnamefont
  {Arovas}}, \bibinfo {author} {\bibfnamefont {J.~R.}\ \bibnamefont
  {Schrieffer}},\ and\ \bibinfo {author} {\bibfnamefont {F.}~\bibnamefont
  {Wilczek}},\ }\bibfield  {title} {\bibinfo {title} {Fractional statistics and
  the quantum {H}all effect},\ }\href
  {https://doi.org/10.1103/physrevlett.53.722} {\bibfield  {journal} {\bibinfo
  {journal} {Phys. Rev. Lett.}\ }\textbf {\bibinfo {volume} {53}},\ \bibinfo
  {pages} {722} (\bibinfo {year} {1984})}\BibitemShut {NoStop}%
\bibitem [{\citenamefont {Pachos}(2012)}]{Pachos2012}%
  \BibitemOpen
  \bibfield  {author} {\bibinfo {author} {\bibfnamefont {J.~K.}\ \bibnamefont
  {Pachos}},\ }\href {https://doi.org/10.1017/cbo9780511792908} {\emph
  {\bibinfo {title} {Introduction to Topological Quantum Computation}}}\
  (\bibinfo  {publisher} {Cambridge University Press},\ \bibinfo {address}
  {Cambridge},\ \bibinfo {year} {2012})\BibitemShut {NoStop}%
\bibitem [{\citenamefont {Rao}(1992)}]{Rao1992}%
  \BibitemOpen
  \bibfield  {author} {\bibinfo {author} {\bibfnamefont {S.}~\bibnamefont
  {Rao}},\ }\bibfield  {title} {\bibinfo {title} {An anyon primer},\
  }\href@noop {} {\bibfield  {journal} {\bibinfo  {journal}
  {arXiv:hep-th/9209066v3}\ } (\bibinfo {year} {1992})}\BibitemShut {NoStop}%
\bibitem [{\citenamefont {Kitaev}(2006)}]{Kitaev2006}%
  \BibitemOpen
  \bibfield  {author} {\bibinfo {author} {\bibfnamefont {A.~Y.}\ \bibnamefont
  {Kitaev}},\ }\bibfield  {title} {\bibinfo {title} {Anyons in an exactly
  solved model and beyond},\ }\bibfield  {journal} {\bibinfo  {journal} {Ann.
  Phys. (N. Y.)}\ }\textbf {\bibinfo {volume} {321}},\ \href
  {https://doi.org/10.1016/j.aop.2005.10.005} {10.1016/j.aop.2005.10.005}
  (\bibinfo {year} {2006})\BibitemShut {NoStop}%
\bibitem [{\citenamefont {Nayak}\ \emph {et~al.}(2008)\citenamefont {Nayak},
  \citenamefont {Simon}, \citenamefont {Stern}, \citenamefont {Freedman},\ and\
  \citenamefont {Sarma}}]{Nayak2008}%
  \BibitemOpen
  \bibfield  {author} {\bibinfo {author} {\bibfnamefont {C.}~\bibnamefont
  {Nayak}}, \bibinfo {author} {\bibfnamefont {S.~H.}\ \bibnamefont {Simon}},
  \bibinfo {author} {\bibfnamefont {A.}~\bibnamefont {Stern}}, \bibinfo
  {author} {\bibfnamefont {M.}~\bibnamefont {Freedman}},\ and\ \bibinfo
  {author} {\bibfnamefont {S.~D.}\ \bibnamefont {Sarma}},\ }\bibfield  {title}
  {\bibinfo {title} {Non-{A}belian anyons and topological quantum
  computation},\ }\bibfield  {journal} {\bibinfo  {journal} {Rev. Mod. Phys.}\
  }\textbf {\bibinfo {volume} {80}},\ \href
  {https://doi.org/10.1103/revmodphys.80.1083} {10.1103/revmodphys.80.1083}
  (\bibinfo {year} {2008})\BibitemShut {NoStop}%
\bibitem [{\citenamefont {Lahtinen}\ and\ \citenamefont
  {Pachos}(2017)}]{Lahtinen2017}%
  \BibitemOpen
  \bibfield  {author} {\bibinfo {author} {\bibfnamefont {V.~T.}\ \bibnamefont
  {Lahtinen}}\ and\ \bibinfo {author} {\bibfnamefont {J.~K.}\ \bibnamefont
  {Pachos}},\ }\bibfield  {title} {\bibinfo {title} {A short introduction to
  topological quantum computation},\ }\bibfield  {journal} {\bibinfo  {journal}
  {SciPost Phys.}\ }\textbf {\bibinfo {volume} {3}},\ \href
  {https://doi.org/10.21468/scipostphys.3.3.021} {10.21468/scipostphys.3.3.021}
  (\bibinfo {year} {2017})\BibitemShut {NoStop}%
\bibitem [{\citenamefont {Doplicher}\ \emph {et~al.}(1971)\citenamefont
  {Doplicher}, \citenamefont {Haag},\ and\ \citenamefont
  {Roberts}}]{Doplicher1971}%
  \BibitemOpen
  \bibfield  {author} {\bibinfo {author} {\bibfnamefont {S.}~\bibnamefont
  {Doplicher}}, \bibinfo {author} {\bibfnamefont {R.}~\bibnamefont {Haag}},\
  and\ \bibinfo {author} {\bibfnamefont {J.~E.}\ \bibnamefont {Roberts}},\
  }\bibfield  {title} {\bibinfo {title} {Local observables and particle
  statistics {I}},\ }\bibfield  {journal} {\bibinfo  {journal} {Commun. Math.
  Phys.}\ }\textbf {\bibinfo {volume} {23}},\ \href
  {https://doi.org/10.1007/bf01646454} {10.1007/bf01646454} (\bibinfo {year}
  {1971})\BibitemShut {NoStop}%
\bibitem [{\citenamefont {Doplicher}\ \emph {et~al.}(1974)\citenamefont
  {Doplicher}, \citenamefont {Haag},\ and\ \citenamefont
  {Roberts}}]{Doplicher1974}%
  \BibitemOpen
  \bibfield  {author} {\bibinfo {author} {\bibfnamefont {S.}~\bibnamefont
  {Doplicher}}, \bibinfo {author} {\bibfnamefont {R.}~\bibnamefont {Haag}},\
  and\ \bibinfo {author} {\bibfnamefont {J.~E.}\ \bibnamefont {Roberts}},\
  }\bibfield  {title} {\bibinfo {title} {Local observables and particle
  statistics {II}},\ }\bibfield  {journal} {\bibinfo  {journal} {Commun. Math.
  Phys.}\ }\textbf {\bibinfo {volume} {35}},\ \href
  {https://doi.org/10.1007/bf01646454} {10.1007/bf01646454} (\bibinfo {year}
  {1974})\BibitemShut {NoStop}%
\bibitem [{\citenamefont {Wang}\ and\ \citenamefont {Levin}(2014)}]{Wang2014}%
  \BibitemOpen
  \bibfield  {author} {\bibinfo {author} {\bibfnamefont {C.}~\bibnamefont
  {Wang}}\ and\ \bibinfo {author} {\bibfnamefont {M.}~\bibnamefont {Levin}},\
  }\bibfield  {title} {\bibinfo {title} {Braiding statistics of loop
  excitations in three dimensions},\ }\bibfield  {journal} {\bibinfo  {journal}
  {Phys. Rev. Lett.}\ }\textbf {\bibinfo {volume} {113}},\ \href
  {https://doi.org/10.1103/physrevlett.113.080403}
  {10.1103/physrevlett.113.080403} (\bibinfo {year} {2014})\BibitemShut
  {NoStop}%
\bibitem [{\citenamefont {Alford}\ \emph {et~al.}(1992)\citenamefont {Alford},
  \citenamefont {Lee}, \citenamefont {March-Russell},\ and\ \citenamefont
  {Preskill}}]{Alford1992}%
  \BibitemOpen
  \bibfield  {author} {\bibinfo {author} {\bibfnamefont {M.~G.}\ \bibnamefont
  {Alford}}, \bibinfo {author} {\bibfnamefont {K.-M.}\ \bibnamefont {Lee}},
  \bibinfo {author} {\bibfnamefont {J.}~\bibnamefont {March-Russell}},\ and\
  \bibinfo {author} {\bibfnamefont {J.}~\bibnamefont {Preskill}},\ }\bibfield
  {title} {\bibinfo {title} {Quantum field theory of non-abelian strings and
  vortices},\ }\bibfield  {journal} {\bibinfo  {journal} {Nucl. Phys. B}\
  }\textbf {\bibinfo {volume} {384}},\ \href
  {https://doi.org/10.1016/0550-3213(92)90468-q} {10.1016/0550-3213(92)90468-q}
  (\bibinfo {year} {1992})\BibitemShut {NoStop}%
\bibitem [{\citenamefont {Ehrenberg}\ and\ \citenamefont
  {Siday}(1949)}]{Ehrenberg1949}%
  \BibitemOpen
  \bibfield  {author} {\bibinfo {author} {\bibfnamefont {W.}~\bibnamefont
  {Ehrenberg}}\ and\ \bibinfo {author} {\bibfnamefont {R.~E.}\ \bibnamefont
  {Siday}},\ }\bibfield  {title} {\bibinfo {title} {The refractive index in
  electron optics and the principles of dynamics},\ }\href
  {https://doi.org/10.1088/0370-1301/62/1/303} {\bibfield  {journal} {\bibinfo
  {journal} {Proc. Phys. Soc. London, Sect. B}\ }\textbf {\bibinfo {volume}
  {62}},\ \bibinfo {pages} {8} (\bibinfo {year} {1949})}\BibitemShut {NoStop}%
\bibitem [{\citenamefont {Aharanov}\ and\ \citenamefont
  {Bohm}(1959)}]{Aharanov1959}%
  \BibitemOpen
  \bibfield  {author} {\bibinfo {author} {\bibfnamefont {Y.}~\bibnamefont
  {Aharanov}}\ and\ \bibinfo {author} {\bibfnamefont {D.}~\bibnamefont
  {Bohm}},\ }\bibfield  {title} {\bibinfo {title} {Significance of
  electromagnetic potentials in the quantum theory},\ }\bibfield  {journal}
  {\bibinfo  {journal} {Phys. Rev.}\ }\textbf {\bibinfo {volume} {115}},\ \href
  {https://doi.org/10.1103/physrev.115.485} {10.1103/physrev.115.485} (\bibinfo
  {year} {1959})\BibitemShut {NoStop}%
\bibitem [{\citenamefont {Lin}\ and\ \citenamefont {Levin}(2014)}]{Lin2014}%
  \BibitemOpen
  \bibfield  {author} {\bibinfo {author} {\bibfnamefont {C.-H.}\ \bibnamefont
  {Lin}}\ and\ \bibinfo {author} {\bibfnamefont {M.}~\bibnamefont {Levin}},\
  }\bibfield  {title} {\bibinfo {title} {Generalizations and limitations of
  string-net models},\ }\bibfield  {journal} {\bibinfo  {journal} {Phys. Rev.
  B}\ }\textbf {\bibinfo {volume} {89}},\ \href
  {https://doi.org/10.1103/physrevb.89.195130} {10.1103/physrevb.89.195130}
  (\bibinfo {year} {2014})\BibitemShut {NoStop}%
\bibitem [{\citenamefont {Lin}\ \emph {et~al.}(2021)\citenamefont {Lin},
  \citenamefont {Levin},\ and\ \citenamefont {Burnell}}]{Lin2021}%
  \BibitemOpen
  \bibfield  {author} {\bibinfo {author} {\bibfnamefont {C.-H.}\ \bibnamefont
  {Lin}}, \bibinfo {author} {\bibfnamefont {M.}~\bibnamefont {Levin}},\ and\
  \bibinfo {author} {\bibfnamefont {F.~J.}\ \bibnamefont {Burnell}},\
  }\bibfield  {title} {\bibinfo {title} {Generalized string-net models: A
  thorough exposition},\ }\bibfield  {journal} {\bibinfo  {journal} {Phys. Rev.
  B}\ }\textbf {\bibinfo {volume} {103}},\ \href
  {https://doi.org/10.1103/physrevb.103.195155} {10.1103/physrevb.103.195155}
  (\bibinfo {year} {2021})\BibitemShut {NoStop}%
\bibitem [{\citenamefont {Levin}\ and\ \citenamefont {Wen}(2005)}]{Levin2005}%
  \BibitemOpen
  \bibfield  {author} {\bibinfo {author} {\bibfnamefont {M.}~\bibnamefont
  {Levin}}\ and\ \bibinfo {author} {\bibfnamefont {X.-G.}\ \bibnamefont
  {Wen}},\ }\bibfield  {title} {\bibinfo {title} {String-net condensation: A
  physical mechanism for topological phases},\ }\href
  {https://doi.org/10.1103/physrevb.71.045110} {\bibfield  {journal} {\bibinfo
  {journal} {Phys. Rev. B}\ }\textbf {\bibinfo {volume} {71}},\ \bibinfo
  {pages} {045110} (\bibinfo {year} {2005})}\BibitemShut {NoStop}%
\bibitem [{\citenamefont {Cheng}\ \emph {et~al.}(2017)\citenamefont {Cheng},
  \citenamefont {Gu}, \citenamefont {Jiang},\ and\ \citenamefont
  {Qi}}]{Cheng2017}%
  \BibitemOpen
  \bibfield  {author} {\bibinfo {author} {\bibfnamefont {M.}~\bibnamefont
  {Cheng}}, \bibinfo {author} {\bibfnamefont {Z.-C.}\ \bibnamefont {Gu}},
  \bibinfo {author} {\bibfnamefont {S.}~\bibnamefont {Jiang}},\ and\ \bibinfo
  {author} {\bibfnamefont {Y.}~\bibnamefont {Qi}},\ }\bibfield  {title}
  {\bibinfo {title} {Exactly solvable models for symmetry-enriched topological
  phases},\ }\href {https://doi.org/10.1103/physrevb.96.115107} {\bibfield
  {journal} {\bibinfo  {journal} {Phys. Rev. B}\ }\textbf {\bibinfo {volume}
  {96}},\ \bibinfo {pages} {115107} (\bibinfo {year} {2017})}\BibitemShut
  {NoStop}%
\bibitem [{\citenamefont {Bais}\ \emph {et~al.}(1992)\citenamefont {Bais},
  \citenamefont {van Driel},\ and\ \citenamefont
  {de~Wild~Propitius}}]{Bais1992}%
  \BibitemOpen
  \bibfield  {author} {\bibinfo {author} {\bibfnamefont {F.~A.}\ \bibnamefont
  {Bais}}, \bibinfo {author} {\bibfnamefont {P.}~\bibnamefont {van Driel}},\
  and\ \bibinfo {author} {\bibfnamefont {M.}~\bibnamefont
  {de~Wild~Propitius}},\ }\bibfield  {title} {\bibinfo {title} {Quantum
  symmetries in discrete gauge theories},\ }\bibfield  {journal} {\bibinfo
  {journal} {Phys. Lett. B}\ }\textbf {\bibinfo {volume} {280}},\ \href
  {https://doi.org/10.1016/0370-2693(92)90773-w} {10.1016/0370-2693(92)90773-w}
  (\bibinfo {year} {1992})\BibitemShut {NoStop}%
\bibitem [{\citenamefont {de~Wild~Propitius}\ and\ \citenamefont
  {Bais}(1999)}]{WildPropitius1999}%
  \BibitemOpen
  \bibfield  {author} {\bibinfo {author} {\bibfnamefont {M.}~\bibnamefont
  {de~Wild~Propitius}}\ and\ \bibinfo {author} {\bibfnamefont {F.~A.}\
  \bibnamefont {Bais}},\ }\bibinfo {title} {Discrete gauge theories},\ in\
  \href {https://doi.org/https://doi.org/10.1007/978-1-4612-1410-6_8} {\emph
  {\bibinfo {booktitle} {Particles and Fields}}},\ \bibinfo {editor} {edited
  by\ \bibinfo {editor} {\bibfnamefont {G.}~\bibnamefont {Semenoff}}\ and\
  \bibinfo {editor} {\bibfnamefont {L.}~\bibnamefont {Vinet}}}\ (\bibinfo
  {publisher} {Springer New York},\ \bibinfo {address} {New York, NY},\
  \bibinfo {year} {1999})\ pp.\ \bibinfo {pages} {353--439}\BibitemShut
  {NoStop}%
\bibitem [{\citenamefont {Fowler}\ \emph {et~al.}(2012)\citenamefont {Fowler},
  \citenamefont {Mariantoni}, \citenamefont {Martinis},\ and\ \citenamefont
  {Cleland}}]{Fowler2012}%
  \BibitemOpen
  \bibfield  {author} {\bibinfo {author} {\bibfnamefont {A.~G.}\ \bibnamefont
  {Fowler}}, \bibinfo {author} {\bibfnamefont {M.}~\bibnamefont {Mariantoni}},
  \bibinfo {author} {\bibfnamefont {J.~M.}\ \bibnamefont {Martinis}},\ and\
  \bibinfo {author} {\bibfnamefont {A.~N.}\ \bibnamefont {Cleland}},\
  }\bibfield  {title} {\bibinfo {title} {Surface codes: Towards practical
  large-scale quantum computation},\ }\bibfield  {journal} {\bibinfo  {journal}
  {Phys. Rev. A}\ }\textbf {\bibinfo {volume} {86}},\ \href
  {https://doi.org/10.1103/physreva.86.032324} {10.1103/physreva.86.032324}
  (\bibinfo {year} {2012})\BibitemShut {NoStop}%
\bibitem [{\citenamefont {Andersen}\ \emph {et~al.}(2020)\citenamefont
  {Andersen}, \citenamefont {Remm}, \citenamefont {Lazar}, \citenamefont
  {Krinner}, \citenamefont {Lacroix}, \citenamefont {Norris}, \citenamefont
  {Gabureac}, \citenamefont {Eichler},\ and\ \citenamefont
  {Wallraff}}]{Andersen2020}%
  \BibitemOpen
  \bibfield  {author} {\bibinfo {author} {\bibfnamefont {C.~K.}\ \bibnamefont
  {Andersen}}, \bibinfo {author} {\bibfnamefont {A.}~\bibnamefont {Remm}},
  \bibinfo {author} {\bibfnamefont {S.}~\bibnamefont {Lazar}}, \bibinfo
  {author} {\bibfnamefont {S.}~\bibnamefont {Krinner}}, \bibinfo {author}
  {\bibfnamefont {N.}~\bibnamefont {Lacroix}}, \bibinfo {author} {\bibfnamefont
  {G.~J.}\ \bibnamefont {Norris}}, \bibinfo {author} {\bibfnamefont
  {M.}~\bibnamefont {Gabureac}}, \bibinfo {author} {\bibfnamefont
  {C.}~\bibnamefont {Eichler}},\ and\ \bibinfo {author} {\bibfnamefont
  {A.}~\bibnamefont {Wallraff}},\ }\bibfield  {title} {\bibinfo {title}
  {Repeated quantum error detection in a surface code},\ }\href
  {https://doi.org/10.1038/s41567-020-0920-y} {\bibfield  {journal} {\bibinfo
  {journal} {Nat. Phys.}\ }\textbf {\bibinfo {volume} {16}},\ \bibinfo {pages}
  {875–880} (\bibinfo {year} {2020})}\BibitemShut {NoStop}%
\bibitem [{\citenamefont {Satzinger}\ \emph {et~al.}(2021)\citenamefont
  {Satzinger}, \citenamefont {Liu}, \citenamefont {Smith} \emph
  {et~al.}}]{Satzinger2021ReducedScience}%
  \BibitemOpen
  \bibfield  {author} {\bibinfo {author} {\bibfnamefont {K.~J.}\ \bibnamefont
  {Satzinger}}, \bibinfo {author} {\bibfnamefont {Y.}~\bibnamefont {Liu}},
  \bibinfo {author} {\bibfnamefont {A.}~\bibnamefont {Smith}}, \emph {et~al.},\
  }\bibfield  {title} {\bibinfo {title} {Realizing topologically ordered states
  on a quantum processor},\ }\bibfield  {journal} {\bibinfo  {journal}
  {Science}\ }\textbf {\bibinfo {volume} {374}},\ \href
  {https://doi.org/10.1126/science.abi8378} {10.1126/science.abi8378} (\bibinfo
  {year} {2021})\BibitemShut {NoStop}%
\bibitem [{\citenamefont {Heinrich}\ \emph {et~al.}(2016)\citenamefont
  {Heinrich}, \citenamefont {Burnell}, \citenamefont {Fidkowski},\ and\
  \citenamefont {Levin}}]{Heinrich2016}%
  \BibitemOpen
  \bibfield  {author} {\bibinfo {author} {\bibfnamefont {C.}~\bibnamefont
  {Heinrich}}, \bibinfo {author} {\bibfnamefont {F.}~\bibnamefont {Burnell}},
  \bibinfo {author} {\bibfnamefont {L.}~\bibnamefont {Fidkowski}},\ and\
  \bibinfo {author} {\bibfnamefont {M.}~\bibnamefont {Levin}},\ }\bibfield
  {title} {\bibinfo {title} {Symmetry enriched string-nets: Exactly solvable
  models for {SET} phases},\ }\bibfield  {journal} {\bibinfo  {journal} {Phys.
  Rev. B}\ }\textbf {\bibinfo {volume} {94}},\ \href
  {https://doi.org/10.1103/physrevb.94.235136} {10.1103/physrevb.94.235136}
  (\bibinfo {year} {2016})\BibitemShut {NoStop}%
\bibitem [{\citenamefont {Hahn}\ and\ \citenamefont {Wolf}(2020)}]{Hahn2020}%
  \BibitemOpen
  \bibfield  {author} {\bibinfo {author} {\bibfnamefont {A.}~\bibnamefont
  {Hahn}}\ and\ \bibinfo {author} {\bibfnamefont {R.}~\bibnamefont {Wolf}},\
  }\bibfield  {title} {\bibinfo {title} {Generalized string-net model for
  unitary fusion categories without tetrahedral symmetry},\ }\bibfield
  {journal} {\bibinfo  {journal} {Phys. Rev. B}\ }\textbf {\bibinfo {volume}
  {102}},\ \href {https://doi.org/10.1103/physrevb.102.115154}
  {10.1103/physrevb.102.115154} (\bibinfo {year} {2020})\BibitemShut {NoStop}%
\bibitem [{\citenamefont {Lake}\ and\ \citenamefont {Wu}(2016)}]{Lake2016}%
  \BibitemOpen
  \bibfield  {author} {\bibinfo {author} {\bibfnamefont {E.}~\bibnamefont
  {Lake}}\ and\ \bibinfo {author} {\bibfnamefont {Y.-S.}\ \bibnamefont {Wu}},\
  }\bibfield  {title} {\bibinfo {title} {Signatures of broken parity and
  time-reversal symmetry in generalized string-net models},\ }\bibfield
  {journal} {\bibinfo  {journal} {Phys. Rev. B}\ }\textbf {\bibinfo {volume}
  {94}},\ \href {https://doi.org/10.1103/physrevb.94.115139}
  {10.1103/physrevb.94.115139} (\bibinfo {year} {2016})\BibitemShut {NoStop}%
\bibitem [{\citenamefont {Runkel}(2020)}]{Runkel2020}%
  \BibitemOpen
  \bibfield  {author} {\bibinfo {author} {\bibfnamefont {I.}~\bibnamefont
  {Runkel}},\ }\bibfield  {title} {\bibinfo {title} {String-net models for
  nonspherical pivotal fusion categories},\ }\href
  {https://doi.org/10.1142/s0218216520500352} {\bibfield  {journal} {\bibinfo
  {journal} {J. Knot Theory Ramif.}\ }\textbf {\bibinfo {volume} {29}},\
  \bibinfo {pages} {2050035} (\bibinfo {year} {2020})}\BibitemShut {NoStop}%
\bibitem [{\citenamefont {Hu}\ \emph {et~al.}(2012)\citenamefont {Hu},
  \citenamefont {Stirling},\ and\ \citenamefont {Wu}}]{Y.Hu2012}%
  \BibitemOpen
  \bibfield  {author} {\bibinfo {author} {\bibfnamefont {Y.}~\bibnamefont
  {Hu}}, \bibinfo {author} {\bibfnamefont {S.~D.}\ \bibnamefont {Stirling}},\
  and\ \bibinfo {author} {\bibfnamefont {Y.-S.}\ \bibnamefont {Wu}},\
  }\bibfield  {title} {\bibinfo {title} {Ground state degeneracy in the
  {L}evin-{W}en model for topological phases},\ }\bibfield  {journal} {\bibinfo
   {journal} {Phys. Rev. B}\ }\textbf {\bibinfo {volume} {85}},\ \href
  {https://doi.org/10.1103/physrevb.85.075107} {10.1103/physrevb.85.075107}
  (\bibinfo {year} {2012})\BibitemShut {NoStop}%
\bibitem [{\citenamefont {Bombin}\ and\ \citenamefont
  {Martin-Delgado}(2008)}]{Bombin2008}%
  \BibitemOpen
  \bibfield  {author} {\bibinfo {author} {\bibfnamefont {H.}~\bibnamefont
  {Bombin}}\ and\ \bibinfo {author} {\bibfnamefont {M.~A.}\ \bibnamefont
  {Martin-Delgado}},\ }\bibfield  {title} {\bibinfo {title} {A family of
  non-{A}belian {K}itaev models on a lattice: Topological condensation and
  confinement},\ }\bibfield  {journal} {\bibinfo  {journal} {Phys. Rev. B}\
  }\textbf {\bibinfo {volume} {78}},\ \href
  {https://doi.org/10.1103/physrevb.78.115421} {10.1103/physrevb.78.115421}
  (\bibinfo {year} {2008})\BibitemShut {NoStop}%
\bibitem [{\citenamefont {Wan}\ \emph {et~al.}(2015)\citenamefont {Wan},
  \citenamefont {Wang},\ and\ \citenamefont {He}}]{Wan2015}%
  \BibitemOpen
  \bibfield  {author} {\bibinfo {author} {\bibfnamefont {Y.}~\bibnamefont
  {Wan}}, \bibinfo {author} {\bibfnamefont {J.~C.}\ \bibnamefont {Wang}},\ and\
  \bibinfo {author} {\bibfnamefont {H.}~\bibnamefont {He}},\ }\bibfield
  {title} {\bibinfo {title} {Twisted gauge theory model of topological phases
  in three dimensions},\ }\bibfield  {journal} {\bibinfo  {journal} {Phys. Rev.
  B}\ }\textbf {\bibinfo {volume} {92}},\ \href
  {https://doi.org/10.1103/physrevb.92.045101} {10.1103/physrevb.92.045101}
  (\bibinfo {year} {2015})\BibitemShut {NoStop}%
\bibitem [{\citenamefont {Wang}\ and\ \citenamefont {Wen}(2015)}]{Wang2015}%
  \BibitemOpen
  \bibfield  {author} {\bibinfo {author} {\bibfnamefont {J.~C.}\ \bibnamefont
  {Wang}}\ and\ \bibinfo {author} {\bibfnamefont {X.-G.}\ \bibnamefont {Wen}},\
  }\bibfield  {title} {\bibinfo {title} {Non-{A}belian string and particle
  braiding in topological order: Modular {SL(3,Z)} representation and {3+1D}
  twisted gauge theory},\ }\href {https://doi.org/10.1103/PhysRevB.91.035134}
  {\bibfield  {journal} {\bibinfo  {journal} {Phys. Rev. B}\ }\textbf {\bibinfo
  {volume} {91}},\ \bibinfo {pages} {035134} (\bibinfo {year}
  {2015})}\BibitemShut {NoStop}%
\bibitem [{\citenamefont {Bullivant}\ and\ \citenamefont
  {Delcamp}(2019)}]{Bullivant2019}%
  \BibitemOpen
  \bibfield  {author} {\bibinfo {author} {\bibfnamefont {A.}~\bibnamefont
  {Bullivant}}\ and\ \bibinfo {author} {\bibfnamefont {C.}~\bibnamefont
  {Delcamp}},\ }\bibfield  {title} {\bibinfo {title} {Tube algebras,
  excitations statistics and compactification in gauge models of topological
  phases},\ }\bibfield  {journal} {\bibinfo  {journal} {J High Energy Phys}\
  }\textbf {\bibinfo {volume} {10}},\ \href
  {https://doi.org/10.1007/jhep10(2019)216} {10.1007/jhep10(2019)216} (\bibinfo
  {year} {2019})\BibitemShut {NoStop}%
\bibitem [{\citenamefont {Dijkgraaf}\ and\ \citenamefont
  {Witten}(1990)}]{Dijkgraaf1990}%
  \BibitemOpen
  \bibfield  {author} {\bibinfo {author} {\bibfnamefont {R.}~\bibnamefont
  {Dijkgraaf}}\ and\ \bibinfo {author} {\bibfnamefont {E.}~\bibnamefont
  {Witten}},\ }\bibfield  {title} {\bibinfo {title} {Topological gauge theories
  and group cohomology},\ }\bibfield  {journal} {\bibinfo  {journal} {Commun.
  Math. Phys.}\ }\textbf {\bibinfo {volume} {129}},\ \href
  {https://doi.org/10.1007/bf02096988} {10.1007/bf02096988} (\bibinfo {year}
  {1990})\BibitemShut {NoStop}%
\bibitem [{\citenamefont {Williamson}\ and\ \citenamefont
  {Wang}(2017)}]{Williamson2017}%
  \BibitemOpen
  \bibfield  {author} {\bibinfo {author} {\bibfnamefont {D.~J.}\ \bibnamefont
  {Williamson}}\ and\ \bibinfo {author} {\bibfnamefont {Z.}~\bibnamefont
  {Wang}},\ }\bibfield  {title} {\bibinfo {title} {Hamiltonian models for
  topological phases of matter in three spatial dimensions},\ }\bibfield
  {journal} {\bibinfo  {journal} {Ann. Phys. (N. Y.)}\ }\textbf {\bibinfo
  {volume} {377}},\ \href {https://doi.org/10.1016/j.aop.2016.12.018}
  {10.1016/j.aop.2016.12.018} (\bibinfo {year} {2017})\BibitemShut {NoStop}%
\bibitem [{\citenamefont {Walker}\ and\ \citenamefont
  {Wang}(2011)}]{Walker2012}%
  \BibitemOpen
  \bibfield  {author} {\bibinfo {author} {\bibfnamefont {K.}~\bibnamefont
  {Walker}}\ and\ \bibinfo {author} {\bibfnamefont {Z.}~\bibnamefont {Wang}},\
  }\bibfield  {title} {\bibinfo {title} {(3+1)-{TQFT}s and topological
  insulators},\ }\href
  {https://doi.org/https://doi.org/10.1007/s11467-011-0194-z} {\bibfield
  {journal} {\bibinfo  {journal} {Front. Phys.}\ }\textbf {\bibinfo {volume}
  {7}},\ \bibinfo {pages} {150} (\bibinfo {year} {2011})}\BibitemShut {NoStop}%
\bibitem [{\citenamefont {von Keyserlingk}\ \emph {et~al.}(2013)\citenamefont
  {von Keyserlingk}, \citenamefont {Burnell},\ and\ \citenamefont
  {Simon}}]{Keyserlingk2013}%
  \BibitemOpen
  \bibfield  {author} {\bibinfo {author} {\bibfnamefont {C.~W.}\ \bibnamefont
  {von Keyserlingk}}, \bibinfo {author} {\bibfnamefont {F.~J.}\ \bibnamefont
  {Burnell}},\ and\ \bibinfo {author} {\bibfnamefont {S.~H.}\ \bibnamefont
  {Simon}},\ }\bibfield  {title} {\bibinfo {title} {Three-dimensional
  topological lattice models with surface anyons},\ }\bibfield  {journal}
  {\bibinfo  {journal} {Phys. Rev. B}\ }\textbf {\bibinfo {volume} {87}},\
  \href {https://doi.org/10.1103/physrevb.87.045107}
  {10.1103/physrevb.87.045107} (\bibinfo {year} {2013})\BibitemShut {NoStop}%
\bibitem [{\citenamefont {Chen}\ \emph {et~al.}(2015)\citenamefont {Chen},
  \citenamefont {Burnell}, \citenamefont {Vishwanath},\ and\ \citenamefont
  {Fidkowski}}]{Chen2015}%
  \BibitemOpen
  \bibfield  {author} {\bibinfo {author} {\bibfnamefont {X.}~\bibnamefont
  {Chen}}, \bibinfo {author} {\bibfnamefont {F.~J.}\ \bibnamefont {Burnell}},
  \bibinfo {author} {\bibfnamefont {A.}~\bibnamefont {Vishwanath}},\ and\
  \bibinfo {author} {\bibfnamefont {L.}~\bibnamefont {Fidkowski}},\ }\bibfield
  {title} {\bibinfo {title} {Anomalous symmetry fractionalization and surface
  topological order},\ }\bibfield  {journal} {\bibinfo  {journal} {Phys. Rev.
  X}\ }\textbf {\bibinfo {volume} {5}},\ \href
  {https://doi.org/10.1103/physrevx.5.041013} {10.1103/physrevx.5.041013}
  (\bibinfo {year} {2015})\BibitemShut {NoStop}%
\bibitem [{\citenamefont {Wang}\ and\ \citenamefont {Chen}(2017)}]{Wang2017}%
  \BibitemOpen
  \bibfield  {author} {\bibinfo {author} {\bibfnamefont {Z.}~\bibnamefont
  {Wang}}\ and\ \bibinfo {author} {\bibfnamefont {X.}~\bibnamefont {Chen}},\
  }\bibfield  {title} {\bibinfo {title} {Twisted gauge theories in
  three-dimensional {Walker-Wang} models},\ }\bibfield  {journal} {\bibinfo
  {journal} {Phys. Rev. B}\ }\textbf {\bibinfo {volume} {95}},\ \href
  {https://doi.org/10.1103/physrevb.95.115142} {10.1103/physrevb.95.115142}
  (\bibinfo {year} {2017})\BibitemShut {NoStop}%
\bibitem [{\citenamefont {Bullivant}\ \emph {et~al.}(2017)\citenamefont
  {Bullivant}, \citenamefont {Calcada}, \citenamefont {Kadar}, \citenamefont
  {Martin},\ and\ \citenamefont {Martins}}]{Bullivant2017}%
  \BibitemOpen
  \bibfield  {author} {\bibinfo {author} {\bibfnamefont {A.}~\bibnamefont
  {Bullivant}}, \bibinfo {author} {\bibfnamefont {M.}~\bibnamefont {Calcada}},
  \bibinfo {author} {\bibfnamefont {Z.}~\bibnamefont {Kadar}}, \bibinfo
  {author} {\bibfnamefont {P.}~\bibnamefont {Martin}},\ and\ \bibinfo {author}
  {\bibfnamefont {J.~F.}\ \bibnamefont {Martins}},\ }\bibfield  {title}
  {\bibinfo {title} {Topological phases from higher gauge symmetry in 3+1{D}},\
  }\href {https://doi.org/https://doi.org/10.1103/PhysRevB.95.155118}
  {\bibfield  {journal} {\bibinfo  {journal} {Phys. Rev. B}\ }\textbf {\bibinfo
  {volume} {95}},\ \bibinfo {pages} {155118} (\bibinfo {year}
  {2017})}\BibitemShut {NoStop}%
\bibitem [{\citenamefont {Delcamp}\ and\ \citenamefont
  {Tiwari}(2018)}]{Delcamp2018}%
  \BibitemOpen
  \bibfield  {author} {\bibinfo {author} {\bibfnamefont {C.}~\bibnamefont
  {Delcamp}}\ and\ \bibinfo {author} {\bibfnamefont {A.}~\bibnamefont
  {Tiwari}},\ }\bibfield  {title} {\bibinfo {title} {From gauge to higher gauge
  models of topological phases},\ }\bibfield  {journal} {\bibinfo  {journal} {J
  High Energy Phys}\ }\textbf {\bibinfo {volume} {10}},\ \href
  {https://doi.org/10.1007/jhep10(2018)049} {10.1007/jhep10(2018)049} (\bibinfo
  {year} {2018})\BibitemShut {NoStop}%
\bibitem [{\citenamefont {Bullivant}\ and\ \citenamefont
  {Delcamp}(2020)}]{Bullivant2020}%
  \BibitemOpen
  \bibfield  {author} {\bibinfo {author} {\bibfnamefont {A.}~\bibnamefont
  {Bullivant}}\ and\ \bibinfo {author} {\bibfnamefont {C.}~\bibnamefont
  {Delcamp}},\ }\bibfield  {title} {\bibinfo {title} {Excitations in strict
  2-group higher gauge models of topological phases},\ }\bibfield  {journal}
  {\bibinfo  {journal} {J High Energy Phys}\ }\textbf {\bibinfo {volume}
  {01}},\ \href {https://doi.org/10.1007/jhep01(2020)107}
  {10.1007/jhep01(2020)107} (\bibinfo {year} {2020})\BibitemShut {NoStop}%
\bibitem [{\citenamefont {Bullivant}\ \emph {et~al.}(2020)\citenamefont
  {Bullivant}, \citenamefont {Calcada}, \citenamefont {Kadar}, \citenamefont
  {Martin},\ and\ \citenamefont {Faria~Martins}}]{Bullivant2020b}%
  \BibitemOpen
  \bibfield  {author} {\bibinfo {author} {\bibfnamefont {A.}~\bibnamefont
  {Bullivant}}, \bibinfo {author} {\bibfnamefont {M.}~\bibnamefont {Calcada}},
  \bibinfo {author} {\bibfnamefont {Z.}~\bibnamefont {Kadar}}, \bibinfo
  {author} {\bibfnamefont {P.}~\bibnamefont {Martin}},\ and\ \bibinfo {author}
  {\bibfnamefont {J.}~\bibnamefont {Faria~Martins}},\ }\bibfield  {title}
  {\bibinfo {title} {Higher lattices, discrete two-dimensional holonomy and
  topological phases in (3+1){D} with higher gauge symmetry},\ }\href
  {https://doi.org/https://doi.org/10.1142/S0129055X20500117} {\bibfield
  {journal} {\bibinfo  {journal} {Rev. Math. Phys.}\ }\textbf {\bibinfo
  {volume} {32}},\ \bibinfo {pages} {2050011} (\bibinfo {year}
  {2020})}\BibitemShut {NoStop}%
\bibitem [{\citenamefont {Yetter}(1993)}]{Yetter1993}%
  \BibitemOpen
  \bibfield  {author} {\bibinfo {author} {\bibfnamefont {D.}~\bibnamefont
  {Yetter}},\ }\bibfield  {title} {\bibinfo {title} {{TQFT's} from homotopy
  2-types},\ }\bibfield  {journal} {\bibinfo  {journal} {J. Knot Theory
  Ramif.}\ }\textbf {\bibinfo {volume} {2}},\ \href
  {https://doi.org/10.1142/s0218216593000076} {10.1142/s0218216593000076}
  (\bibinfo {year} {1993})\BibitemShut {NoStop}%
\bibitem [{\citenamefont {Else}\ and\ \citenamefont {Nayak}(2017)}]{Else2017}%
  \BibitemOpen
  \bibfield  {author} {\bibinfo {author} {\bibfnamefont {D.~V.}\ \bibnamefont
  {Else}}\ and\ \bibinfo {author} {\bibfnamefont {C.}~\bibnamefont {Nayak}},\
  }\bibfield  {title} {\bibinfo {title} {Cheshire charge in {(3+1)-D}
  topological phases},\ }\href
  {https://doi.org/https://doi.org/10.1103/PhysRevB.96.045136} {\bibfield
  {journal} {\bibinfo  {journal} {Phys. Rev. B}\ }\textbf {\bibinfo {volume}
  {96}},\ \bibinfo {pages} {045136} (\bibinfo {year} {2017})}\BibitemShut
  {NoStop}%
\bibitem [{\citenamefont {Jiang}\ \emph {et~al.}(2014)\citenamefont {Jiang},
  \citenamefont {Mesaros},\ and\ \citenamefont {Ran}}]{Jiang2014}%
  \BibitemOpen
  \bibfield  {author} {\bibinfo {author} {\bibfnamefont {S.}~\bibnamefont
  {Jiang}}, \bibinfo {author} {\bibfnamefont {A.}~\bibnamefont {Mesaros}},\
  and\ \bibinfo {author} {\bibfnamefont {Y.}~\bibnamefont {Ran}},\ }\bibfield
  {title} {\bibinfo {title} {Generalized modular transformations in 3+1{D}
  topologically ordered phases and triple linking invariant of loop braiding},\
  }\href {https://doi.org/https://doi.org/10.1103/PhysRevX.4.031048} {\bibfield
   {journal} {\bibinfo  {journal} {Phys. Rev. X}\ }\textbf {\bibinfo {volume}
  {4}},\ \bibinfo {pages} {031048} (\bibinfo {year} {2014})}\BibitemShut
  {NoStop}%
\bibitem [{\citenamefont {Pfeiffer}(2003)}]{Pfeiffer2003}%
  \BibitemOpen
  \bibfield  {author} {\bibinfo {author} {\bibfnamefont {H.}~\bibnamefont
  {Pfeiffer}},\ }\bibfield  {title} {\bibinfo {title} {Higher gauge theory and
  a non-{A}belian generalization of 2-form electrodynamics},\ }\href
  {https://doi.org/10.1016/s0003-4916(03)00147-7} {\bibfield  {journal}
  {\bibinfo  {journal} {Annals of Physics}\ }\textbf {\bibinfo {volume}
  {308}},\ \bibinfo {pages} {447} (\bibinfo {year} {2003})}\BibitemShut
  {NoStop}%
\bibitem [{\citenamefont {Baez}\ and\ \citenamefont {Huerta}(2010)}]{Baez2010}%
  \BibitemOpen
  \bibfield  {author} {\bibinfo {author} {\bibfnamefont {J.~C.}\ \bibnamefont
  {Baez}}\ and\ \bibinfo {author} {\bibfnamefont {J.}~\bibnamefont {Huerta}},\
  }\bibfield  {title} {\bibinfo {title} {An invitation to higher gauge
  theory},\ }\href {https://doi.org/10.1007/s10714-010-1070-9} {\bibfield
  {journal} {\bibinfo  {journal} {General Relativity and Gravitation}\ }\textbf
  {\bibinfo {volume} {43}},\ \bibinfo {pages} {2335} (\bibinfo {year}
  {2010})}\BibitemShut {NoStop}%
\bibitem [{\citenamefont {Gukov}\ and\ \citenamefont
  {Kapustin}(2013)}]{Gukov2013}%
  \BibitemOpen
  \bibfield  {author} {\bibinfo {author} {\bibfnamefont {S.}~\bibnamefont
  {Gukov}}\ and\ \bibinfo {author} {\bibfnamefont {A.}~\bibnamefont
  {Kapustin}},\ }\bibfield  {title} {\bibinfo {title} {Topological quantum
  field theory, nonlocal operators, and gapped phases of gauge theories},\
  }\href@noop {} {\bibfield  {journal} {\bibinfo  {journal} {arXiv:1307.4793}\
  } (\bibinfo {year} {2013})},\ \Eprint {https://arxiv.org/abs/1307.4793}
  {1307.4793} \BibitemShut {NoStop}%
\bibitem [{\citenamefont {Kapustin}\ and\ \citenamefont
  {Thorngren}(2014)}]{Kapustin2014}%
  \BibitemOpen
  \bibfield  {author} {\bibinfo {author} {\bibfnamefont {A.}~\bibnamefont
  {Kapustin}}\ and\ \bibinfo {author} {\bibfnamefont {R.}~\bibnamefont
  {Thorngren}},\ }\bibfield  {title} {\bibinfo {title} {Topological field
  theory on a lattice, discrete theta-angles and confinement},\ }\bibfield
  {journal} {\bibinfo  {journal} {Adv. Theor. Math. Phys.}\ }\textbf {\bibinfo
  {volume} {18}},\ \href {https://doi.org/10.4310/atmp.2014.v18.n5.a4}
  {10.4310/atmp.2014.v18.n5.a4} (\bibinfo {year} {2014}),\ \Eprint
  {https://arxiv.org/abs/1308.2926} {1308.2926} \BibitemShut {NoStop}%
\bibitem [{\citenamefont {Kapustin}\ and\ \citenamefont
  {Thorngren}(2017)}]{Kapustin2017}%
  \BibitemOpen
  \bibfield  {author} {\bibinfo {author} {\bibfnamefont {A.}~\bibnamefont
  {Kapustin}}\ and\ \bibinfo {author} {\bibfnamefont {R.}~\bibnamefont
  {Thorngren}},\ }\bibinfo {title} {Higher symmetry and gapped phases of gauge
  theories},\ in\ \href {https://doi.org/10.1007/978-3-319-59939-7_5} {\emph
  {\bibinfo {booktitle} {Algebra, Geometry, and Physics in the 21st Century:
  Kontsevich Festschrift}}},\ \bibinfo {editor} {edited by\ \bibinfo {editor}
  {\bibfnamefont {D.}~\bibnamefont {Auroux}}, \bibinfo {editor} {\bibfnamefont
  {L.}~\bibnamefont {Katzarkov}}, \bibinfo {editor} {\bibfnamefont
  {T.}~\bibnamefont {Pantev}}, \bibinfo {editor} {\bibfnamefont
  {Y.}~\bibnamefont {Soibelman}},\ and\ \bibinfo {editor} {\bibfnamefont
  {Y.}~\bibnamefont {Tschinkel}}}\ (\bibinfo  {publisher} {Springer
  International Publishing},\ \bibinfo {address} {Cham},\ \bibinfo {year}
  {2017})\ pp.\ \bibinfo {pages} {177--202},\ \Eprint
  {https://arxiv.org/abs/1309.4721} {1309.4721} \BibitemShut {NoStop}%
\bibitem [{\citenamefont {Bullivant}\ \emph {et~al.}(2019)\citenamefont
  {Bullivant}, \citenamefont {Martins},\ and\ \citenamefont
  {Martin}}]{Bullivant2018}%
  \BibitemOpen
  \bibfield  {author} {\bibinfo {author} {\bibfnamefont {A.}~\bibnamefont
  {Bullivant}}, \bibinfo {author} {\bibfnamefont {J.~F.}\ \bibnamefont
  {Martins}},\ and\ \bibinfo {author} {\bibfnamefont {P.}~\bibnamefont
  {Martin}},\ }\bibfield  {title} {\bibinfo {title} {Representations of the
  loop braid group and aharonov{\textendash}bohm like effects in discrete
  (3+1)-dimensional higher gauge theory},\ }\href
  {https://doi.org/10.4310/atmp.2019.v23.n7.a1} {\bibfield  {journal} {\bibinfo
   {journal} {Advances in Theoretical and Mathematical Physics}\ }\textbf
  {\bibinfo {volume} {23}},\ \bibinfo {pages} {1685} (\bibinfo {year}
  {2019})}\BibitemShut {NoStop}%
\bibitem [{\citenamefont {Baez}\ \emph {et~al.}(2007)\citenamefont {Baez},
  \citenamefont {Crans},\ and\ \citenamefont {Wise}}]{Baez2007a}%
  \BibitemOpen
  \bibfield  {author} {\bibinfo {author} {\bibfnamefont {J.~C.}\ \bibnamefont
  {Baez}}, \bibinfo {author} {\bibfnamefont {A.~S.}\ \bibnamefont {Crans}},\
  and\ \bibinfo {author} {\bibfnamefont {D.~K.}\ \bibnamefont {Wise}},\
  }\bibfield  {title} {\bibinfo {title} {Exotic statistics for strings in 4d
  {BF} theory},\ }\href
  {https://doi.org/https://dx.doi.org/10.4310/ATMP.2007.v11.n5.a1} {\bibfield
  {journal} {\bibinfo  {journal} {Advances in Theoretical and Mathematical
  Physics}\ }\textbf {\bibinfo {volume} {11}},\ \bibinfo {pages} {707}
  (\bibinfo {year} {2007})}\BibitemShut {NoStop}%
\bibitem [{\citenamefont {Damiani}(2017)}]{Damiani2017}%
  \BibitemOpen
  \bibfield  {author} {\bibinfo {author} {\bibfnamefont {C.}~\bibnamefont
  {Damiani}},\ }\bibfield  {title} {\bibinfo {title} {A journey through loop
  braid groups},\ }\bibfield  {journal} {\bibinfo  {journal} {Expo. Math.}\
  }\textbf {\bibinfo {volume} {35}},\ \href
  {https://doi.org/10.1016/j.exmath.2016.12.003} {10.1016/j.exmath.2016.12.003}
  (\bibinfo {year} {2017})\BibitemShut {NoStop}%
\bibitem [{\citenamefont {Huxford}\ and\ \citenamefont
  {Simon}(2022{\natexlab{a}})}]{HuxfordPaper2}%
  \BibitemOpen
  \bibfield  {author} {\bibinfo {author} {\bibfnamefont {J.}~\bibnamefont
  {Huxford}}\ and\ \bibinfo {author} {\bibfnamefont {S.~H.}\ \bibnamefont
  {Simon}},\ }\bibfield  {title} {\bibinfo {title} {Excitations in the higher
  lattice gauge theory model for topological phases {II}: the 2+1d case},\
  }\href@noop {} {\bibfield  {journal} {\bibinfo  {journal} {arXiv:2204.05341}\
  } (\bibinfo {year} {2022}{\natexlab{a}})}\BibitemShut {NoStop}%
\bibitem [{\citenamefont {Huxford}\ and\ \citenamefont
  {Simon}(2022{\natexlab{b}})}]{HuxfordPaper3}%
  \BibitemOpen
  \bibfield  {author} {\bibinfo {author} {\bibfnamefont {J.}~\bibnamefont
  {Huxford}}\ and\ \bibinfo {author} {\bibfnamefont {S.~H.}\ \bibnamefont
  {Simon}},\ }\bibfield  {title} {\bibinfo {title} {Excitations in the higher
  lattice gauge theory model for topological phases {III}: the 3+1d case},\
  }\href@noop {} {\bibfield  {journal} {\bibinfo  {journal} {arXiv:2206.09941}\
  } (\bibinfo {year} {2022}{\natexlab{b}})}\BibitemShut {NoStop}%
\bibitem [{\citenamefont {Montvay}\ and\ \citenamefont
  {Münster}(1994)}]{Montvay1994}%
  \BibitemOpen
  \bibfield  {author} {\bibinfo {author} {\bibfnamefont {I.}~\bibnamefont
  {Montvay}}\ and\ \bibinfo {author} {\bibfnamefont {G.}~\bibnamefont
  {Münster}},\ }\href {https://doi.org/10.1017/cbo9780511470783} {\emph
  {\bibinfo {title} {Quantum Fields on a Lattice}}},\ Cambridge Monographs on
  Mathematical Physics\ (\bibinfo  {publisher} {Cambridge University Press},\
  \bibinfo {address} {Cambridge},\ \bibinfo {year} {1994})\BibitemShut
  {NoStop}%
\bibitem [{\citenamefont {Wilson}(1964)}]{Wilson1964}%
  \BibitemOpen
  \bibfield  {author} {\bibinfo {author} {\bibfnamefont {K.~G.}\ \bibnamefont
  {Wilson}},\ }\bibfield  {title} {\bibinfo {title} {Confinement of quarks},\
  }\bibfield  {journal} {\bibinfo  {journal} {Phys. Rev. D}\ }\textbf {\bibinfo
  {volume} {10}},\ \href {https://doi.org/10.1103/physrevd.10.2445}
  {10.1103/physrevd.10.2445} (\bibinfo {year} {1964})\BibitemShut {NoStop}%
\bibitem [{\citenamefont {Durhuus}(1980)}]{Durhuus1980}%
  \BibitemOpen
  \bibfield  {author} {\bibinfo {author} {\bibfnamefont {B.}~\bibnamefont
  {Durhuus}},\ }\bibfield  {title} {\bibinfo {title} {On the structure of gauge
  invariant classical observables in lattice gauge theories},\ }\bibfield
  {journal} {\bibinfo  {journal} {Lett. Math. Phys.}\ }\textbf {\bibinfo
  {volume} {4}},\ \href {https://doi.org/10.1007/bf00943439}
  {10.1007/bf00943439} (\bibinfo {year} {1980})\BibitemShut {NoStop}%
\bibitem [{\citenamefont {Baez}(2002)}]{Baez2002}%
  \BibitemOpen
  \bibfield  {author} {\bibinfo {author} {\bibfnamefont {J.~C.}\ \bibnamefont
  {Baez}},\ }\bibfield  {title} {\bibinfo {title} {Higher {Yang–Mills}
  theory},\ }\href@noop {} {\bibfield  {journal} {\bibinfo  {journal}
  {arXiv:hep-th/0206130v2}\ } (\bibinfo {year} {2002})}\BibitemShut {NoStop}%
\bibitem [{\citenamefont {McCool}(1986)}]{McCool1986}%
  \BibitemOpen
  \bibfield  {author} {\bibinfo {author} {\bibfnamefont {J.}~\bibnamefont
  {McCool}},\ }\bibfield  {title} {\bibinfo {title} {On basis-conjugating
  automorphisms of free groups},\ }\bibfield  {journal} {\bibinfo  {journal}
  {Can. J. Math.}\ }\textbf {\bibinfo {volume} {XXXVIII}},\ \href
  {https://doi.org/10.4153/cjm-1986-073-3} {10.4153/cjm-1986-073-3} (\bibinfo
  {year} {1986})\BibitemShut {NoStop}%
\bibitem [{\citenamefont {Savushkina}(1996)}]{Savushkina1996}%
  \BibitemOpen
  \bibfield  {author} {\bibinfo {author} {\bibfnamefont {A.~G.}\ \bibnamefont
  {Savushkina}},\ }\bibfield  {title} {\bibinfo {title} {On the group of
  conjugating automorphisms of a free group},\ }\bibfield  {journal} {\bibinfo
  {journal} {Math Notes+}\ }\href {https://doi.org/10.1007/bf02308881}
  {10.1007/bf02308881} (\bibinfo {year} {1996})\BibitemShut {NoStop}%
\bibitem [{\citenamefont {Fenn}\ \emph {et~al.}(1997)\citenamefont {Fenn},
  \citenamefont {Rimanyi},\ and\ \citenamefont {Rourke}}]{Fenn1997}%
  \BibitemOpen
  \bibfield  {author} {\bibinfo {author} {\bibfnamefont {R.}~\bibnamefont
  {Fenn}}, \bibinfo {author} {\bibfnamefont {R.}~\bibnamefont {Rimanyi}},\ and\
  \bibinfo {author} {\bibfnamefont {C.}~\bibnamefont {Rourke}},\ }\bibfield
  {title} {\bibinfo {title} {The braid-permutation group},\ }\bibfield
  {journal} {\bibinfo  {journal} {Topology}\ }\textbf {\bibinfo {volume}
  {36}},\ \href {https://doi.org/10.1016/0040-9383(95)00072-0}
  {10.1016/0040-9383(95)00072-0} (\bibinfo {year} {1997})\BibitemShut {NoStop}%
\bibitem [{\citenamefont {Dahm}(1962)}]{Dahm1962}%
  \BibitemOpen
  \bibfield  {author} {\bibinfo {author} {\bibfnamefont {D.~M.}\ \bibnamefont
  {Dahm}},\ }\href@noop {} {\emph {\bibinfo {title} {A generalization of braid
  theory}}},\ \bibinfo {type} {Ph.D. Thesis}\ (\bibinfo  {institution}
  {Princeton University},\ \bibinfo {year} {1962})\BibitemShut {NoStop}%
\bibitem [{\citenamefont {Goldsmith}(1981)}]{Goldsmith1981}%
  \BibitemOpen
  \bibfield  {author} {\bibinfo {author} {\bibfnamefont {D.~L.}\ \bibnamefont
  {Goldsmith}},\ }\bibfield  {title} {\bibinfo {title} {The theory of motion
  groups},\ }\bibfield  {journal} {\bibinfo  {journal} {Michigan Math. J.}\
  }\textbf {\bibinfo {volume} {28}},\ \href
  {https://doi.org/10.1307/mmj/1029002454} {10.1307/mmj/1029002454} (\bibinfo
  {year} {1981})\BibitemShut {NoStop}%
\bibitem [{\citenamefont {Preskill}(2004)}]{Preskill2004}%
  \BibitemOpen
  \bibfield  {author} {\bibinfo {author} {\bibfnamefont {J.}~\bibnamefont
  {Preskill}},\ }\href
  {http://theory.caltech.edu/\~preskill/ph219/topological.pdf} {\bibinfo
  {title} {Lecture notes for physics 219: Quantum computation}} (\bibinfo
  {year} {2004}),\ \bibinfo {note} {accessed 08/07/2022}\BibitemShut {NoStop}%
\bibitem [{\citenamefont {Bais}(1980)}]{Bais1980}%
  \BibitemOpen
  \bibfield  {author} {\bibinfo {author} {\bibfnamefont {F.~A.}\ \bibnamefont
  {Bais}},\ }\bibfield  {title} {\bibinfo {title} {Flux metamorphosis},\
  }\bibfield  {journal} {\bibinfo  {journal} {Nucl. Phys. B}\ }\textbf
  {\bibinfo {volume} {170}},\ \href
  {https://doi.org/10.1016/0550-3213(80)90474-5} {10.1016/0550-3213(80)90474-5}
  (\bibinfo {year} {1980})\BibitemShut {NoStop}%
\bibitem [{\citenamefont {Bucher}\ \emph {et~al.}(1992)\citenamefont {Bucher},
  \citenamefont {Lee},\ and\ \citenamefont {Preskill}}]{Bucher1992}%
  \BibitemOpen
  \bibfield  {author} {\bibinfo {author} {\bibfnamefont {M.}~\bibnamefont
  {Bucher}}, \bibinfo {author} {\bibfnamefont {K.-M.}\ \bibnamefont {Lee}},\
  and\ \bibinfo {author} {\bibfnamefont {J.}~\bibnamefont {Preskill}},\
  }\bibfield  {title} {\bibinfo {title} {On detecting discrete {Cheshire}
  charge},\ }\bibfield  {journal} {\bibinfo  {journal} {Nucl. Phys. B}\
  }\textbf {\bibinfo {volume} {386}},\ \href
  {https://doi.org/10.1016/0550-3213(92)90174-a} {10.1016/0550-3213(92)90174-a}
  (\bibinfo {year} {1992})\BibitemShut {NoStop}%
\bibitem [{\citenamefont {Komar}\ and\ \citenamefont
  {Landon-Cardinal}(2017)}]{Komar2017}%
  \BibitemOpen
  \bibfield  {author} {\bibinfo {author} {\bibfnamefont {A.}~\bibnamefont
  {Komar}}\ and\ \bibinfo {author} {\bibfnamefont {O.}~\bibnamefont
  {Landon-Cardinal}},\ }\bibfield  {title} {\bibinfo {title} {Anyons are not
  energy eigenspaces of quantum double {Hamiltonians}},\ }\bibfield  {journal}
  {\bibinfo  {journal} {Phys. Rev. B}\ }\textbf {\bibinfo {volume} {96}},\
  \href {https://doi.org/10.1103/physrevb.96.195150}
  {10.1103/physrevb.96.195150} (\bibinfo {year} {2017})\BibitemShut {NoStop}%
\bibitem [{\citenamefont {Clifford}(1937)}]{Clifford1937}%
  \BibitemOpen
  \bibfield  {author} {\bibinfo {author} {\bibfnamefont {A.~H.}\ \bibnamefont
  {Clifford}},\ }\bibfield  {title} {\bibinfo {title} {Representations induced
  in an invariant subgroup},\ }\bibfield  {journal} {\bibinfo  {journal} {Ann.
  Math.}\ }\textbf {\bibinfo {volume} {38}},\ \href
  {https://doi.org/10.2307/1968599} {10.2307/1968599} (\bibinfo {year}
  {1937})\BibitemShut {NoStop}%
\bibitem [{\citenamefont {Bais}\ and\ \citenamefont
  {Slingerland}(2009)}]{Bais2009}%
  \BibitemOpen
  \bibfield  {author} {\bibinfo {author} {\bibfnamefont {F.~A.}\ \bibnamefont
  {Bais}}\ and\ \bibinfo {author} {\bibfnamefont {J.~K.}\ \bibnamefont
  {Slingerland}},\ }\bibfield  {title} {\bibinfo {title} {Condensate induced
  transitions between topologically ordered phases},\ }\bibfield  {journal}
  {\bibinfo  {journal} {Phys. Rev. B}\ }\textbf {\bibinfo {volume} {79}},\
  \href {https://doi.org/10.1103/physrevb.79.045316}
  {10.1103/physrevb.79.045316} (\bibinfo {year} {2009})\BibitemShut {NoStop}%
\bibitem [{\citenamefont {Burnell}(2018)}]{Burnell2018}%
  \BibitemOpen
  \bibfield  {author} {\bibinfo {author} {\bibfnamefont {F.~J.}\ \bibnamefont
  {Burnell}},\ }\bibfield  {title} {\bibinfo {title} {Anyon condensation and
  its applications},\ }\bibfield  {journal} {\bibinfo  {journal} {Annu. Rev.
  Condens. Matter Phys.}\ }\textbf {\bibinfo {volume} {9}},\ \href
  {https://doi.org/10.1146/annurev-conmatphys-033117-054154}
  {10.1146/annurev-conmatphys-033117-054154} (\bibinfo {year}
  {2018})\BibitemShut {NoStop}%
\bibitem [{\citenamefont {Neupert}\ \emph {et~al.}(2016)\citenamefont
  {Neupert}, \citenamefont {He}, \citenamefont {von Keyserlingk}, \citenamefont
  {Sierra},\ and\ \citenamefont {Bernevig}}]{Neupert2016}%
  \BibitemOpen
  \bibfield  {author} {\bibinfo {author} {\bibfnamefont {T.}~\bibnamefont
  {Neupert}}, \bibinfo {author} {\bibfnamefont {H.}~\bibnamefont {He}},
  \bibinfo {author} {\bibfnamefont {C.}~\bibnamefont {von Keyserlingk}},
  \bibinfo {author} {\bibfnamefont {G.}~\bibnamefont {Sierra}},\ and\ \bibinfo
  {author} {\bibfnamefont {B.~A.}\ \bibnamefont {Bernevig}},\ }\bibfield
  {title} {\bibinfo {title} {Boson condensation in topologically ordered
  quantum liquids},\ }\bibfield  {journal} {\bibinfo  {journal} {Phys. Rev. B}\
  }\textbf {\bibinfo {volume} {93}},\ \href
  {https://doi.org/10.1103/physrevb.93.115103} {10.1103/physrevb.93.115103}
  (\bibinfo {year} {2016})\BibitemShut {NoStop}%
\bibitem [{\citenamefont {Bais}\ \emph {et~al.}(2003)\citenamefont {Bais},
  \citenamefont {Schroers},\ and\ \citenamefont {Slingerland}}]{Bais2003}%
  \BibitemOpen
  \bibfield  {author} {\bibinfo {author} {\bibfnamefont {F.~A.}\ \bibnamefont
  {Bais}}, \bibinfo {author} {\bibfnamefont {B.~J.}\ \bibnamefont {Schroers}},\
  and\ \bibinfo {author} {\bibfnamefont {J.~K.}\ \bibnamefont {Slingerland}},\
  }\bibfield  {title} {\bibinfo {title} {Hopf symmetry breaking and confinement
  in (2+1)-dimensional gauge theory},\ }\bibfield  {journal} {\bibinfo
  {journal} {J High Energy Phys}\ }\textbf {\bibinfo {volume} {05}},\ \href
  {https://doi.org/10.1088/1126-6708/2003/05/068}
  {10.1088/1126-6708/2003/05/068} (\bibinfo {year} {2003})\BibitemShut
  {NoStop}%
\bibitem [{\citenamefont {Eli{\"e}ns}\ \emph {et~al.}(2014)\citenamefont
  {Eli{\"e}ns}, \citenamefont {Romers},\ and\ \citenamefont
  {Bais}}]{Eliens2014}%
  \BibitemOpen
  \bibfield  {author} {\bibinfo {author} {\bibfnamefont {I.~S.}\ \bibnamefont
  {Eli{\"e}ns}}, \bibinfo {author} {\bibfnamefont {J.~C.}\ \bibnamefont
  {Romers}},\ and\ \bibinfo {author} {\bibfnamefont {F.~A.}\ \bibnamefont
  {Bais}},\ }\bibfield  {title} {\bibinfo {title} {Diagrammatics for {Bose}
  condensation in anyon theories},\ }\bibfield  {journal} {\bibinfo  {journal}
  {Phys. Rev. B}\ }\textbf {\bibinfo {volume} {90}},\ \href
  {https://doi.org/10.1103/physrevb.90.195130} {10.1103/physrevb.90.195130}
  (\bibinfo {year} {2014})\BibitemShut {NoStop}%
\bibitem [{\citenamefont {Burnell}\ \emph {et~al.}(2013)\citenamefont
  {Burnell}, \citenamefont {von Keyserlingk},\ and\ \citenamefont
  {Simon}}]{Burnell2013}%
  \BibitemOpen
  \bibfield  {author} {\bibinfo {author} {\bibfnamefont {F.~J.}\ \bibnamefont
  {Burnell}}, \bibinfo {author} {\bibfnamefont {C.~W.}\ \bibnamefont {von
  Keyserlingk}},\ and\ \bibinfo {author} {\bibfnamefont {S.~H.}\ \bibnamefont
  {Simon}},\ }\bibfield  {title} {\bibinfo {title} {Phase transitions in
  three-dimensional topological lattice models with surface anyons},\
  }\bibfield  {journal} {\bibinfo  {journal} {Phys. Rev. B}\ }\textbf {\bibinfo
  {volume} {88}},\ \href {https://doi.org/10.1103/physrevb.88.235120}
  {10.1103/physrevb.88.235120} (\bibinfo {year} {2013})\BibitemShut {NoStop}%
\bibitem [{\citenamefont {Ye}\ \emph {et~al.}(2016)\citenamefont {Ye},
  \citenamefont {Hughes}, \citenamefont {Maciejko},\ and\ \citenamefont
  {Fradkin}}]{Ye2016}%
  \BibitemOpen
  \bibfield  {author} {\bibinfo {author} {\bibfnamefont {P.}~\bibnamefont
  {Ye}}, \bibinfo {author} {\bibfnamefont {T.~L.}\ \bibnamefont {Hughes}},
  \bibinfo {author} {\bibfnamefont {J.}~\bibnamefont {Maciejko}},\ and\
  \bibinfo {author} {\bibfnamefont {E.}~\bibnamefont {Fradkin}},\ }\bibfield
  {title} {\bibinfo {title} {Composite particle theory of three-dimensional
  gapped fermionic phases: Fractional topological insulators and charge-loop
  excitation symmetry},\ }\bibfield  {journal} {\bibinfo  {journal} {Phys. Rev.
  B}\ }\textbf {\bibinfo {volume} {94}},\ \href
  {https://doi.org/10.1103/physrevb.94.115104} {10.1103/physrevb.94.115104}
  (\bibinfo {year} {2016})\BibitemShut {NoStop}%
\bibitem [{\citenamefont {Atiyah}(1988)}]{Atiyah1988}%
  \BibitemOpen
  \bibfield  {author} {\bibinfo {author} {\bibfnamefont {M.~F.}\ \bibnamefont
  {Atiyah}},\ }\bibfield  {title} {\bibinfo {title} {Topological quantum field
  theory},\ }\bibfield  {journal} {\bibinfo  {journal} {Publ. Math. IHÉS}\
  }\textbf {\bibinfo {volume} {68}},\ \href
  {https://doi.org/10.1007/bf02698547} {10.1007/bf02698547} (\bibinfo {year}
  {1988})\BibitemShut {NoStop}%
\bibitem [{\citenamefont {Lan}\ \emph {et~al.}(2018)\citenamefont {Lan},
  \citenamefont {Kong},\ and\ \citenamefont {Wen}}]{Lan2018}%
  \BibitemOpen
  \bibfield  {author} {\bibinfo {author} {\bibfnamefont {T.}~\bibnamefont
  {Lan}}, \bibinfo {author} {\bibfnamefont {L.}~\bibnamefont {Kong}},\ and\
  \bibinfo {author} {\bibfnamefont {X.-G.}\ \bibnamefont {Wen}},\ }\bibfield
  {title} {\bibinfo {title} {Classification of {(3+1)D} bosonic topological
  orders: The case when pointlike excitations are all bosons},\ }\href
  {https://doi.org/https://doi.org/10.1103/PhysRevX.8.021074} {\bibfield
  {journal} {\bibinfo  {journal} {Phys. Rev. X}\ }\textbf {\bibinfo {volume}
  {8}},\ \bibinfo {pages} {021074} (\bibinfo {year} {2018})}\BibitemShut
  {NoStop}%
\bibitem [{\citenamefont {Koppen}\ \emph {et~al.}(2021)\citenamefont {Koppen},
  \citenamefont {Martins},\ and\ \citenamefont {Martin}}]{Koppen2021}%
  \BibitemOpen
  \bibfield  {author} {\bibinfo {author} {\bibfnamefont {V.}~\bibnamefont
  {Koppen}}, \bibinfo {author} {\bibfnamefont {J.~F.}\ \bibnamefont
  {Martins}},\ and\ \bibinfo {author} {\bibfnamefont {P.~P.}\ \bibnamefont
  {Martin}},\ }\bibfield  {title} {\bibinfo {title} {Exactly solvable models
  for 2+1{D} topological phases derived from crossed modules of semisimple
  {H}opf algebras},\ }\href@noop {} {\bibfield  {journal} {\bibinfo  {journal}
  {arXiv:2104.02766}\ } (\bibinfo {year} {2021})}\BibitemShut {NoStop}%
\end{thebibliography}%
\end{document}